\documentclass[traditabstract]{aa} 
\usepackage{graphicx}
\usepackage{txfonts}
\usepackage{natbib}
\bibpunct{(}{)}{;}{a}{,}

\newcommand\lesim{\lower.5ex\hbox{$\; \buildrel < \over \sim \;$}}
\newcommand\gesim{\lower.5ex\hbox{$\; \buildrel > \over \sim \;$}}
\newcommand\HI{H{\small I}}
\newcommand\HII{H{\small II}}
\newcommand\kms{km s$^{-1}$}
\newcommand\Ha{H$_{\alpha }$}
\newcommand\mjybeam{mJy beam$^{-1}$}
\newcommand\atomscmsq{atoms cm$^{-2}$}

\newcommand\magarcsecsq{mag arcsec$^{-2}$}

\begin{document}
   \title{The dark matter halo shape of edge-on disk galaxies}

\subtitle{III. Modelling the HI observations: results}

\titlerunning{The dark matter halo shape of edge-on disk galaxies III}
   \author{J.C. O'Brien
          \inst{1},
K.C. Freeman\inst{1}
          \and
          P.C. van der Kruit\inst{2}}

   \institute{Research School of Astronomy and Astrophysics, Australian National
University, Mount Stromlo Observatory, Cotter Road, ACT 2611, Australia\\
              \email{jesscobrien@gmail.com; kcf@mso.anu.edu.au$^\star$}
         \and
             Kapteyn Astronomical Institute, University of Groningen, P.O. Box 800,
9700 AV Groningen, the Netherlands\\
             \email{vdkruit@astro.rug.nl}\thanks{For 
correspondence contact Ken Freeman or Piet van der Kruit}
             }
\authorrunning{J.C. O'Brien et al.}

   \date{Received Xxxxxxx 00, 2010; accepted Xxxxxxx 00, 2010}

 
  \abstract
   {This is the third paper in a series in which we attempt to put
constraints on the flattening of dark halos in disk galaxies. We
observed for this purpose the \HI\ in edge-on galaxies, where it is in principle
possible to measure the force field in the halo vertically and radially
from gas layer flaring and rotation curve decomposition respectively.  
For this purpose we need to analyse the observed XV diagrams in such a
way as to accurately measure all three functions that describe the
planar kinematics and distribution of a galaxy: the radial \HI\ surface
density, the rotation curve and the \HI\ velocity dispersion. 
In this  paper, we first present the results of the modelling
of our \HI\ observations of 8 \HI\ rich, late-type, edge-on galaxies. We find that
in all of these we find differential rotation. Most systems display \HI\ velocity
dispersions of 6.5 to 7.5 km s$^{-1}$ and all except one show radial structure in 
this property. There is an increase in the mean \HI\ velocity dispersion with 
maximum rotation velocity, at least up to 120 km s$^{-1}$. 

Next we analyse the \HI\ observations to derive the radial variation of
the thickness of the \HI\ layer. The combination of these gas flaring 
measurements with the \HI\ kinematics measurements allow 
us to measure the total vertical force of each galaxy assuming
hydrostatic equilibrium. We find that with the exception of the 
asymmetric IC5052,
all of the galaxies in our sample are good candidates for 3D mass
modelling to measure the dark halo shape. The flaring profiles are symmetric
with respect to the galactic centres and have a common shape, increasing
linearly inside the stellar disks and exponential outside where the gravitational
potential is dominated by the dark halo. In the best example, UGC7321, we find
in the inner regions
small deviations from the midplane and accompanying increases in thickness of the
\HI\ layer that are  possibly a result of perturbations of the
gravitational field by a relatively strong bar.
}
   {}


   \maketitle
%

\section{Introduction}

In paper I in this series \citep{ofk2008a}
we presented \HI\ observations of a sample of 8
edge-on, \HI\ rich, late-type galaxies. The aim of the project has been
described there in detail. Briefly, we attempt to put 
constraints on the flattening of dark halos around disk galaxies by
measuring the force field of the halo vertically from the flaring of
the \HI\ layer and radially from rotation curve decomposition. For the
vertical force field we need to determine in these galaxies both the
velocity dispersion of the \HI\ gas (preferably as a function of
height from the central plane of the disk) and the thickness of the \HI\
layer, all of these as a function of galactocentric radius. In addition
we also need to extract information on the rotation curve of the galaxy
and the deprojected
\HI\ surface density, also as a function of galactocentric radius. 
To adequately constrain the halo shape these parameters needed to 
be measured out to as low surface densities as possible.Here
we present the determination of these properties from our \HI\ 
observations.


\begin{figure*}[t]
 \centering
\includegraphics[width=16cm]{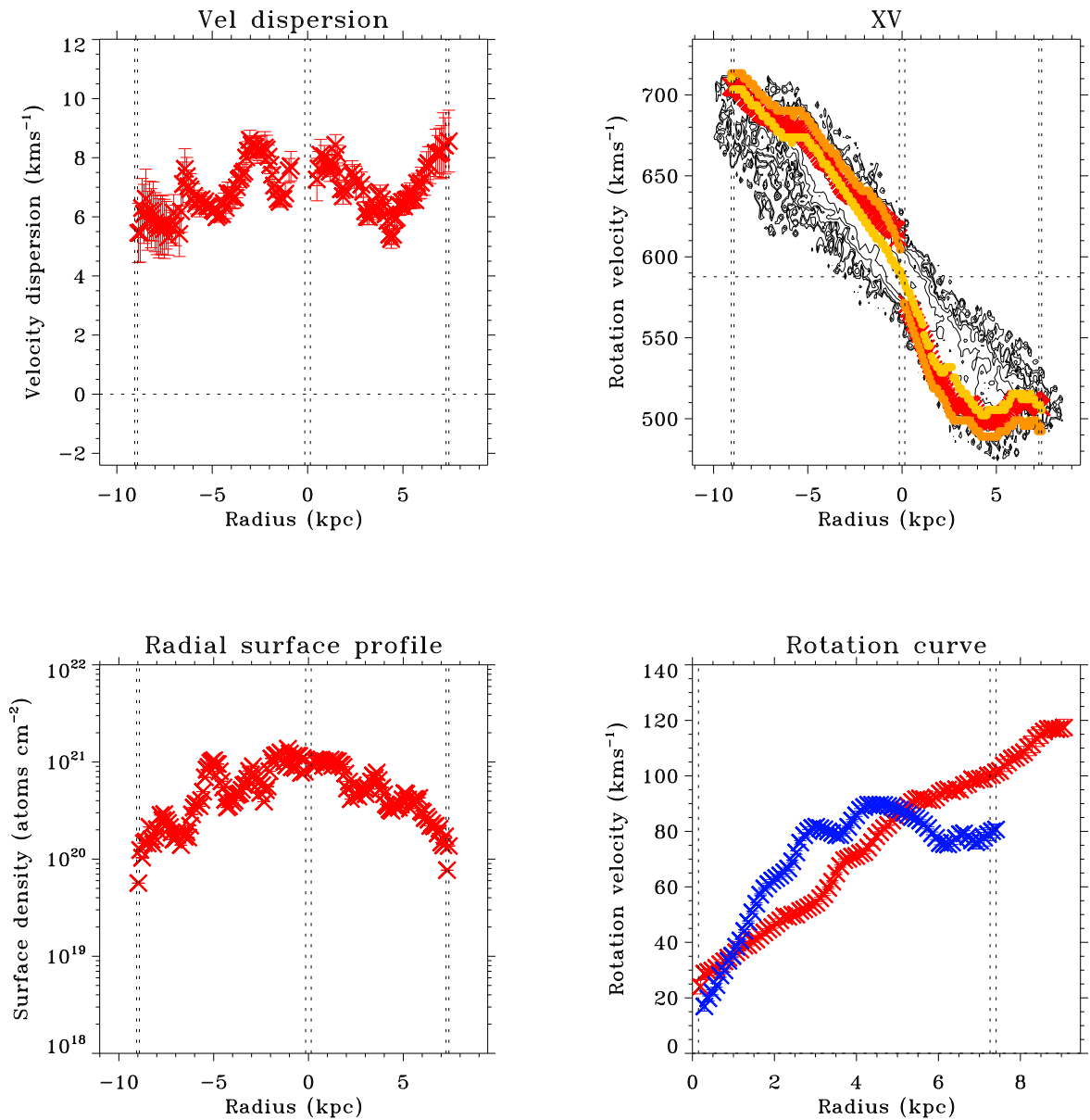}
  \caption[ESO074-G015 (IC5052): Rotation curve, \HI\ velocity
  dispersion and face-on \HI\ surface density]{ESO074-G015 (IC5052).
    This figure shows the results of the modelling of the kinematics.
The best fit measurements  for each kinematic function are shown in as (red)
crosses for the the \HI\ velocity dispersion in the top left-hand panel
and in the lower left-hand plot for the deprojected \HI\
    surface density. 
  The top right-hand panel shows the derived
    rotation curve (dark/red) overlaid on the observed XV diagram. Also
    shown on the XV diagram are the rotation fits obtained using the
    less accurate envelope tracing method (dark grey/orange) and the peak flux
    method (light grey/yellow). The latter always shows a lower rotation
    velocity than the other two fits.
 The lower right-hand plot displays the derived rotation
    curve with the receding and approaching sides separately 
(in the electronic version red and blue respectively). 
The FWHM synthesised
    beam has dimensions $292\times292$ pc.}
  \label{fig:ch5-e074-g015-kine}
\end{figure*}

 In order to resolve the vertical structure, we selected a sample of 
 nearby, \HI\ rich, late-type edge-on galaxies which are likely to exhibit 
 large \HI\ flaring expected from high dark matter fractions. From 
 paper I, we recall that the gradient of the force field in the vertical
direction for a Gaussian gas layer can be written as 
\begin{equation}
\label{eq:ch1-hydro}
\frac{\partial K_z}{\partial z} = -
\frac{\sigma_{v,g}^2}{({\rm FWHM}_{z,g}/2.35)^2},
\end{equation}
where $\sigma_{v,g}^2$ is the vertical velocity dispersion of the gas and
FWHM$_{z,g}$ the gas layer thickness. 
Using the flaring of the \HI\ distribution 
and the derived density distribution
of the gas and stellar distributions, it is possible to measure the
halo shape over the \HI\ extent of the luminous disk.  For
a given vertical gas velocity dispersion, the flaring will decrease
and the gas surface density will increase, for increasingly flattened
dark halos. Thus, in gas-rich late-type galaxies it is possible to
measure the force field of the halo vertically from the gas layer 
flaring and radially from rotation curve decomposition.

In the second paper \citep{ofk2008b}, paper II,
we presented a new method to accurately determine the
rotation curve, deprojected \HI\ surface density and velocity
dispersion at all radii in a edge-on gas disk.  The
superposition of velocity profiles from many radii in each sightline
through an edge-on \HI\ disk tends to cause an overestimate of the
velocity dispersion with most measurement methods (see paper II).
Measuring the radial flaring profile requires a model of the galaxy
rotation and face-on surface density; this also necessitates high
accuracy rotation curve measurement and \HI\ surface density
deprojection.

In this third paper we present the rotation curve, 
\HI\ velocity dispersion
and deprojected \HI\ surface density from the XV maps measured using the
iterative XV diagram modelling program (see paper II). 
Previously, measurement of these parameters in edge-on galaxies has been 
difficult. Indeed, the gas velocity dispersion has been measured
as a function of radius in only one edge-on galaxy NGC4244
\citep{olling1996a}. If the data have suffiently high S/N, then
our iterative XV modelling program is capable of deriving the
variation of rotation, velocity dispersion and density as a function
of both radius and height $z$ above the equatorial plane of these
edge-on galaxies. For the data described in paper I, the S/N is not
high enough to derive the variations in both $z$ and $R$. We use
the XV diagrams integrated over $z$, and measure the gas surface
density distribution and kinematics as a function of radius only,
assuming that the gas velocity dispersion is isotropic and vertically
isothermal.

\section{Results for kinematic modelling}
\label{sec:ch5-results}


\begin{figure*}[t]
  \centering
\includegraphics[width=16cm]{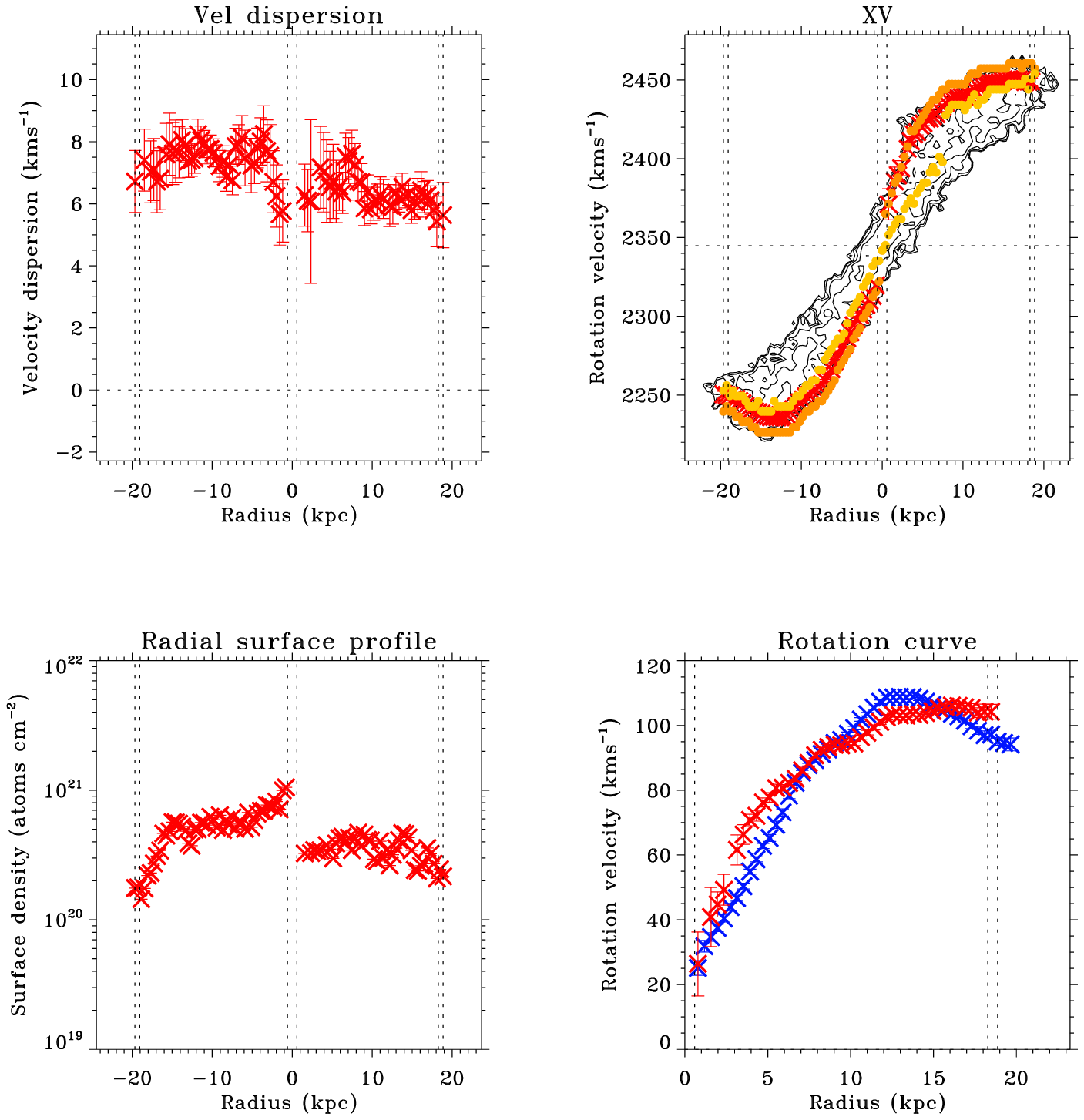}
  \caption[ESO109-G021 (IC5249): Rotation curve, \HI\ velocity
  dispersion and face-on \HI\ surface density]{ESO109-G021 (IC5249). 
This figure shows the results of the modelling of the kinematics. For an
explanation of the figure see the caption of Fig.~\ref{fig:ch5-e074-g015-kine}.
The FWHM synthesised beam has dimensions $1179\times1179$
    pc.}
  \label{fig:ch5-e109-g021-kine}
\end{figure*}

For each galaxy in our sample we present the best fit measurements of the
radial kinematics, and show the map of the residuals formed by comparing the
model XV diagram to the observed XV map. 
These XV maps were formed by integrating the observed \HI\ channel maps 
over the $z$ axis.

\subsection{ESO074-G015 (IC5052)}

Although ESO074-G015 is one of the nearest edge-on low mass galaxies, it is 
a galaxy less suitable for dark halo fitting because it
is asymmetric in both the \HI\ and optical distribution. Furthermore, it shows a
bright star formation region $2$ kpc west of the galactic center with
associated radio continuum emission, indicating a possible regional deviation from 
hydrostatic equilibrium. 
The measured rotation curve (see Fig.~\ref{fig:ch5-e074-g015-kine}) also shows a 
large asymmetry which is unlikely 
to be solely due to the \HI\ warp. Given the irregular
optical distribution, it is more likely to indicate a non-axisymmetric
mass distribution commonly found in Magellanic dwarf galaxies.  
The centre of mass was poorly determined in IC5052 due to the irregular 
structure of the XV distribution.

Despite the apparent mass asymmetry, the \HI\ velocity
dispersion is roughly constant at 6-7 \kms. Similarly the \HI\
radial profile is reasonably symmetric, although it displays a broad
plateau in the bright star formation region on the west side. As the
channel maps clearly show vertical flaring of the \HI\ layer, we will
proceed with the flaring analysis for this galaxy. However, the gross
kinematic asymmetry precludes it from use in the hydrostatic analysis
to determine the shape of the dark matter distribution.


\begin{figure*}[t]
  \centering
\includegraphics[width=16cm]{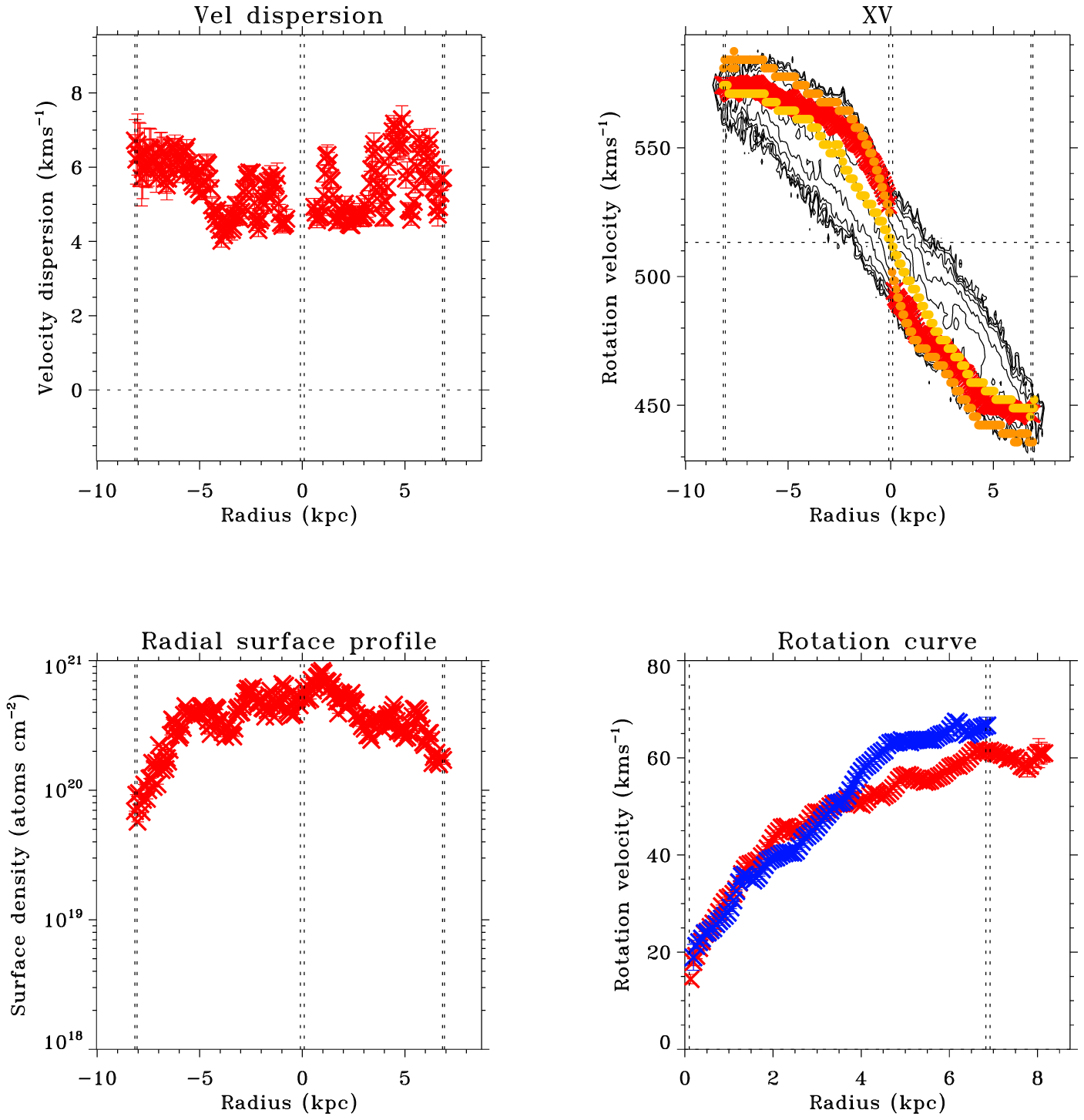}
  \caption[ESO115-G021: Rotation curve, \HI\ velocity
  dispersion and face-on \HI\ surface density]{ESO115-G021. 
This figure shows the results of the modelling of the kinematics. For an
explanation of the figure see the caption of Fig.~\ref{fig:ch5-e074-g015-kine}.
The FWHM synthesised beam has dimensions $194\times194$
    pc.}
  \label{fig:ch5-e115-g021-kine}
\end{figure*}


\begin{figure*}[t]
  \centering
\includegraphics[width=16cm]{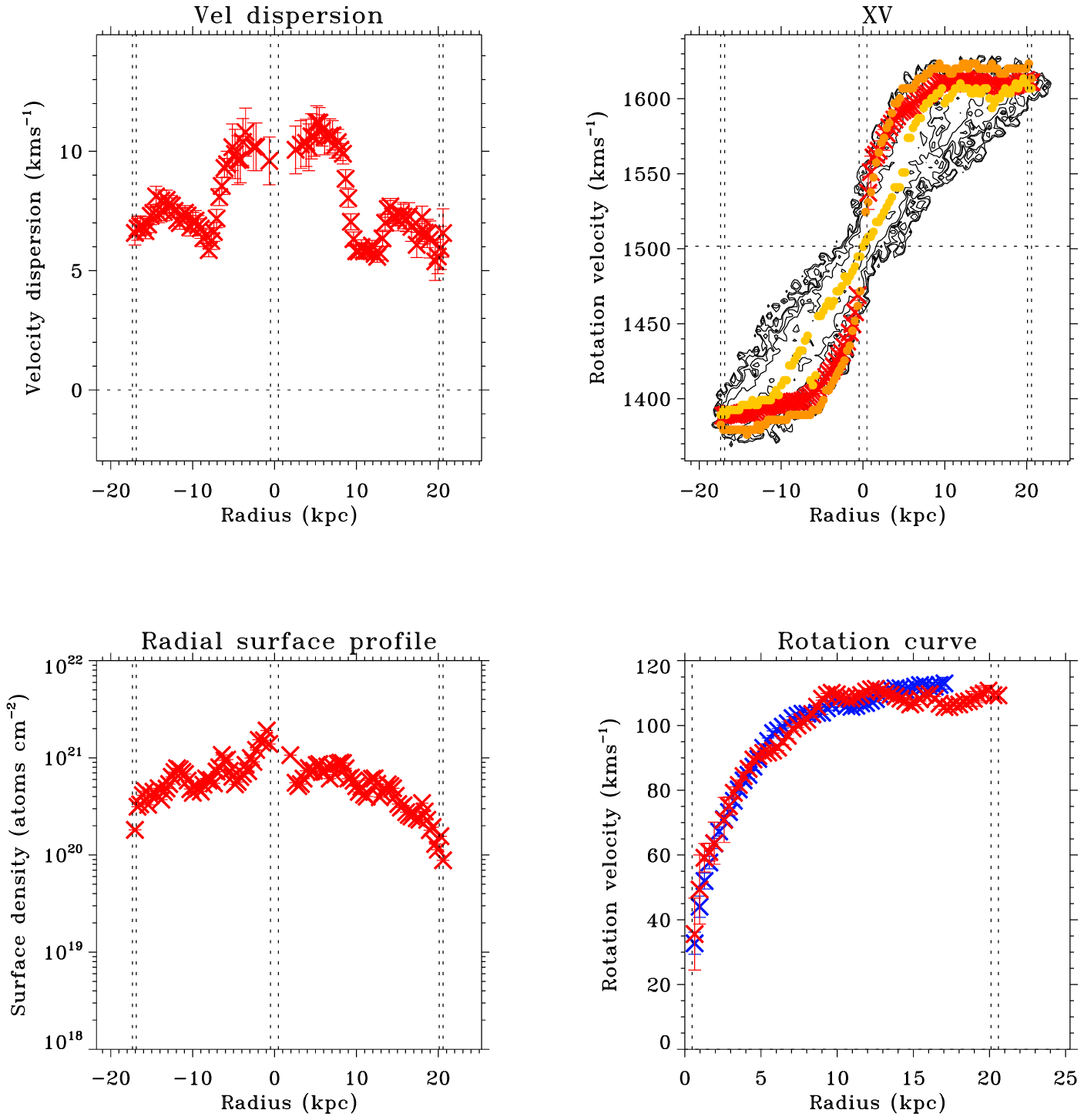}
  \caption[ESO138-G014: Rotation curve, \HI\ velocity
  dispersion and face-on \HI\ surface density]{ESO138-G014.
This figure shows the results of the modelling of the kinematics. For an
explanation of the figure see the caption of Fig.~\ref{fig:ch5-e074-g015-kine}.
The
    FWHM synthesised beam has dimensions $965\times965$ pc.}
  \label{fig:ch5-e138-g14-kine}
\end{figure*}

\begin{figure*}[t]
  \centering
\includegraphics[width=16cm]{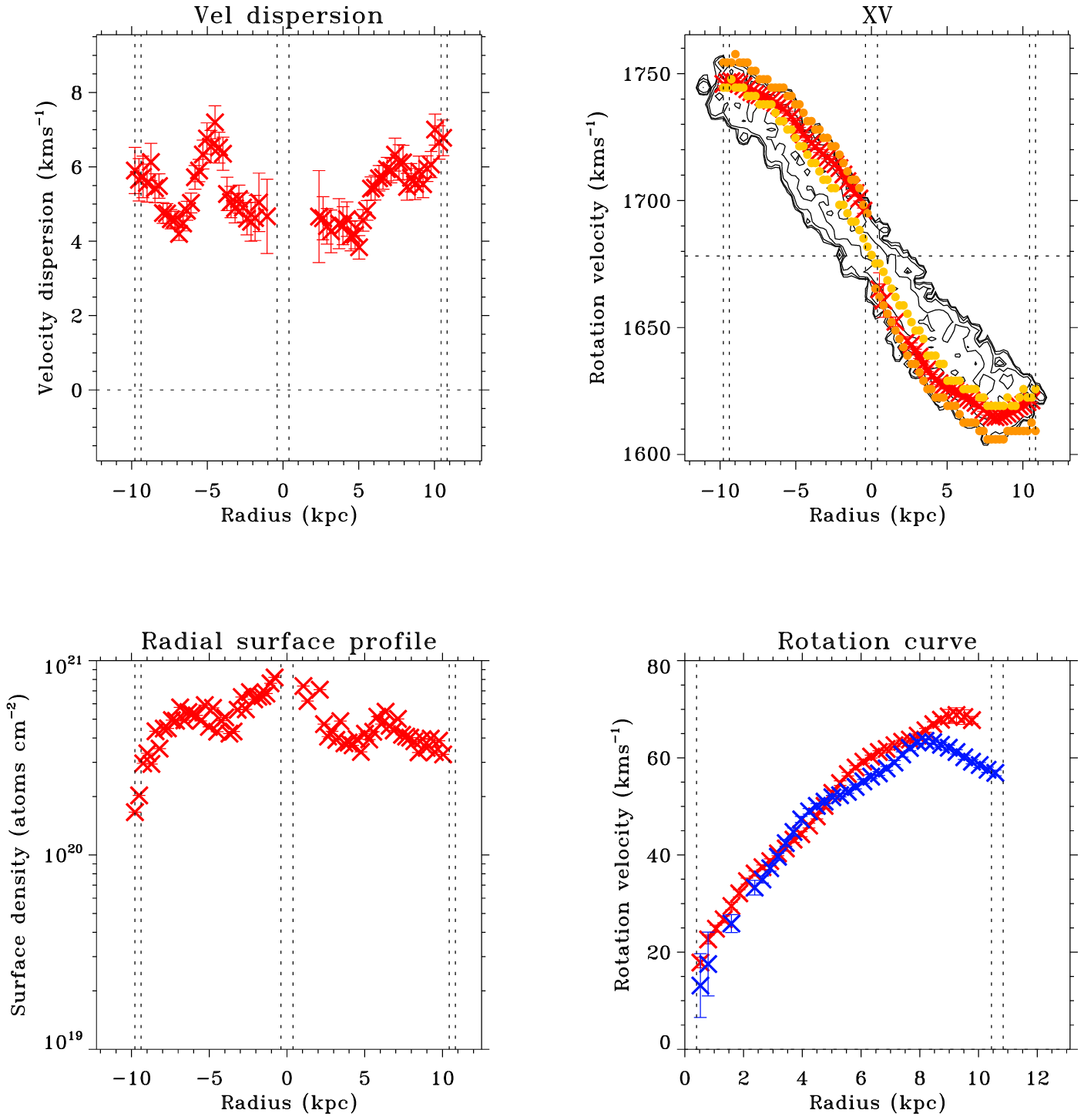}
  \caption[ESO146-G014: Rotation curve, \HI\ velocity
  dispersion and face-on \HI\ surface density]{ESO146-G014.
This figure shows the results of the modelling of the kinematics. For an
explanation of the figure see the caption of Fig.~\ref{fig:ch5-e074-g015-kine}.
The
    FWHM synthesised beam has dimensions $793\times793$ pc.}
  \label{fig:ch5-e146-g14-kine}
\end{figure*}

\subsection{ESO109-G021 (IC5249)}

ESO109-G021 (IC5249) is one of several well-known ``superthin''
galaxies, renowned for the large optical axial ratio.  The rotation
curve shown in Fig.~\ref{fig:ch5-e109-g021-kine} shows that
ESO109-G021 has a maximum rotation speed of $110$ \kms. ESO109-G021 is
the only one of the galaxies in our sample with rotation speeds
larger than $100$ \kms\ that does not show the ``figure-8'' signature
in the \HI\ distribution indicative of a possible bar.

The measured \HI\ radial surface density has a shallow gradient with
dips at around 11-12 kpc in agreement with the features in the
deprojected surface density obtained by \citet{vdkjvkf2001}. However
these authors also measured a central \HI\ depression, while we
find a small central peak when measuring from the left-hand side and
a plateaued central surface density when measuring from the right-hand
side. The deprojected surface density was measured after folding the
observed surface density calculated from the XV map about the galactic
center.

One possible explanation for the discrepancy between our measured
surface density and that of \citet{vdkjvkf2001}, is the lower
sensitivity of their observations. In addition to the $12$ hr 6.0C
ATCA observation used by all parties, we acquired an additional $30$
hr of \HI\ synthesis observations spread over the ATCA EW352, 750D and
6.0A array configurations. This resulted in a peak
signal-to-noise of $46.1$ and a synthesised beam FWHM$_{\theta}$ of
$8$\arcsec\ in the XV map, which appears to be a significantly higher sensitivity and spatial resolution than that acquired by \citet{vdkjvkf2001}. 

The measured rotation curve is reasonably symmetrical, in
agreement with that obtained by \citet{vdkjvkf2001} which was derived
by modelling the XV distribution using the
fixed velocity dispersion XV modelling method of
\citet[][see also Sect.2.5 in paper II]{vdkruit1981}. Like that of
\citet{vdkjvkf2001}, our measurement shows a steep inner
rise --- as opposed to the earlier less accurate measurements by
\citet{abeetal1999} which appear to have been measured from the peak
flux ridge of the XV map.

The measured \HI\ velocity dispersion varies from $5.5$ to $8.5$ \kms, falling 
to lower values on the right-hand side of the galaxy. The mean \HI\ velocity 
dispersion of $6.9\pm0.8$ \kms is in agreement with the best fit value of $7$ \kms\ 
found by \citet{vdkjvkf2001}.


\begin{figure*}[t]
  \centering
\includegraphics[width=16cm]{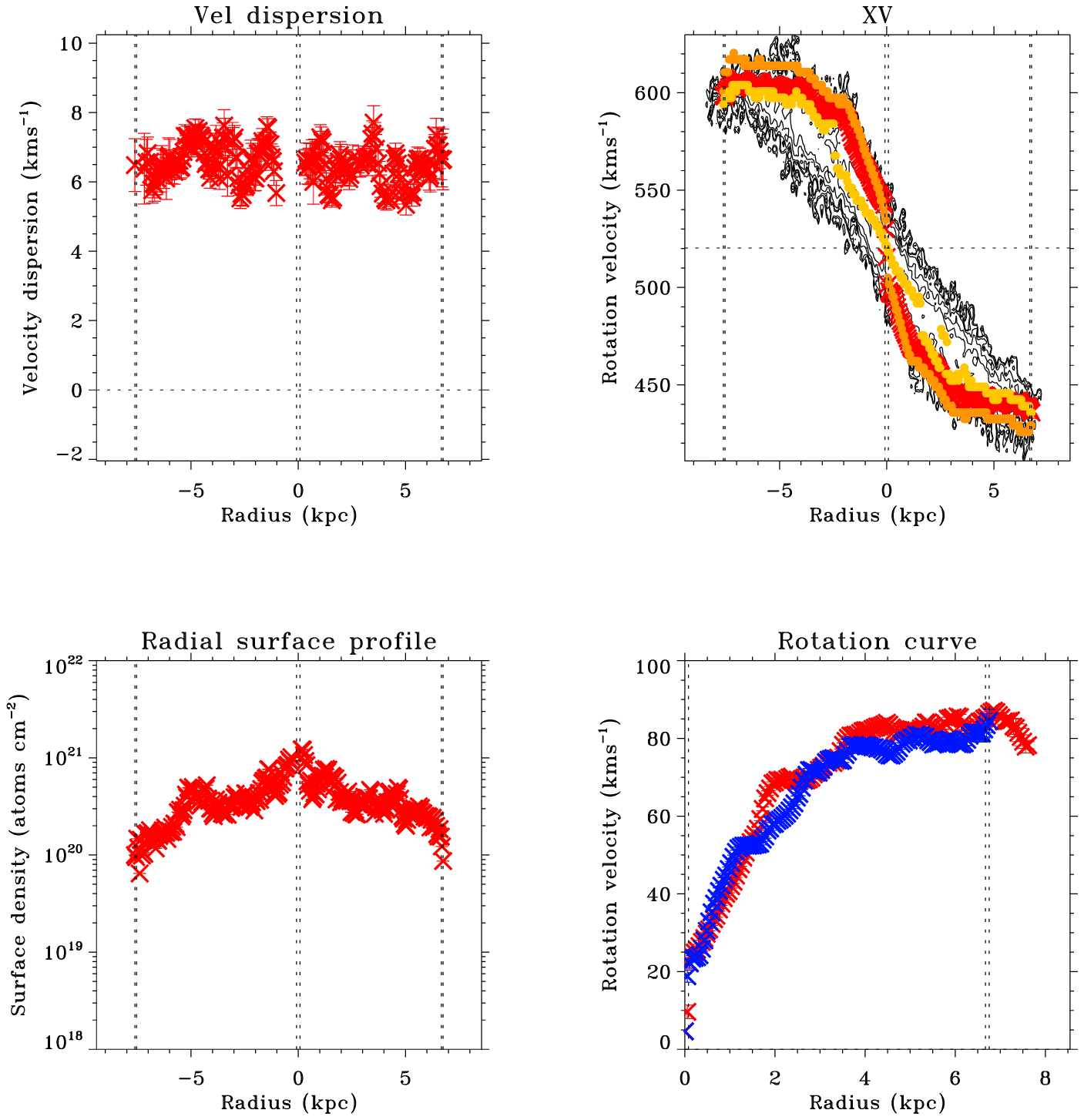}
  \caption[ESO274-G001: Rotation curve, \HI\ velocity
  dispersion and face-on \HI\ surface density]{ESO274-G001.
This figure shows the results of the modelling of the kinematics. For an
explanation of the figure see the caption of Fig.~\ref{fig:ch5-e074-g015-kine}.
The
    FWHM synthesised beam has dimensions $162\times162$ pc.}
  \label{fig:ch5-e274-g1-kine}
\end{figure*}
\begin{figure*}[t]
  \centering
\includegraphics[width=16cm]{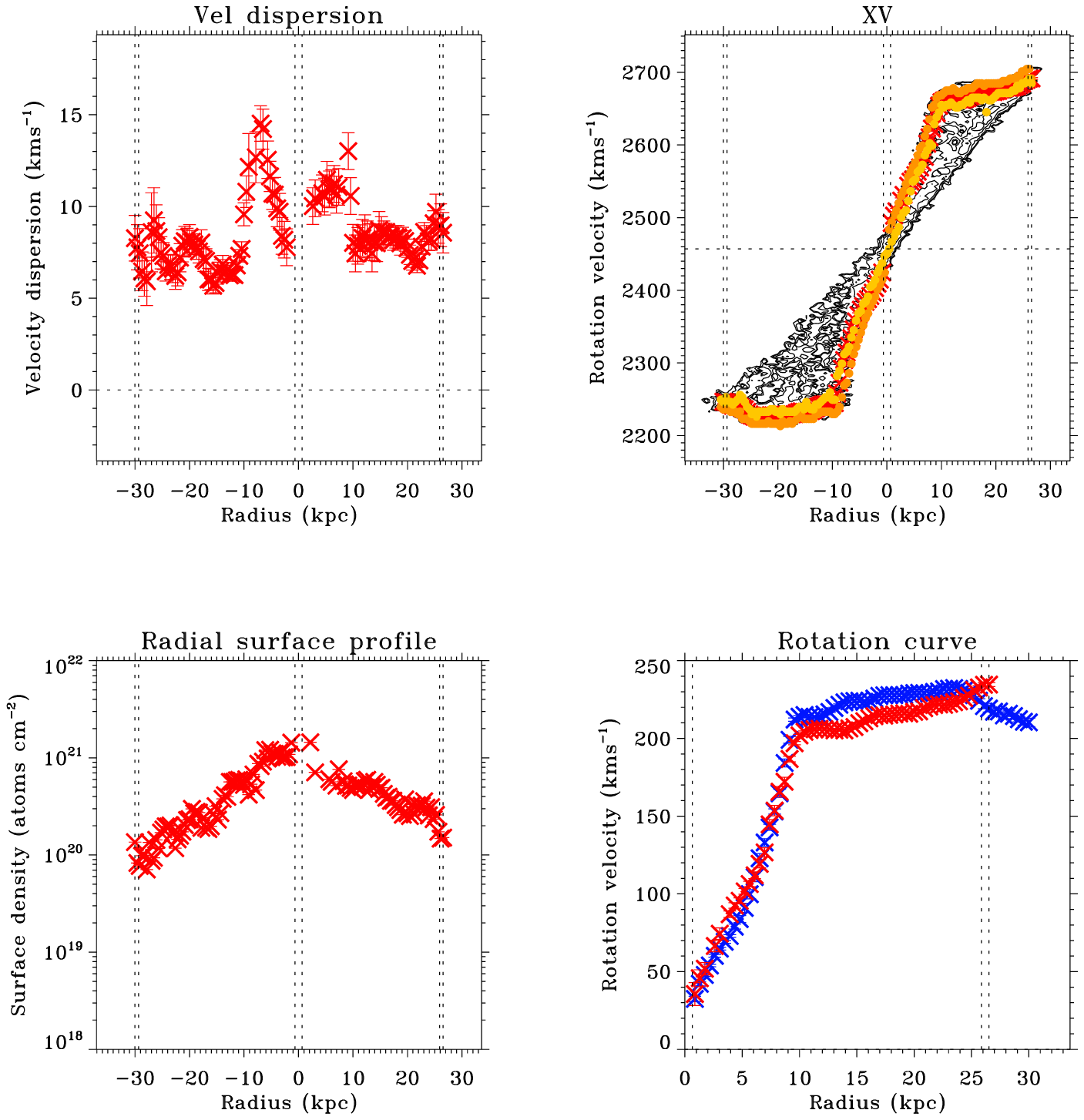}
  \caption[ESO435-G025 (IC2531): Rotation curve, \HI\ velocity
  dispersion and face-on \HI\ surface density]{ESO435-G025 (IC2531).
This figure shows the results of the modelling of the kinematics. For an
explanation of the figure see the caption of Fig.~\ref{fig:ch5-e074-g015-kine}.
The FWHM synthesised beam has dimensions $1305\times1305$
    pc.}
  \label{fig:ch5-e435-g025-kine}
\end{figure*}


\begin{subfigures}
\begin{figure*}[t]
  \centering
\includegraphics[width=16cm]{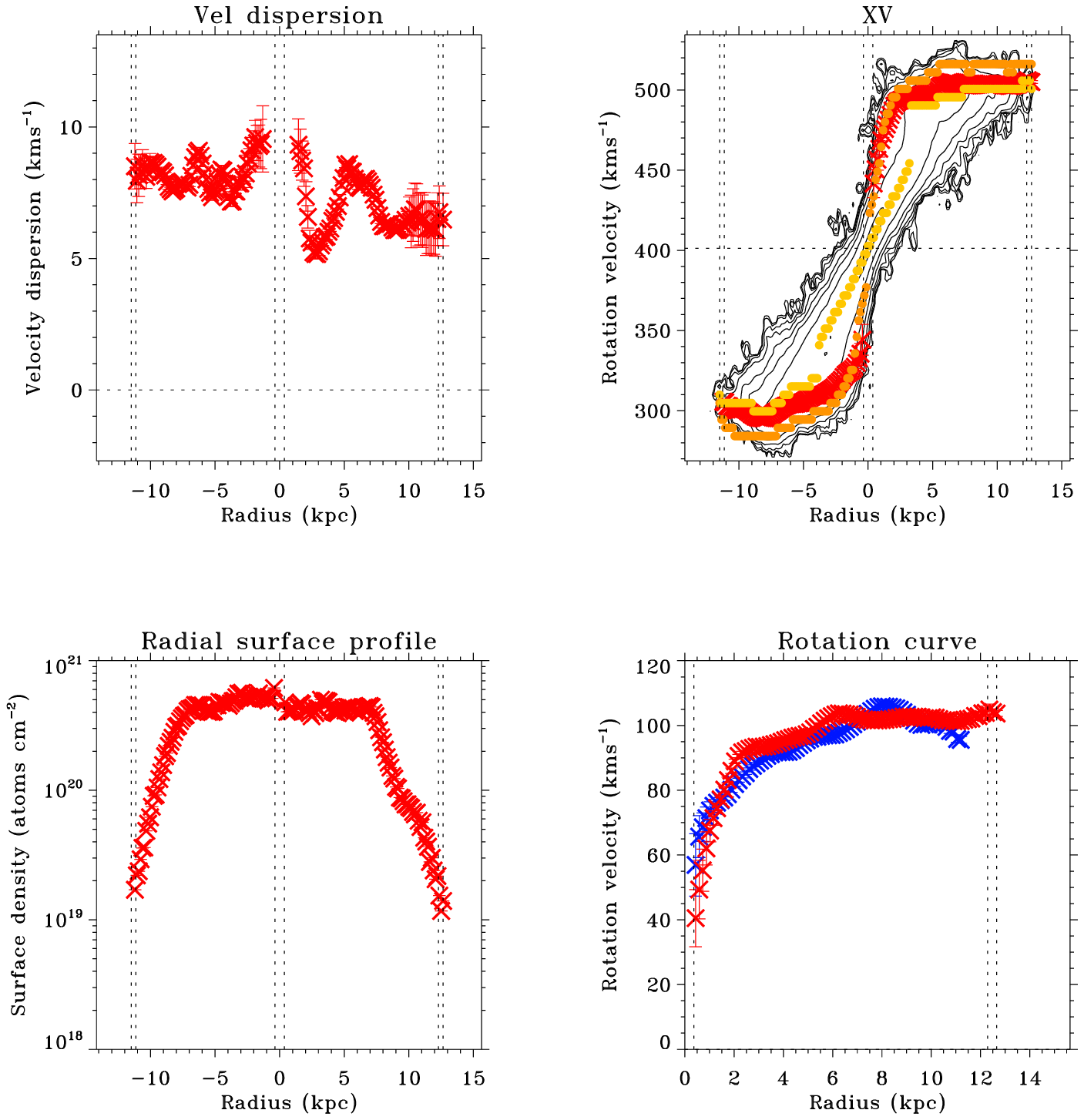}
  \caption[UGC7321: Rotation curve, \HI\ velocity
  dispersion and face-on \HI\ surface density]{UGC7321.
This figure shows the results of the modelling of the kinematics. For an
explanation of the figure see the caption of Fig.~\ref{fig:ch5-e074-g015-kine}.
The
    FWHM synthesised beam has dimensions $727\times727$ pc.}
  \label{fig:ch5-ugc7321-kine}
\end{figure*}

\setcounter{figure}{1}
\begin{figure*}[t]
  \centering
\includegraphics[width=16cm]{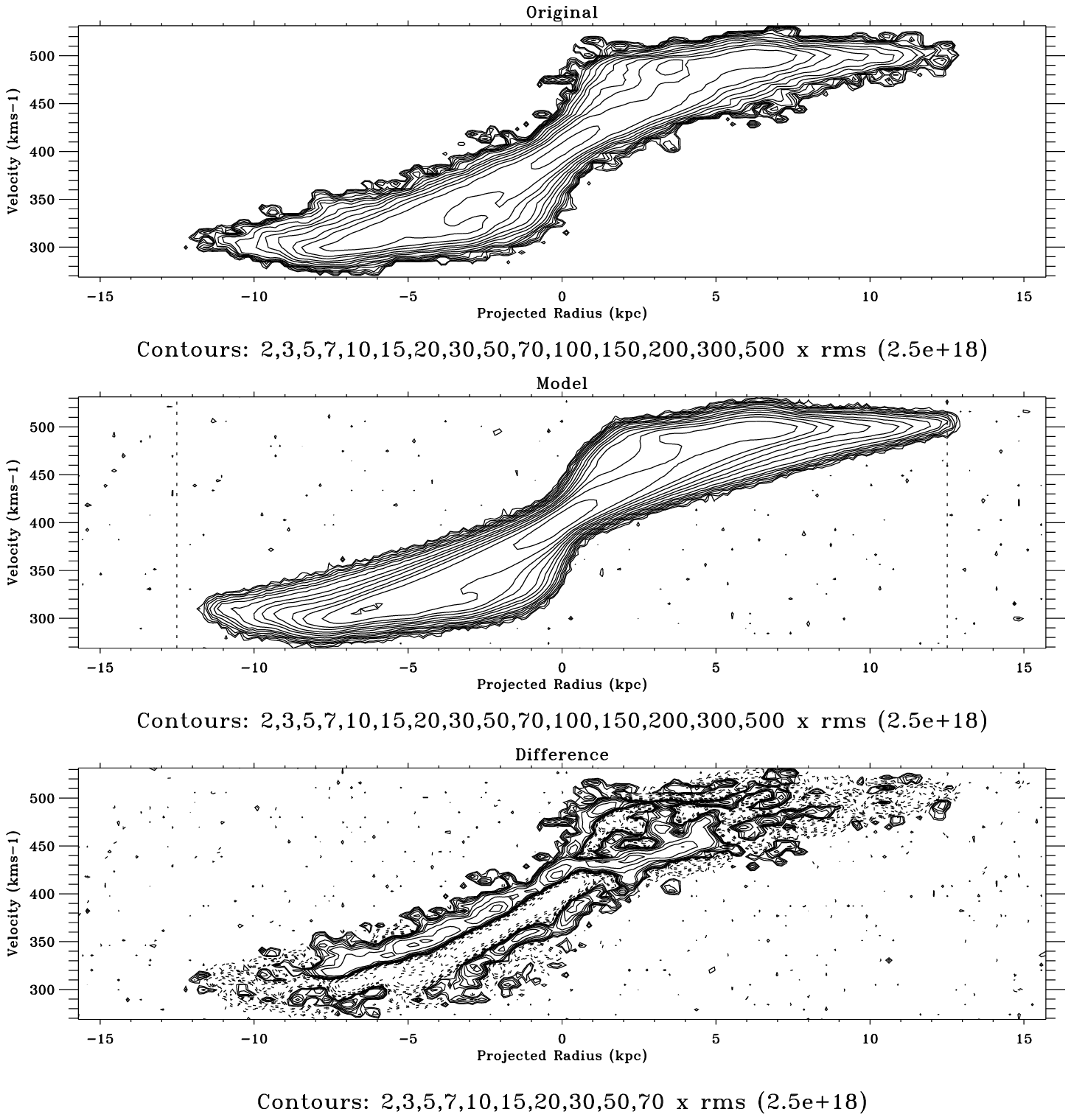}
  \caption[UGC7321: Observed XV, model XV formed
  from the measured kinematics, and residual XV map]{UGC7321.
This figure shows the observed XV diagram (upper panel),
    the model XV diagram formed from the best fit kinematics (middle panel), 
and the
    map with residuals between the two (lower panel).
The contours are in units of the rms
    noise, which is $\sigma = 0.524$ \mjybeam.  The FWHM
    of the \HI\ synthesised beam is $727$ pc. To the left of the
    galactic centre the residual map clearly shows the gas traced by
   the orbits in the bar. This is less clearly seen on the right-hand side
 due to
    the small warp.  If one excludes the gas in the bar,
    the fitted kinematics depict a very good representation of
    the observed XV map.}
  \label{fig:ch5-ugc7321-xvmodel}
\end{figure*}
\end{subfigures}

\subsection{ESO115-G021}

The high resolution \HI\ imaging of ESO115-G21 yields an almost solid-body
rotation curve flattening out at a radius of 5-6 kpc at
a peak rotation speed of approximately $65$ \kms\ (see
Fig.~\ref{fig:ch5-e115-g021-kine}).  Like the other small galaxies
in our sample it shows steep inner rotation unlike the solid
body rotation commonly obtained from low resolution studies of small
galaxies.  Despite an apparently accurate
rotation curve fit there is still significant flux in the residual in
the part of the XV diagram containing flux at large radii. This appears to be
due to an underestimate of the \HI\ surface density and the velocity
dispersion at large radii.

The measured \HI\ velocity dispersion is surprisingly low at only
4.5-6 \kms\ at radii within $5$ kpc from the galactic centre. This
galaxy displays particularily bright \HI\ emission in the inner 2-3
kpc, with a brightness temperature $T_B \sim 150$ K. The low measured 
velocity dispersion and high brightness temperature are consistent with a partially
opaque inner \HI\ disk. In the face-on galaxy NGC1058,
\citet{pr2006} found regions of \HI\ velocity dispersion as low as
$4$ \kms\ --- indicating cold low velocity dispersion \HI\ on
scales of at least $500$ pc which is quite large compared to typical star
formation scales in the ISM. Given the 
large beamwidth of $1.5$kpc of the NGC1058 \HI\ observations, it is likely 
that smaller regions of even lower velocity dispersion gas exist in 
NGC1058. In ESO115-G021, regions with a \HI\ velocity
dispersion of $4.5$ \kms\ occur on radial scales of $1$ kpc, similar to the cold gas 
distribution in NGC1058.

\subsection{ESO138-G014}

ESO138-G014 is a relatively large Sd galaxy with a \HI\ disk extending
to $20$ kpc, and a maximum rotation speed of $112$ \kms\ (see
Fig.~\ref{fig:ch5-e138-g14-kine}). The rotation curve is clearly
differential, rising steeply in the innermost $5$ kpc, then starts to flatten 
outside of $10$ kpc from the
galactic centre. As noted in paper I, the XV map
displays a faint ``figure-8'' signature, indicating a possible bar. 

The \HI\ velocity dispersion is approximately $7$ \kms\ at radii outside
of $9$ kpc. At radii inside of $9$ kpc, the \HI\ velocity dispersion
rises steeply to a plateau of 10-11 \kms\ within $5$ kpc radii.

The \HI\ radial profile is roughly constant within $\sim$9 kpc from the
centre, except for central peak within $\sim$2 kpc radius. The large
central plateau spans the same radial domain as the increased
velocity dispersion and could be explained by radial streaming in the bar
causing redistribution of the gas. As explained in 
Sect.~\ref{subsection:ESO138-G014}, we will leave this galaxy out of the
discussion because of missing short spacings and a poorly determined flaring
profile of the \HI\ layer.

\subsection{ESO146-G014}

ESO146-G014 is another small Sd galaxy similar to ESO115-G021 with a
particularly low \HI\ velocity dispersion of $\sim$4.5 \kms\ at radii
less than $4$ kpc. At larger radii, the mean velocity dispersion is
$6$ \kms, with a peak of $7.5$ \kms\ at $5$ kpc west of the
galactic centre. This high velocity dispersion region on the west side
corresponds to a $1$ kpc wide bright overdensity in the B-band optical
image, and to a $1$ kpc gap between bright \HI\ regions in the \HI\
total intensity map (see paper I, Fig.~5a).

The rotation curve of ESO146-G014 peaks at only $70$ \kms\ and is quite
symmetric about the galactic centre, particularly within radii of $4.5$
kpc. The \HI\ surface density shows a bright central peak, a plateau from $3$ to $6$ kpc
radius, after which it declines steeply on the left-hand side, and less so on
the right-hand side.

\subsection{ESO274-G001}

ESO274-G001 is the closest galaxy in our sample at a distance of only
$3.4$ Mpc. This allowed the $6$ km maximum baseline of the ATCA to
yield an exceptionally small \HI\ synthesised beam of $162$ pc. Thus
despite the small radius of only 7-8 kpc, each side of the galaxy is
spanned by over 47 independent beams. Using a beam sampling of 3
pixels per beam, approximately 150 kinematic fits are obtained on each
side of the galactic centre.

The measured kinematics shown in Fig.~\ref{fig:ch5-e274-g1-kine}
display a differentially rotating galaxy with a maximum rotation
speed of only $86$ \kms.  This rotation velocity shows that
ESO274-G001 is a small galaxy with a rotation curve
that continues to rise until the outermost kpc. The $20\%$-level envelope fitting method
overestimates the rotation speed by up to 10-15 \kms, while the peak
flux method dramatically underestimates it by up to
$50\%$.  The \HI\ surface density is reasonably symmetric,
with a central peak within $2$ kpc, and a plateau to $5.5$ kpc
on the left-hand side and $4.5$ kpc on the right-hand side, declining 
to an \HI\
surface density of 7-8$\times 10^{19}$ \atomscmsq. The \HI\ velocity
dispersion at $6.5$ \kms shows no systematic trend with radius, with small
oscillations of only $1$ \kms\ around this value, throughout the \HI\ disk.

\subsection{ESO435-G025 (IC2531)}

ESO435-G025 is the only Sc galaxy in our sample and more massive
than the next largest galaxy by a factor of four. The rotation curve displays an
unusual concave shape (see
Fig.~\ref{fig:ch5-e435-g025-kine}) indicating a dip in the gradient of rotation. 
ESO435-G025 was one of the first detections of a bar in an 
edge-on galaxy \citep{bf1999}. 
Despite the bar,  the
measured kinematics are an excellent representation of the galaxy. The
inner rotation of IC2531 is similar to that of NGC891, however lacking
gas tracing an inner ring, seen in both \HI\ \citep{ssvdh1997} and CO (Tony
Wong, private communication). It is quite possible that
like NGC891, the inner galaxy of ESO435-G025 is largely populated
by molecular gas in the inner disk \citep{wbvdh2004}.

ESO435-G025 is the most distant galaxy in our sample and with a low
peak signal-to-noise of $41.4$ in the XV map, and particularly low
signal-to-noise in the parts of the XV diagram with radii less than
$7$ kpc. Unfortunately, the signal-to-noise in the inner galaxy was
too low to deconvolve for the effect of the \HI\ beam
(FWHM$_{\theta}=1.3$ kpc) while modelling the XV diagram. As a result 
the \HI\
surface density is probably overestimated at radii within $10$ kpc
where the gradient of rotation over each beamwidth
is large. The measured \HI\ velocity dispersion displays a significant peak
and central trough, which is largely an effect of beam smearing.
At radii outside $10$ kpc, where the beam smearing effect is minimal, 
the \HI\ velocity dispersion has a
mean of $7$ \kms. To compensate for the inability to properly measure
the velocity dispersion inside of $10$ kpc, we adopt a smooth central
velocity dispersion rising from $8$ \kms\ at $10$ kpc radius to $10$
\kms\ in the centre. Inside of $10$ kpc we adopt a
central plateau for the \HI\ surface density, but we flag that the derived
properties within 10 kpc should be treated with caution.

\subsection{UGC7321}
\label{sec:ch5-results_ugc7321}

The \HI\ bar of UGC7321 causes a significant distortion in the peak
flux of the XV distribution. The effect of non-circular motions in the
bar causes a deviation in the projected rotation of gas rotating in
the bar. The XV modelling is able to recover the azimuthally averaged
surface density, and rotation (Fig.~\ref{fig:ch5-ugc7321-kine});
however, it is unable to fit the non-axisymmetric gas in the bar, 
as can been appreciated from
Fig.~\ref{fig:ch5-ugc7321-xvmodel}, which shows the observed, model and 
residual XV diagrams. Orbits in a
barred potential have been studied by \citet{cm1977}, who found that
there are two families, where the main orbit family is designated 
$x_1$ \citep[see also][]{bcsbu1991}. On the left-hand 
side of the galactic centre, the $x_1$ orbit showing the near and far
side of the bar is clearly apparent in the residual map. 
The residual structure on
the right-hand side of the galactic centre is less clear.

The peak flux in the $x_1$ orbits of the residual is $\approx15\%$ of the
peak flux of the observed XV diagram. To ascertain the likely surface mass
density perturbation caused by the bar, we inspected the K band
azimuthal flux profile of the bars in several late-type face-on
galaxies with weak bars. If one assumes a significant fraction of the
brightness variation is biased by young stars contributing to
the bar, galaxies with a relatively weak, diffuse bar are
unlikely to have a difference of stellar surface density between bar and 
inter-bar regions greater than $50\%$ (for example, the Sc
galaxy NGC2525). Photometry of UGC7321 by \citet{pbld2003}
shows that the stellar bar is very faint and thin, indicating that
minimal buckling of the bar has occured. The thin stellar bar of
UGC7321, and the small \HI\ mass fraction in the bar suggest that the
inner mass density perturbation is quite low, and
unlikely to significantly effect the halo mass modelling.

\section{The flaring of the \HI\ layer}

There have been few attempts to directly measure the flaring of the
\HI\ layer in galaxies. With the exception of Olling's study of NGC4244
\citep{olling1996a} and an earlier less sensitive study of NGC5023
\citep{bsvdk1986}, all the flaring measurements were undertaken on massive
Sb-Sc galaxies. It is particularly difficult to measure the flaring in 
large galaxies,
as such galaxies are expected to have higher disk surface densities and 
consequently thinner gas layers. For gas in hydrostatic
equilibrium the FWHM$_z(R)$ of the \HI\ is inversely proportional to
the maximum rotation velocity $V_{max}$ \citep[Eqn.~(D31) of
][]{olling1995}
\begin{equation}
\label{eq:ch6-fwhmz-disp-vmax}
FWHM_{z,\HI}(R) \propto \frac{\sigma_{v,\HI}(R)}{V_{max}},
\end{equation}
so large galaxies with max rotation speeds similar to the Milky Way,
will only flare to $1/3$ of the height of small galaxies like
ESO146-G014, given similar \HI\ vertical velocity dispersions
$\sigma_{v,\HI}(R)$. Most of these previous \HI\ flaring measurements of 
large galaxies were undertaken around 20 years ago, with larger telescope 
beams and poorer sensitivity resulting in partial resolution of the FWHM$_z$ 
vertical thickness of the gas layer.

One of the first measurements of the gas layer thickness in an
external galaxy was the early low resolution WSRT study of NGC891 by
\citet{sa1979}. This data were re-analyzed by \citet{vdkruit1981}.
These authors found that the vertical thickness of the \HI\
layer was unresolved by the $1.4$ kpc beam in the inner parts ($R < 8$
kpc) and increased to $2$ kpc at the outermost measured point of $R
\approx 14$ kpc. Using deeper higher resolution VLA data with
synthesised beam FWHM$_{\theta}$ of $0.4$ and $0.9$ kpc respectively
for NGC4565 and NGC891, \citet{rupen1989} found that the gas thickness
increased exponentially with radius, although NGC891 displays
substantial scatter. This is due to a low channel map peak
signal-to-noise of $\approx13$ in the NGC891 \HI\ cube, compared to
$\approx64$ in the NGC4565 dataset. This result is consistent with the
thickness of the gas layer in our Galaxy,
which is also found to increase exponentially
from $200$ pc near the Galactic Centre to $1.3$ kpc at $\sim$20 kpc
\citep{merrifield1992}. Meanwhile, \citet{braun1991} fitted the gas
thickness of M31 with an exponential $z$ profile and found that the
gas thickness increases roughly linearly from a scale height of $200$
pc at $5$ kpc radius to $600$ pc at $25$ kpc.

The most accurate measurement of \HI\ flaring was obtained for the
small Scd galaxy NGC4244 by \citet{olling1996a} who found that the
FWHM$_z$ thickness increased from $400$ pc at $5$ kpc radius to $1.8$
kpc at $14$ kpc radius, with uncertainty ranging from $\approx 100$ pc
at inner radii to 200-500 pc at the outermost measured point. These
were by far the most sensitive high resolution observations obtained,
with $14$ hrs of VLA data spread over B, C and D arrays, resulting in
a FWHM$_{\theta}$ of $170$ pc. The only other small galaxy that has been
observed is the Scd NGC5023 imaged at the WSRT with a spatial
resolution of $\sim$650 pc \citep{bsvdk1986}. These authors
find a constant thickness of $1.2\pm 0.3$ kpc despite the fact that
their Fig.~4 appears to show an increase of thickness from $1.0$ to
$1.7$ kpc, particularly on the east side. However the spectral
resolution of $16.6$ \kms\ available on the WSRT at the time was
a little coarse given the low maximum rotation speed ($86$ \kms) of
NGC5023. This low spectral resolution and the low signal-to-noise of
the observations would have increased the uncertainty in their radial
flaring profile.

\subsection{Methods used to measure the vertical gas thickness}
\label{sec:prev-flare-methods}

Several methods have been used to measure the vertical flaring of high
inclination galaxies. The gas disk thickness must be measured from
individual channel maps, rather than from the total \HI\ column density
map, as the latter measures the thickness of the integrated flux.
If the \HI\ distribution is well resolved by the telescope beam,
measuring the gas thickness from channel maps is elementary. The
challenge lies in deprojecting the measurements from the projected
radius of each measurement to the galactocentric radius. In each
method below, the measured FWHM$_z$ vertical thicknesses are corrected
for the spatial resolution of the beam (see
Eqn.~(\ref{eq:ch6-beamcorr}) below).

The most common method used to determine the radial flaring profile is
to fit the vertical thickness of just the extreme velocity channel
maps along each line of sight. This provides an accurate measurement of the flaring in the outer disk if 
the rotation curve is flat over this radial domain.  If gas in the extreme velocity channel maps is emitted from a
narrow azimuthal range around the line of nodes, then the gas fitted
by a vertical profile at each projected radius $R^{\prime}$ is emitted
from a narrow radial range, allowing the radius of each thickness
measurement to be approximated by the major axis distance
$R^{\prime}$. This method was used in the early flaring
measurements of NGC891 \citep{sa1979,rupen1989}, NGC4565
\citep{rupen1989} and NGC5023 \citep{bsvdk1986}.

Unfortunately this deprojection approximation can often miscalculate
the radial shape of the flaring profile. This occurs in less
differentially rotating galaxies, as the radial range of gas emission
is larger along each line of sight through the extreme channel maps.
Also, less massive galaxies with shallower rotation curves will have a
smaller major axis extent spanned in the outer velocity channels and the
radial domain where the rotation curve is most flattened may not be
where the gas disk flares. The other disadvantage is that the vertical
thickness at each radius is measured only twice, once on either side
of the galaxy. Another problem is that the gas thickness is only
measured at radii with flux in the edge channel maps. In a large
galaxy with a flat rotation curve, these channel maps will probe most
of the radial extent of the galaxy, however in a less massive galaxy
with a less steep inner rotation curve, the radial span of the outer
channel maps is much less. 

The analysis of \citet{vdkruit1981} of the NGC 891 data of \citet{sa1979}
involves matching the observed XV map to a simulated XV formed with different
assumed values of gas thickness and flaring. This study also involved an investigation into
the effects of small residual inclinations away from purely edge-on.

The technique used by \citet{braun1991} to measure the flaring of the
inclined galaxy M31 was more complicated. Braun used the
fitted inclination and rotation curve to fit the vertical thickness of
\HI\ features in the spiral arms.

In the next section, we present the method which we have developed to
accurately deproject the projected radius to the 
galactocentric radius for each vertical thickness measurement, and
correct the measured thickess for the spatial resolution of the
telescope along the $z$ axis. As our method fits the vertical gas
thickness in all channel maps with sufficient flux, the number of
measurements at each radius produces a more reliable radial flaring
profile, and also allows the flaring profile to be measured at radii
much closer to the galactic center. Our method is very similar to that
used by \citet{olling1996a}; however, we only allow for warps of the
\HI\ disk in the plane of the sky.


\begin{figure*}[t]
  \centering
\includegraphics[width=12cm]{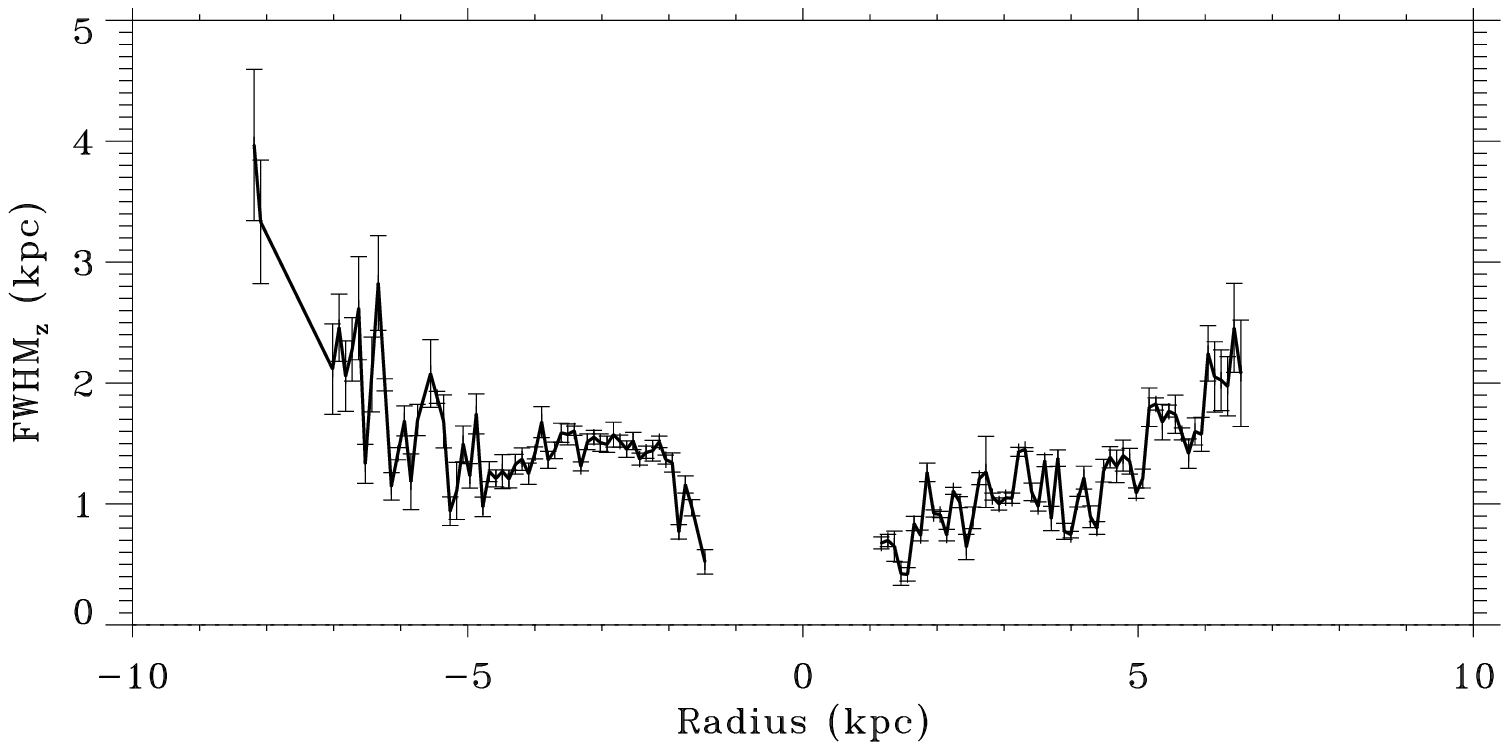}
\includegraphics[width=12cm]{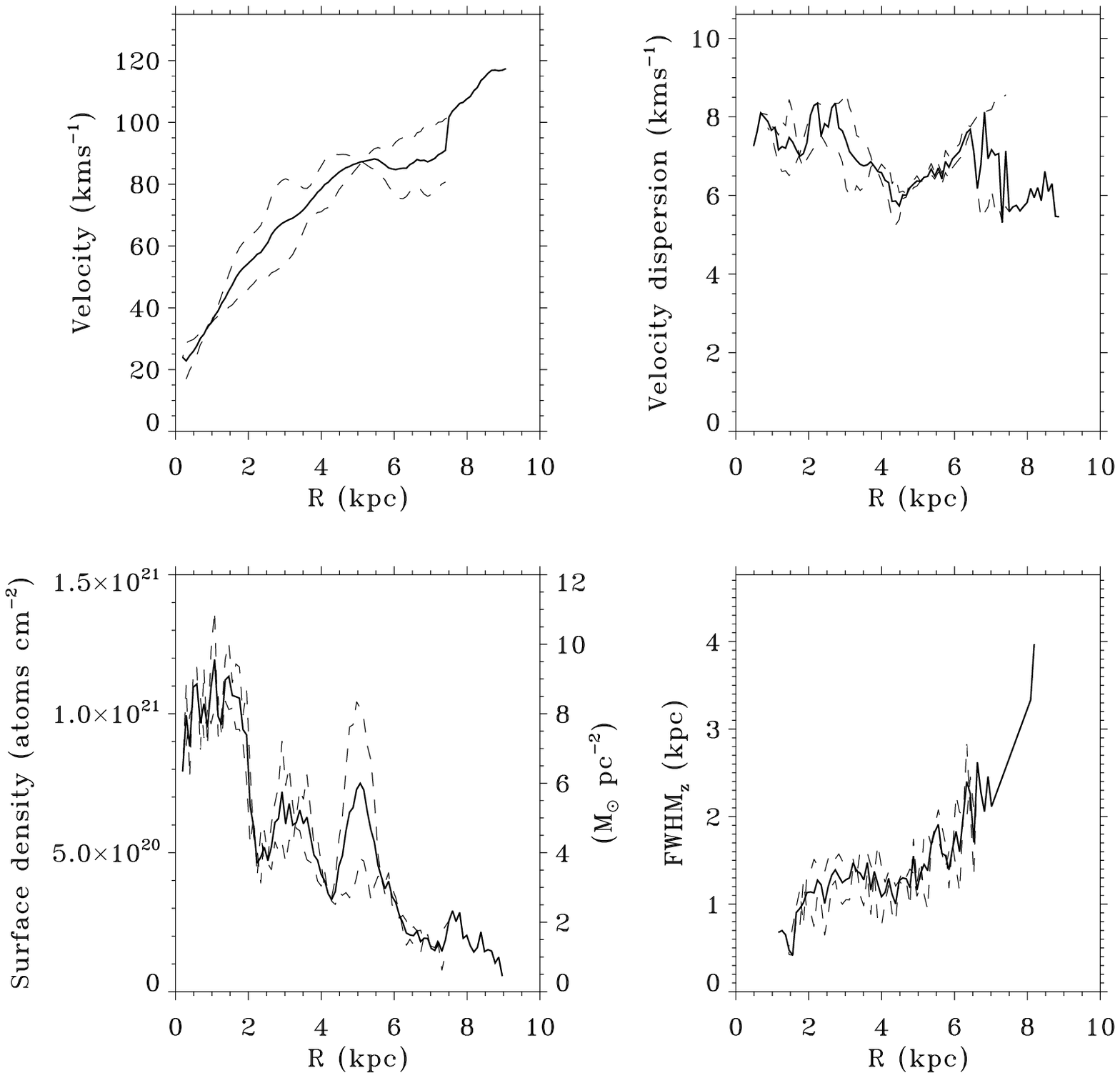}
  \caption[ESO074-G015 (IC5052): \HI\ FWHM$_z$ thickness, rotation
  curve, \HI\ velocity dispersion and face-on \HI\ surface
  density]{ESO074-G015 (IC5052). Top panel: Measured FWHM$_z$ thickness of
    the \HI\ gas disks. 
Lower panels: Each of the kinematic functions of the \HI\
    reflected about the galactic centre to show both sides of the
    galaxy, and the average of both sides. Middle left: rotation
    curve; middle right: \HI\ velocity dispersion; lower left: \HI\
    surface density; lower right: \HI\ vertical thickness FWHM$_z$.
}
  \label{fig:ch6-e074-g015-flaring}
\end{figure*}


\begin{figure*}[t]
  \centering
\includegraphics[width=12cm]{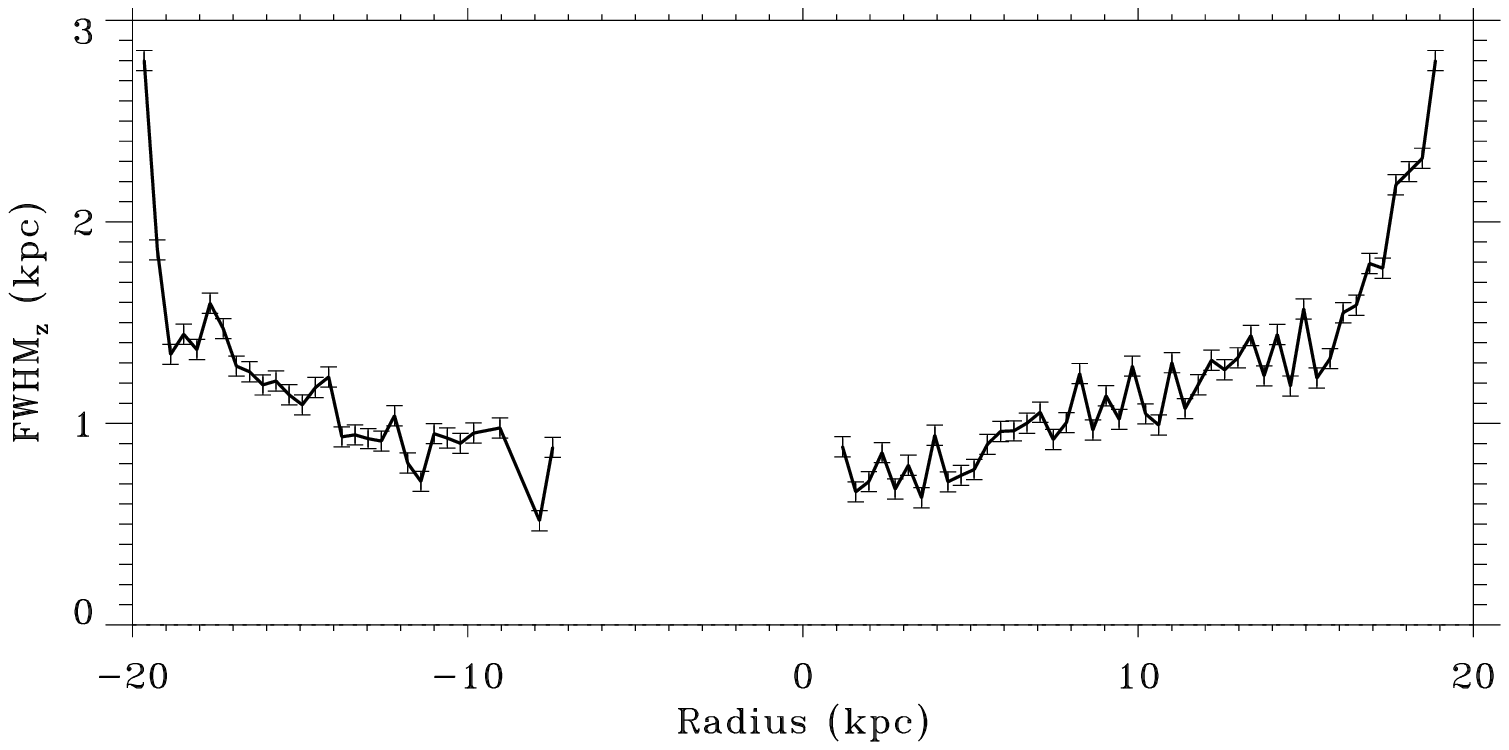}
\includegraphics[width=12cm]{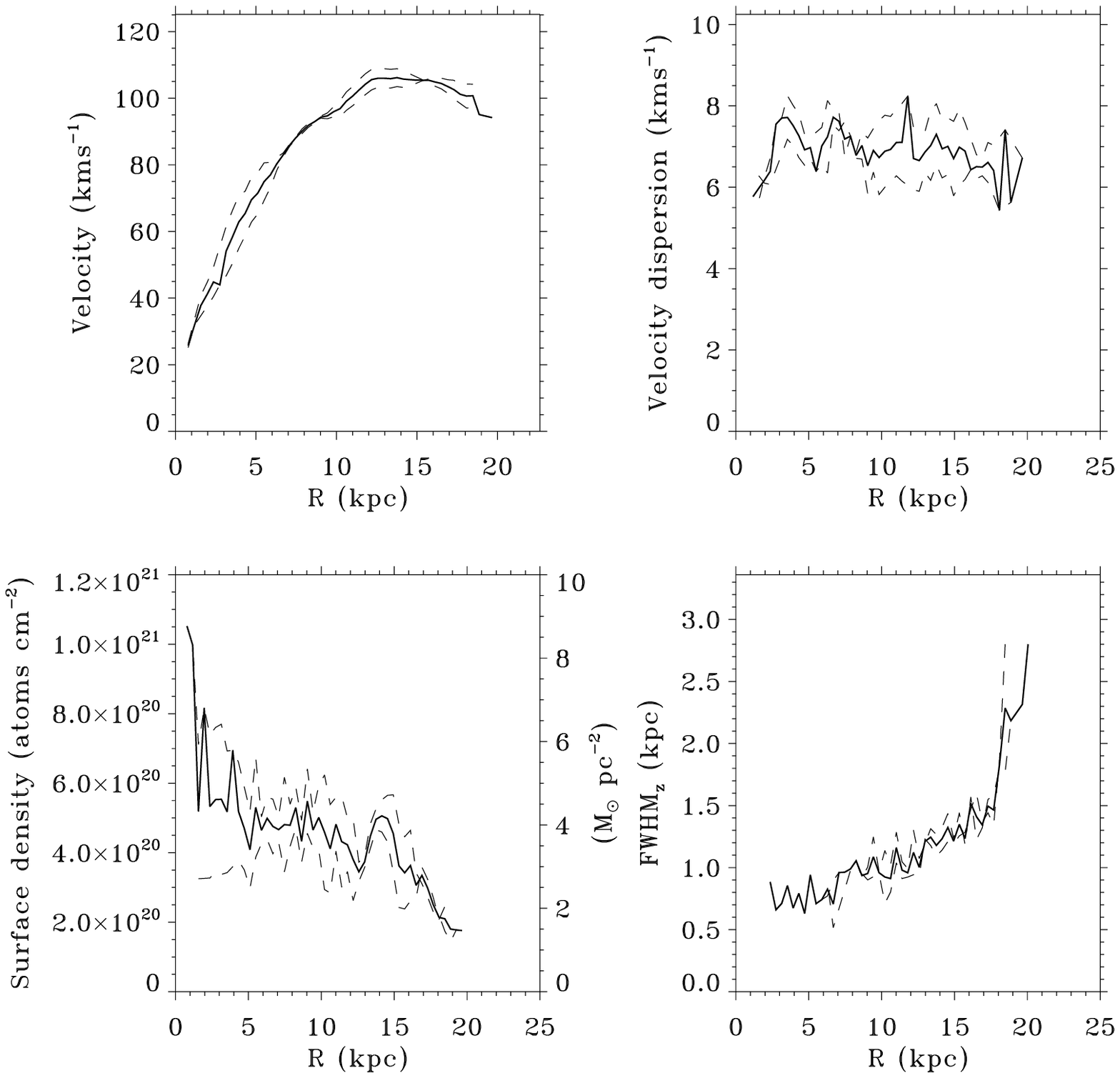}
  \caption[ESO109-G021 (IC5249): \HI\ FWHM_z$ thickness, rotation
  curve, \HI\ velocity dispersion and face-on \HI\ surface
  density]{ESO109-G021 (IC5249). Measured FWHM$_z$ thickness of
    the \HI\ gas disk and the kinematic functions of the \HI.
For details see the caption of Fig.~\ref{fig:ch6-e074-g015-flaring}.}
  \label{fig:ch6-e109-g021-flaring}
\end{figure*}


\begin{figure*}[t]
  \centering
\includegraphics[width=12cm]{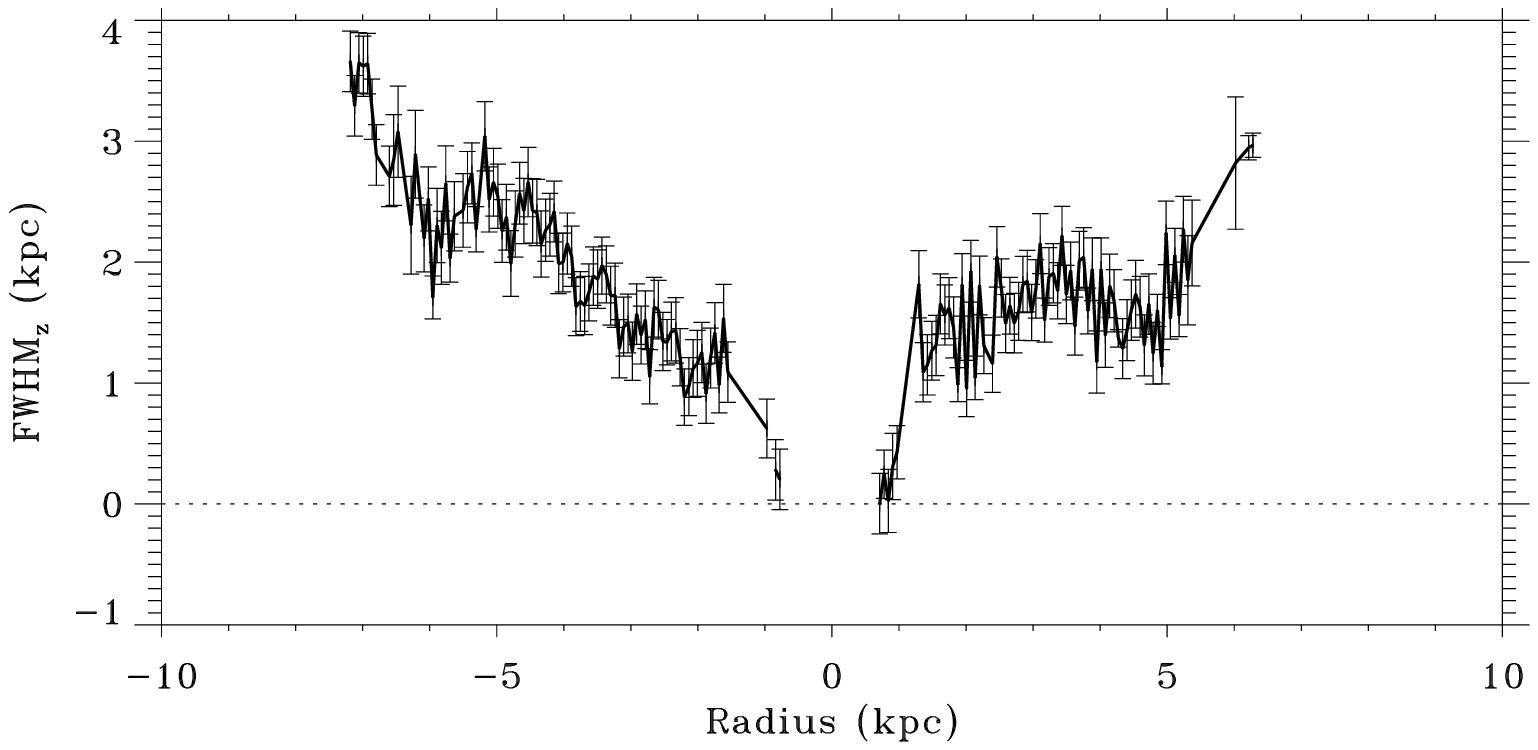}
\includegraphics[width=12cm]{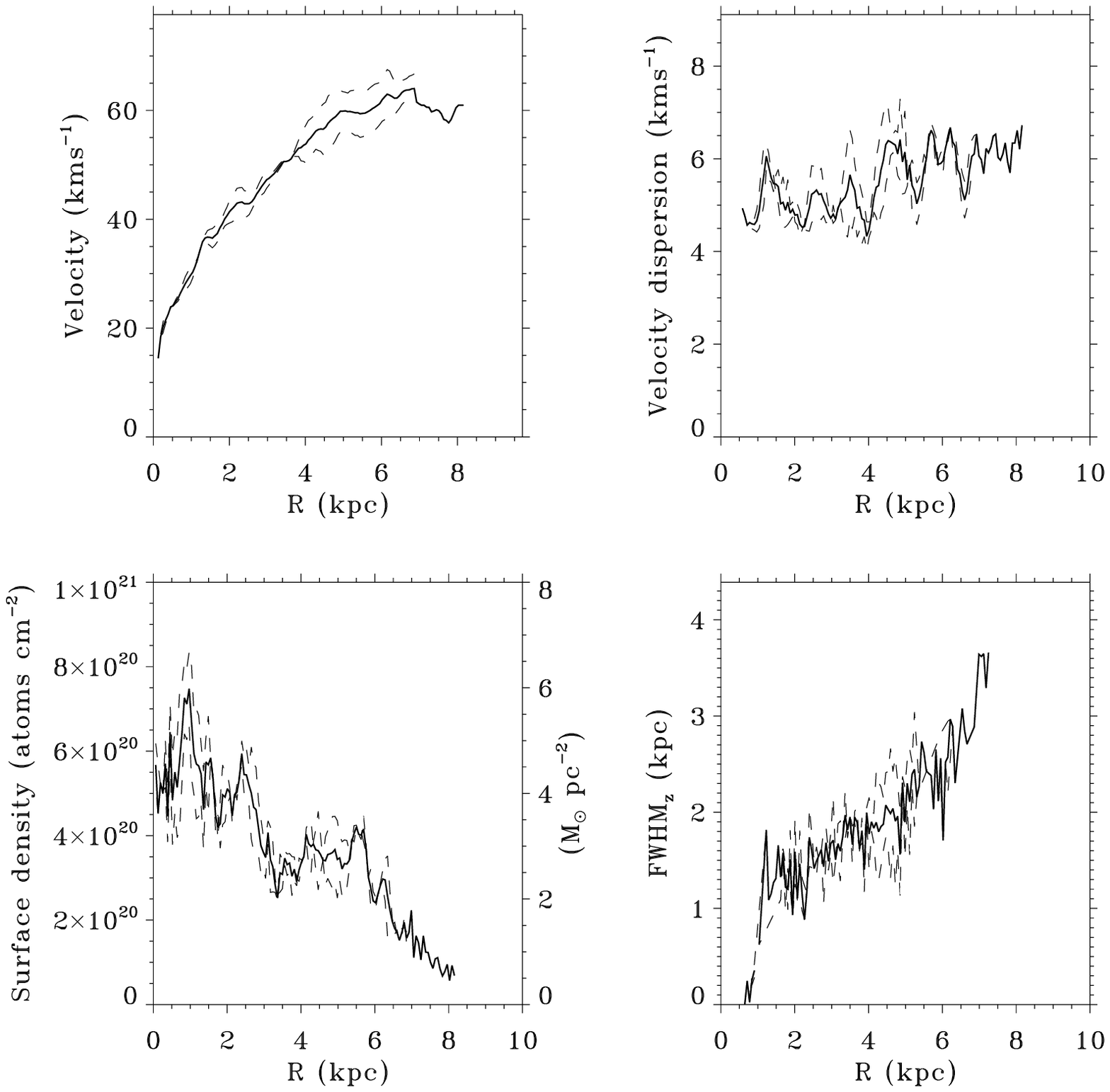}
  \caption[ESO115-G021: \HI\ FWHM$_z$ thickness, rotation
  curve, \HI\ velocity dispersion and face-on \HI\ surface
  density]{ESO115-G021. Measured FWHM$_z$ thickness of
    the \HI\ gas disk and the kinematic functions of the \HI.
For details see the caption of Fig.~\ref{fig:ch6-e074-g015-flaring}.}
  \label{fig:ch6-e115-g021-flaring}
\end{figure*}


\begin{figure*}[t]
  \centering
\includegraphics[width=12cm]{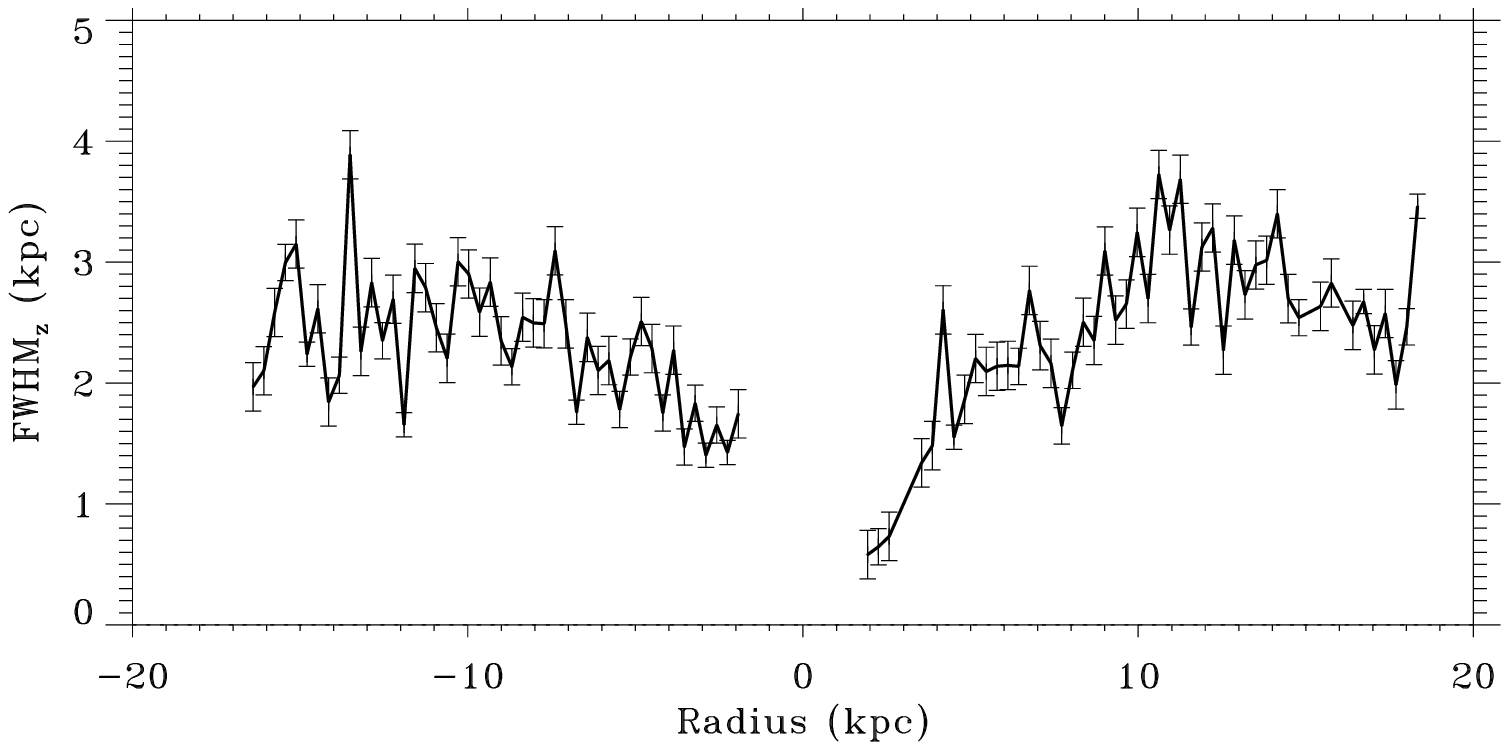}
\includegraphics[width=12cm]{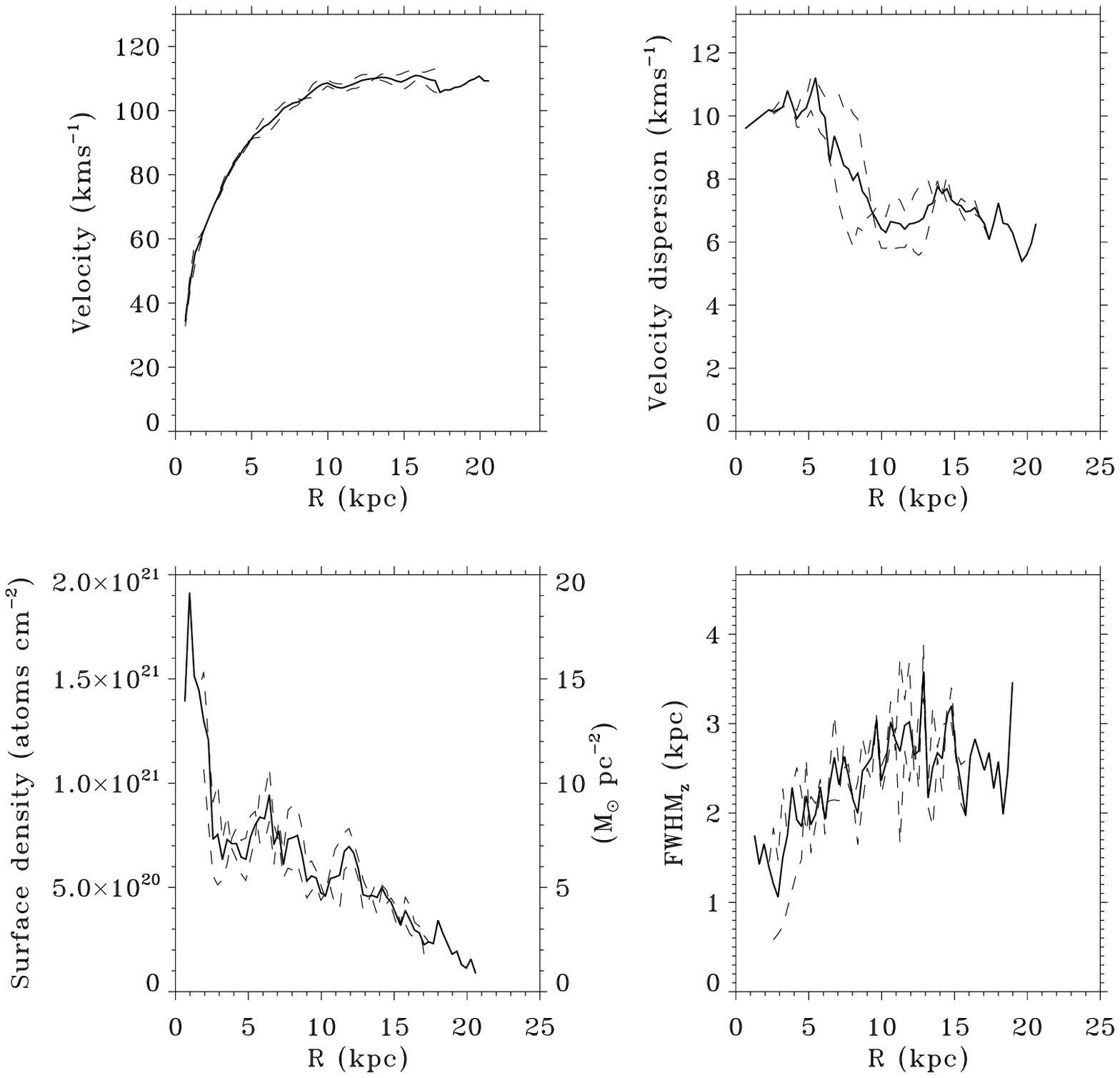}
  \caption[ESO138-G014: \HI\ FWHM$_z$ thickness, rotation
  curve, \HI\ velocity dispersion and face-on \HI\ surface
  density]{ESO138-G014. Measured FWHM$_z$ thickness of
    the \HI\ gas disk and the kinematic functions of the \HI.
For details see the caption of Fig.~\ref{fig:ch6-e074-g015-flaring}.}
  \label{fig:ch6-e138-g14-flaring}
\end{figure*}

\subsection{Gas flaring measurement by iterative \HI\ cube
  modelling}
\label{sec:ch6-method}

The FWHM thickness of the \HI\ disk was determined by fitting a
Gaussian to each vertical slice across the galaxy major axis in each 
\HI\ channel map, enabling the gas layer thickness to be measured independently on
each side of the galactic centre. To facilitate vertical thickness
measurements, each \HI\ cube was rotated in the plane of
the sky to align the galaxy major axis with the $x$ axis, ie. cube
axes $(x,z,v)$, (see Sect.~6 of
paper I for rotated \HI\ cubes for each galaxy).
The measured FWHM thickness is corrected for the effect of the beam by
\begin{equation}
\label{eq:ch6-beamcorr}
{\rm FWHM}_{z,\HI} = \sqrt{{\rm FWHM}_{z,tot}^2 - {\rm FWHM}_{\theta}^2 }
\end{equation}
to obtain the actual gas thickness FWHM$_{z,\HI}$, where the FWHM of
the telescope beam is denoted by FWHM$_{\theta}$. For simplicity,
hereafter we use FWHM$_z$ to refer to the beam-corrected \HI\ gas
thickness.

The projected radius was deprojected to the galactocentric radius
using the kinematics measured from the XV map 
in Sect.~\ref{sec:ch5-results}
to identify the galactocentric
radius that contributes emission at the projected radius 
and line-of-sight velocity of the channel map being measured. This was
facilitated by building a ``face-on'' \HI\ cube that
showed the \HI\ distribution in the equatorial plane $(R^\prime,s)$
(where $R^\prime$ is the projected radius and $s$ is the distance in
the line-of-sight) of the emission in each observed line-of-sight
velocity. For each galaxy a face-on \HI\
cube was built using the radial kinematics determined in
Sect.~\ref{sec:ch5-results}.

To simulate the observational effects of the telescope, the face-on
\HI\ cube was convolved by the FWHM of the telescope beam along the
$x$ axis. In order to account for the intrinsic \HI\ velocity
dispersion and the finite spectral resolution, the spectral axis of
the face-on \HI\ cube was convolved by a 1D point-spread-function
with total velocity dispersion $\sigma_{v,tot}(R)$ equal to
\begin{equation}
\label{eq:ch6-totdisp}
\sigma_{v,tot}(R) = \sqrt{\sigma_{v,\HI}(R)^2 +
  \left(\frac{{\rm FWHM}_{v,ch}}{2\sqrt{2 \log 2 }}\right)^2}
\end{equation}
where $\sigma_{v,\HI}(R)$ is the radial \HI\ velocity dispersion
measured in Sect.~\ref{sec:ch5-results} and FWHM$_{v,ch}$ is the spectral
resolution of the telescope correlator. The telescope signal response
function is a sinc function, which has a FWHM$_{v,ch}$ of $\approx
1.2$ times the width of the correlator channels.

The intensity-weighted velocity fields of the face-on \HI\ cubes
were calculated for each galaxy. These velocity fields provide a
representation of the galactocentric radii contributing flux to each
major axis slice of the actual observed \HI\ cubes.

Each channel map of the face-on \HI\ cube was masked by the
maximum radius of \HI\ emission on each side of the galaxy. Next the
galactocentric radius at each projected radius of a channel map was
derived by taking the intensity-weighted mean of the
galactocentric radii of all points along that projected radius in the
face-on channel map, where the intensity weights are the \HI\ flux at
those positions. This process was used to find the weighted mean
galactocentric radius
corresponding to the major axis position of each vertical thickness
measured from the actual observed \HI\ cubes.

The FWHM$_z(R)$ was determined from
the average of all the FWHM$_z$ fitted at that galactocentric radius. The
number of channel maps with emission from any one galactocentric
radius usually ranged from one to $20$, as dictated by the galaxy
kinematics and the resolution of the observations. Galaxies with
flatter rotation curves enabled a greater number of measurements for
each galactocentric radius. Higher spectral resolution relative to the
\HI\ velocity dispersion caused the \HI\ emission at each
galactocentric radius to be spread over more channel maps. However, 
increasing the spectral resolution resulted in lower signal-to-noise and fewer fits, 
as fitting was only possible on vertical 
slices with at least 4 consecutive points with flux greater than
two times the rms noise.

To correct for the scatter in the vertical thickness measurements due
to beam-smearing and low signal-to-noise, the whole \HI\
cube fitting process was iterated in a similar manner as that used for
the XV modelling algorithm. The known kinematics (from
Sect.~\ref{sec:ch5-results}) and the newly measured FWHM$_z(R)$
flaring fits formed the first model which was used to build a new \HI
$(x,z,v)$ \HI\ cube. The effect of the telescope observation was
simulated by convolving the \HI\ cube spatially by the telescope
synthesised beam and spectrally by total velocity dispersion
$\sigma_{v,tot}(R)$ (defined in Eqn.~(\ref{eq:ch6-totdisp})). The model
\HI\ cube was then fitted, and the fits to the model \HI\
cube compared to the fits of the actual observed \HI\ 
cube.  This cycle was iterated, with all kinematics except the
flaring held fixed, until FWHM$_z(R)$ fits to the model \HI\
cube matched the FWHM$_z(R)$ fits of the actual observed \HI\ 
cube to within $10\%$ of the FWHM of the telescope beam along
the $z$ axis or until the total difference in fits stopped decreasing.
The best fits were usually obtained after 3-4 iterations.

This iterative flare fitting method was tested under various conditions on
synthetic galaxy observations. Simulations showed that high
signal-to-noise observations could be corrected for beam smearing to
within 5-10\% of the FWHM beam size along $z$; however, lower
signal-to-noise observations with large beams were somewhat limited
with a FWHM$_z$ uncertainty of $\approx 20\%$ of the FWHM beam size
along $z$. Although the flare fitting algorithm does accurately
correct for the projection and beam-smearing of the \HI\ distribution,
the final FWHM$_z$ accuracy is naturally still dependent on the
adequate signal-to-noise of the \HI\ data cube.

\section{Results for flaring profiles}
\label{sec:ch6-galaxies}

For each galaxy in our sample we present the best fit measurements of
the FWHM$_z$ thickness of the \HI\ disk as a function of radius,
derived independently on each side of the galaxy. The measured
thickness along each major axis slice of the \HI\ channel maps was
undertaken using the kinematics measured by XV modelling in
Sect.~\ref{sec:ch5-results}.
The error bars shown on the flaring profiles in this section are the
1$\sigma$ errors of the Gaussian fits. 

We were able to measure the \HI\ flaring to radii within 1-2 kpc of
the maximum radius at which the other \HI\ kinematics were measured.
This was a particularily impressive result considering that the
signal-to-noise of most vertical slices was only 3-5. The average \HI\ cube peak 
signal-to-noise was only $13.6$, except for
UGC7321, which had almost $\sim$6 times the signal-to-noise of the
other galaxies. Analysis of the vertical shape of the \HI\ distribution in UGC7321, 
facilitated by the high sensitivity observations, found unexpected azimuthal variations 
in the gas thickness and midplane displacement of this edge-on galaxy. 
These surprising results for UGC7321 are presented in Sect.~\ref{sec:ch6-u7321}.


\begin{figure*}[t]
  \centering
\includegraphics[width=12cm]{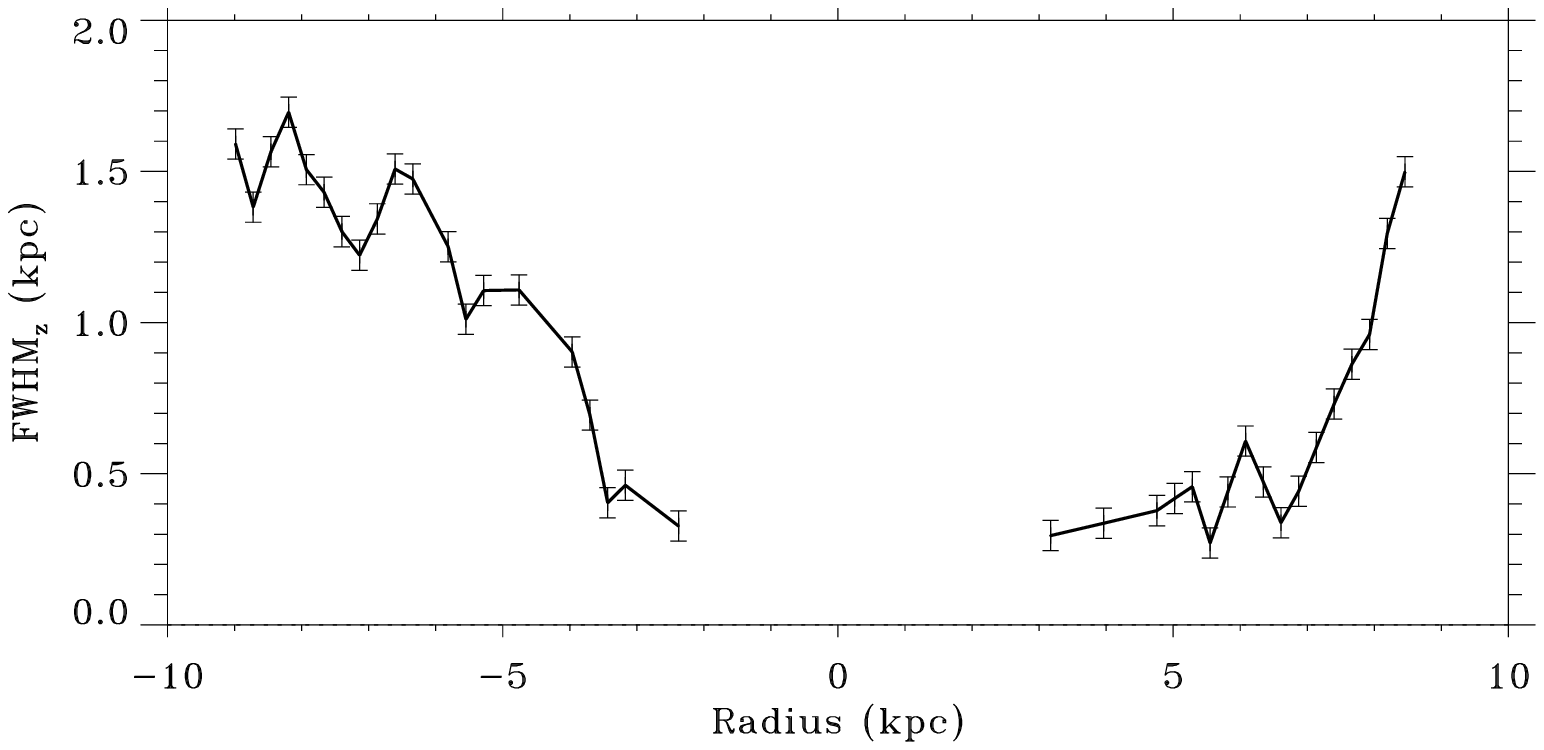}
\includegraphics[width=12cm]{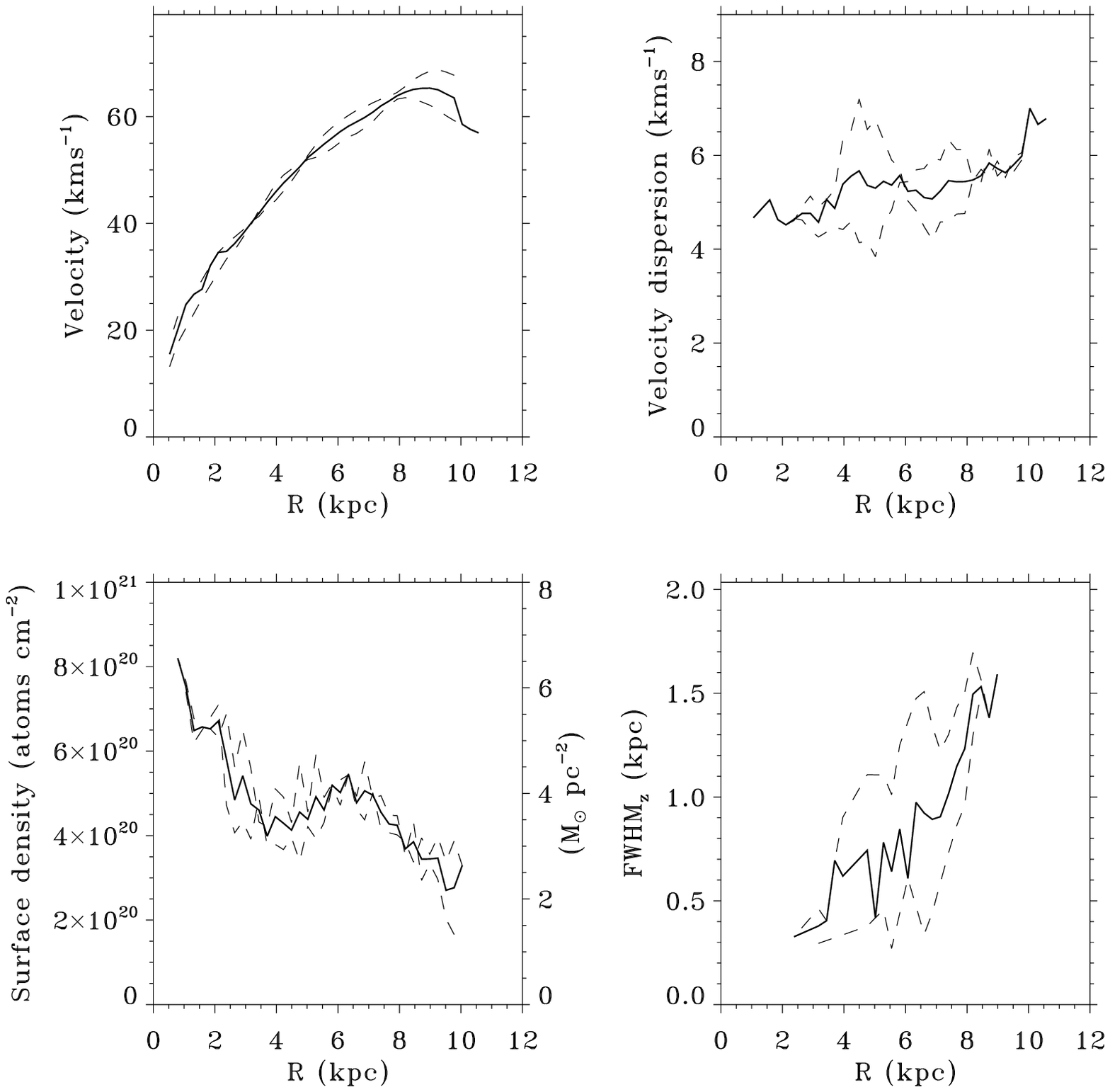}
  \caption[ESO146-G014: \HI\ FWHM$_z$ thickness, rotation
  curve, \HI\ velocity dispersion and face-on \HI\ surface
  density]{ESO146-G014. Measured FWHM$_z$ thickness of
    the \HI\ gas disk and the kinematic functions of the \HI.
For details see the caption of Fig.~\ref{fig:ch6-e074-g015-flaring}.}
  \label{fig:ch6-e146-g014-flaring}
\end{figure*}


\begin{figure*}[t]
  \centering
\includegraphics[width=12cm]{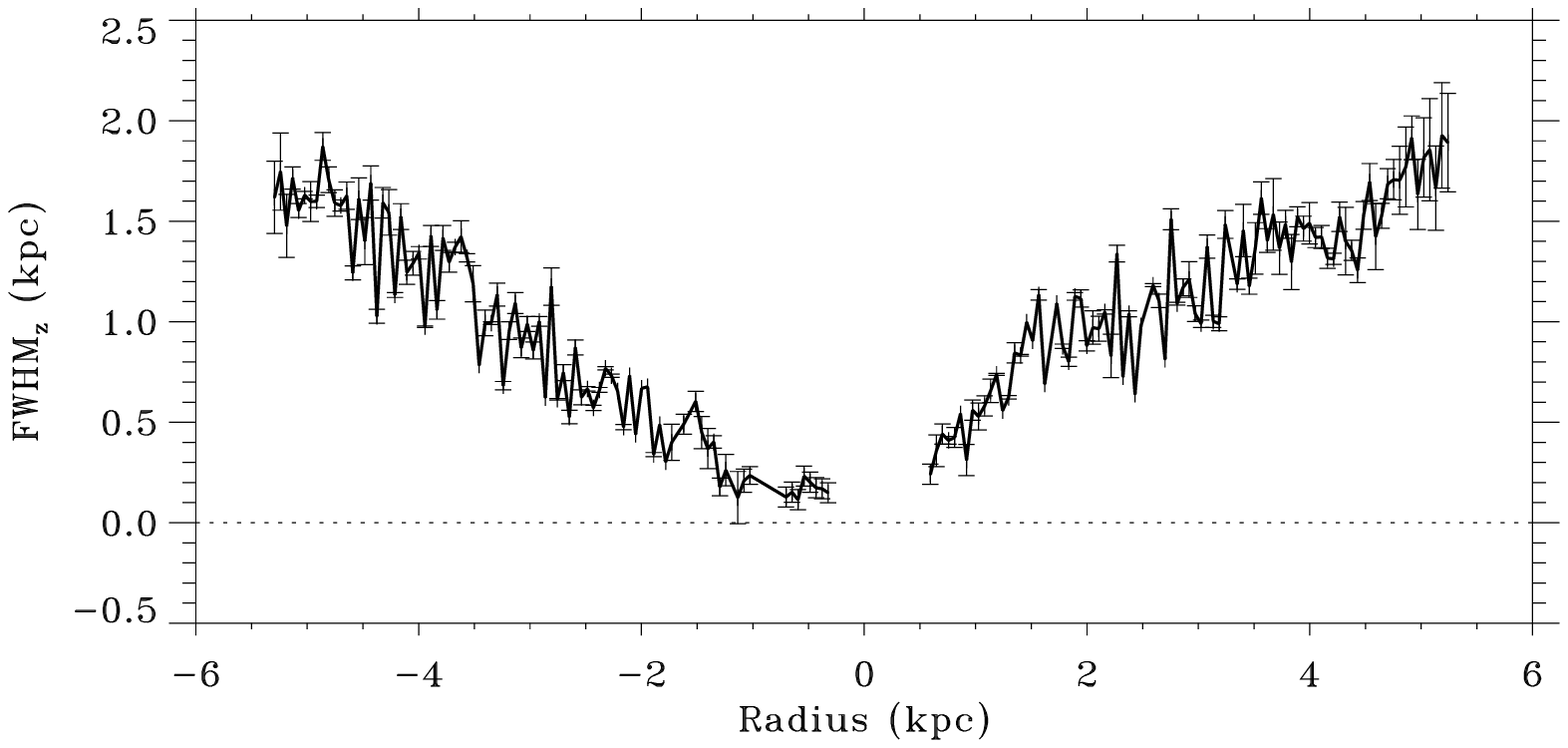}
\includegraphics[width=12cm]{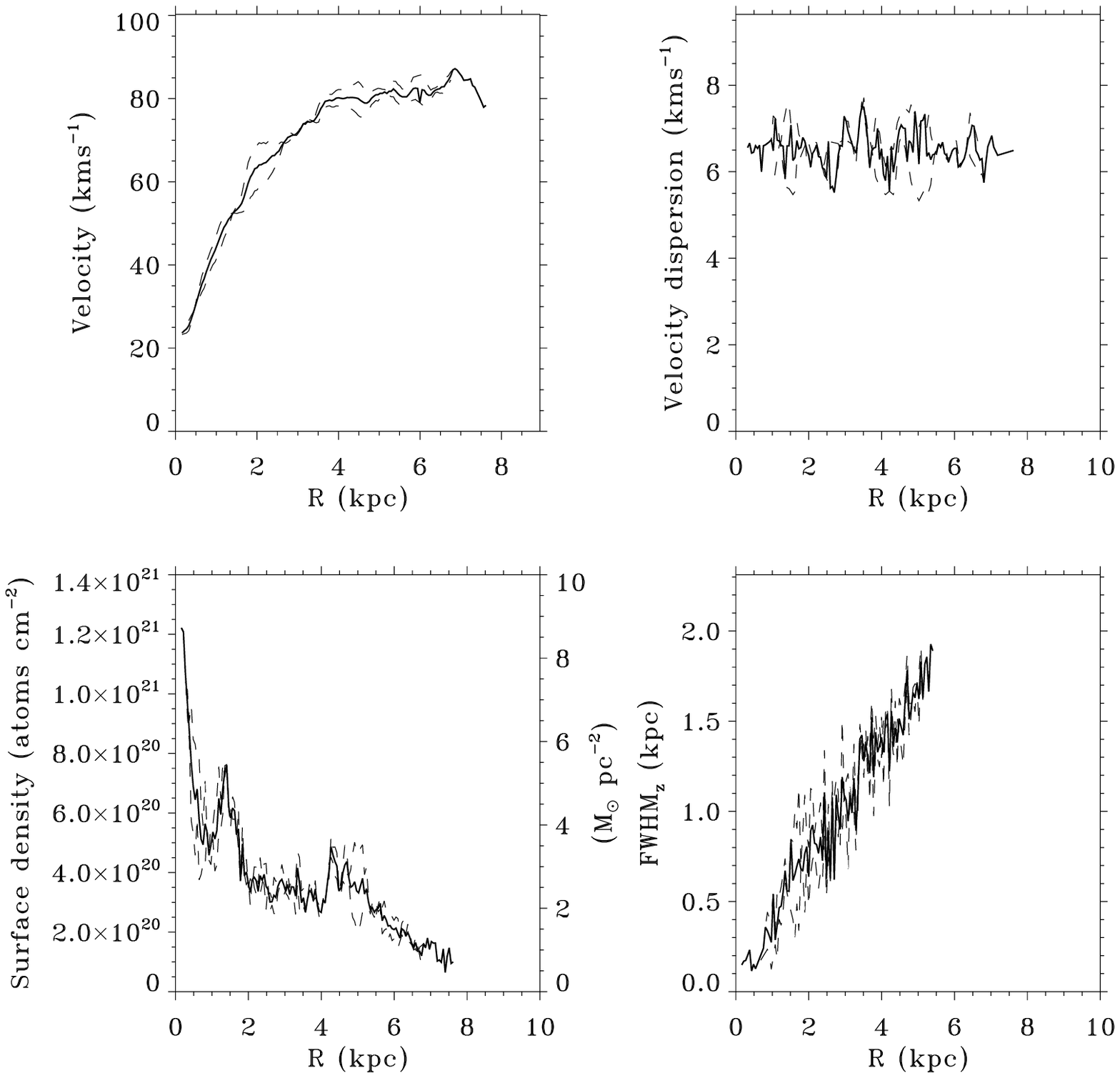}
  \caption[ESO274-G001: \HI\ FWHM$_z$ thickness, rotation
  curve, \HI\ velocity dispersion and face-on \HI\ surface
  density]{ESO274-G001. Measured FWHM$_z$ thickness of
    the \HI\ gas disk and the kinematic functions of the \HI.
For details see the caption of Fig.~\ref{fig:ch6-e074-g015-flaring}.}
  \label{fig:ch6-e274-g001-flaring}
\end{figure*}


\begin{figure*}[t]
  \centering
\includegraphics[width=12cm]{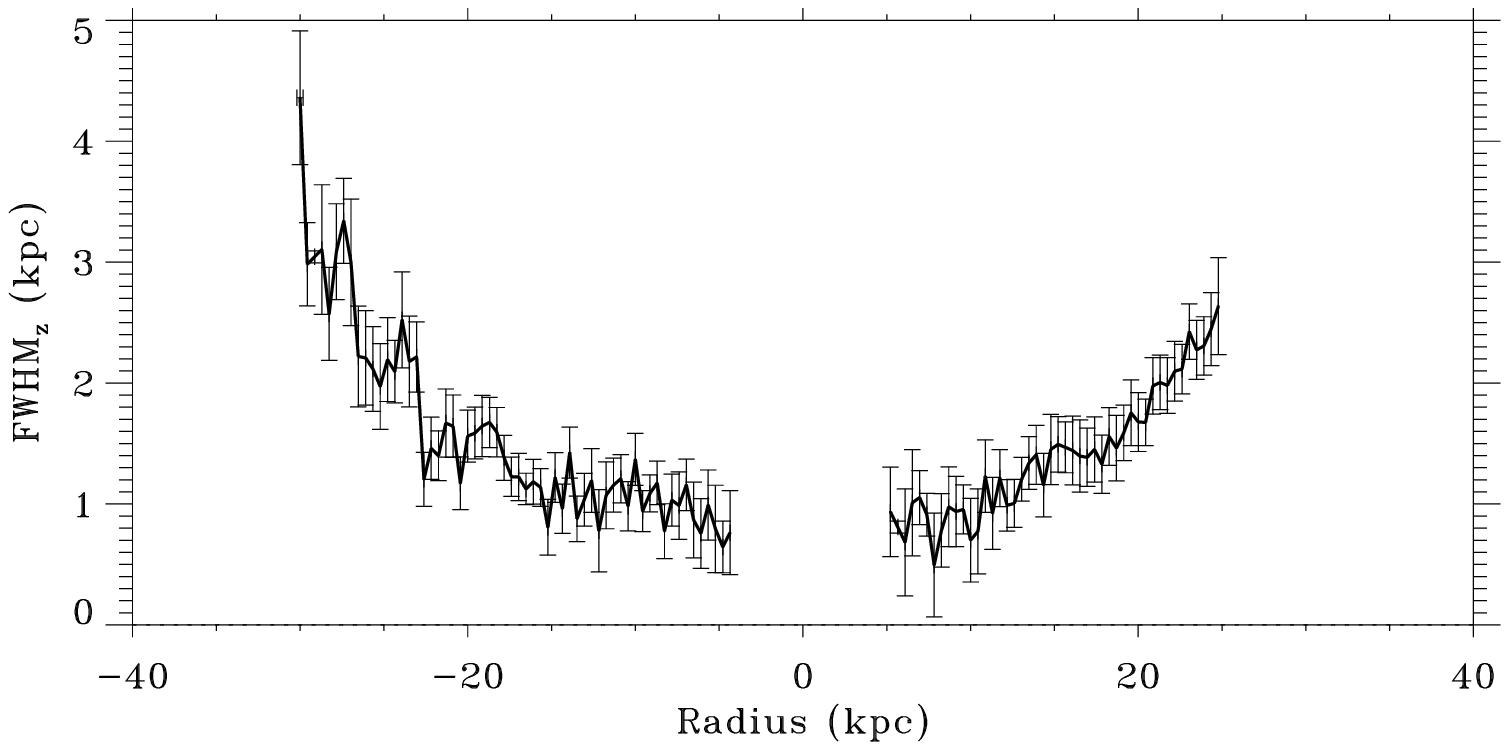}
\includegraphics[width=12cm]{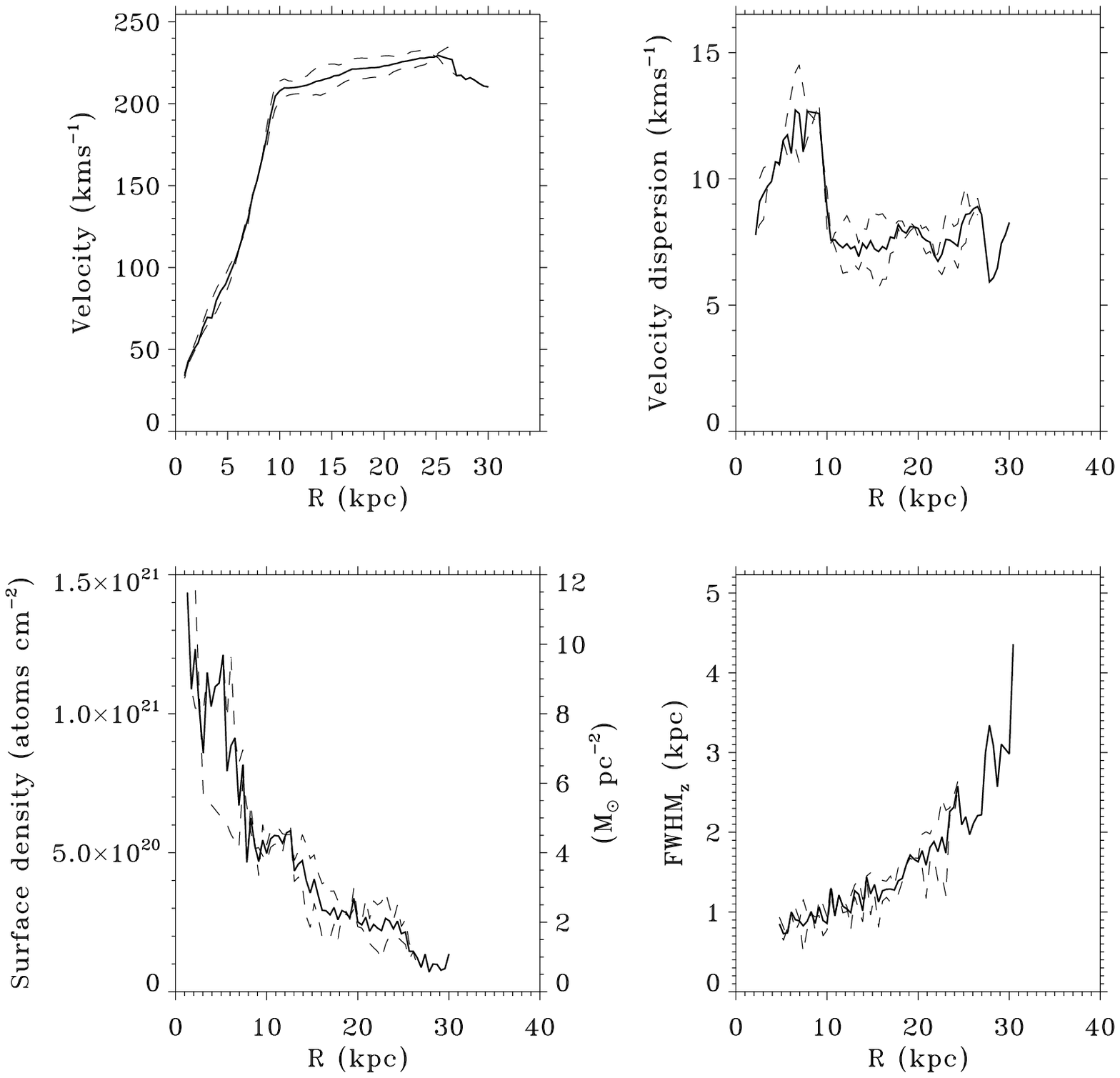}
  \caption[ESO435-G025 (IC2531): \HI\ FWHM$_z$ thickness, rotation
  curve, \HI\ velocity dispersion and face-on \HI\ surface
  density]{ESO435-G025 (IC2531). Measured FWHM$_z$ thickness of
    the \HI\ gas disk and the kinematic functions of the \HI.
For details see the caption of Fig.~\ref{fig:ch6-e074-g015-flaring}.}
  \label{fig:ch6-e435-g025-flaring}
\end{figure*}


\begin{figure*}[t]
\centering
\includegraphics[width=12cm]{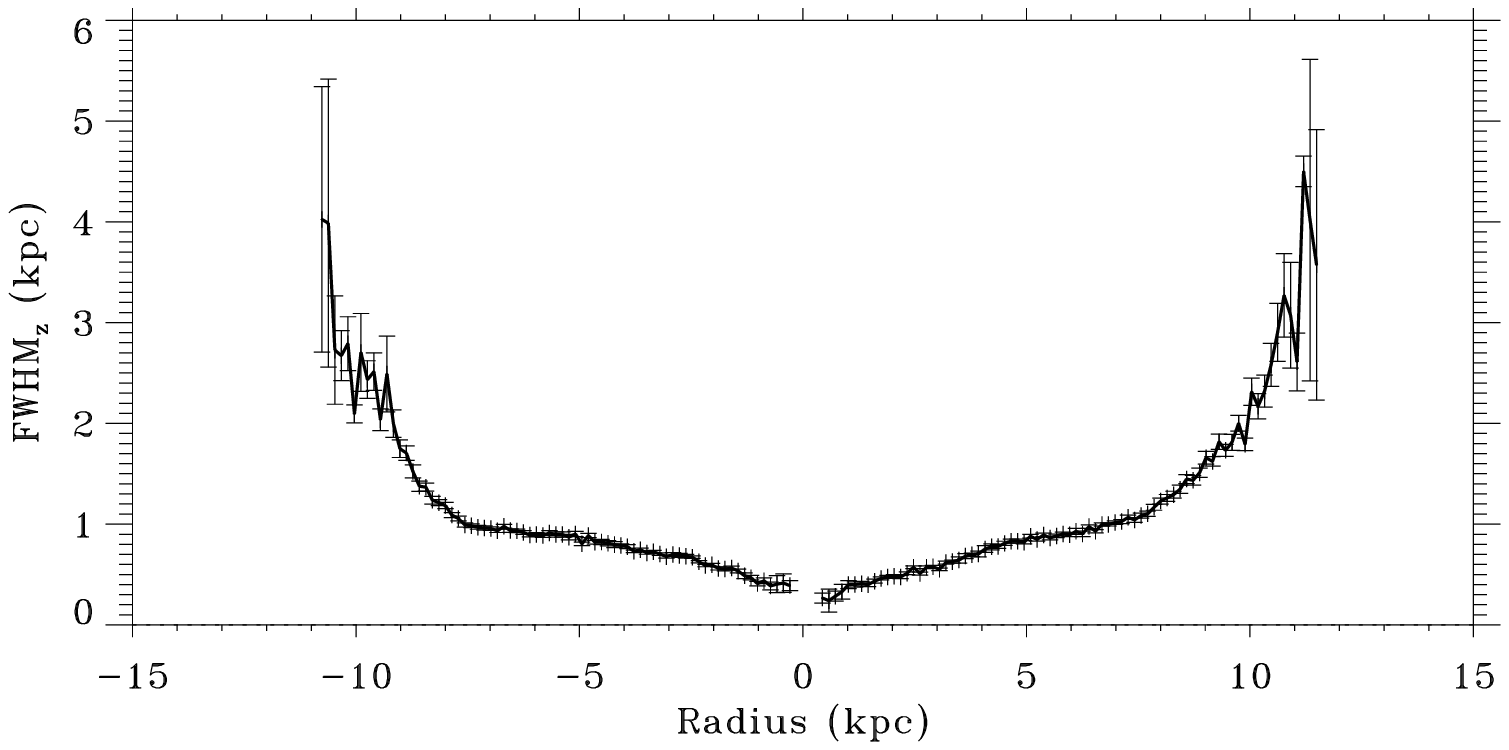}
\includegraphics[width=12cm]{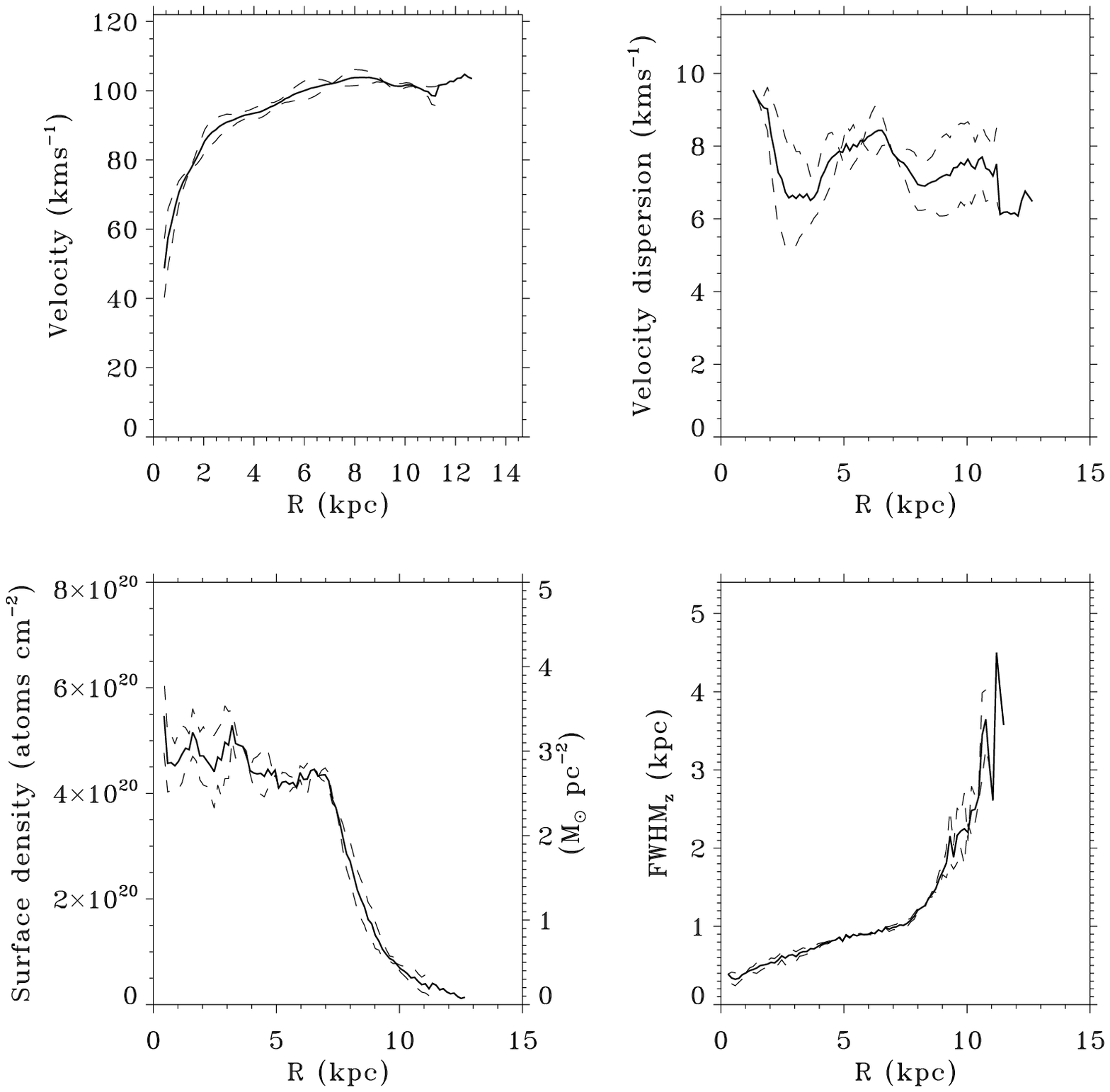}
  \caption[UGC7321: \HI\ FWHM$_z$ thickness, rotation
  curve, \HI\ velocity dispersion and face-on \HI\ surface
  density]{UGC7321. Measured FWHM$_z$ thickness of
    the \HI\ gas disk and the kinematic functions of the \HI.
For details see the caption of Fig.~\ref{fig:ch6-e074-g015-flaring}.}
  \label{fig:ch6-ugc7321-flaring}
\end{figure*}

\begin{figure*}[t]
  \centering
\includegraphics[width=9cm]{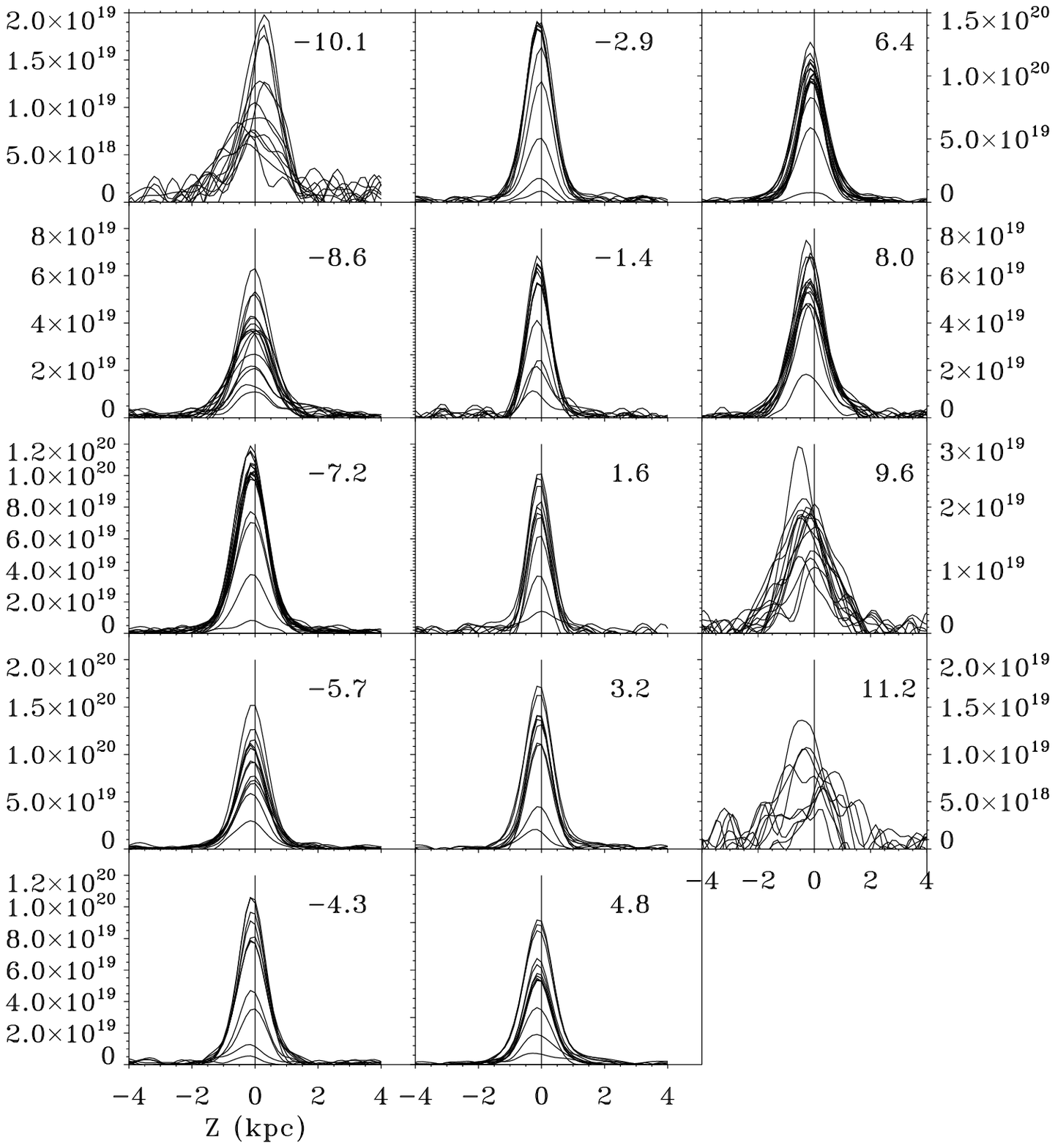}
\includegraphics[width=9cm]{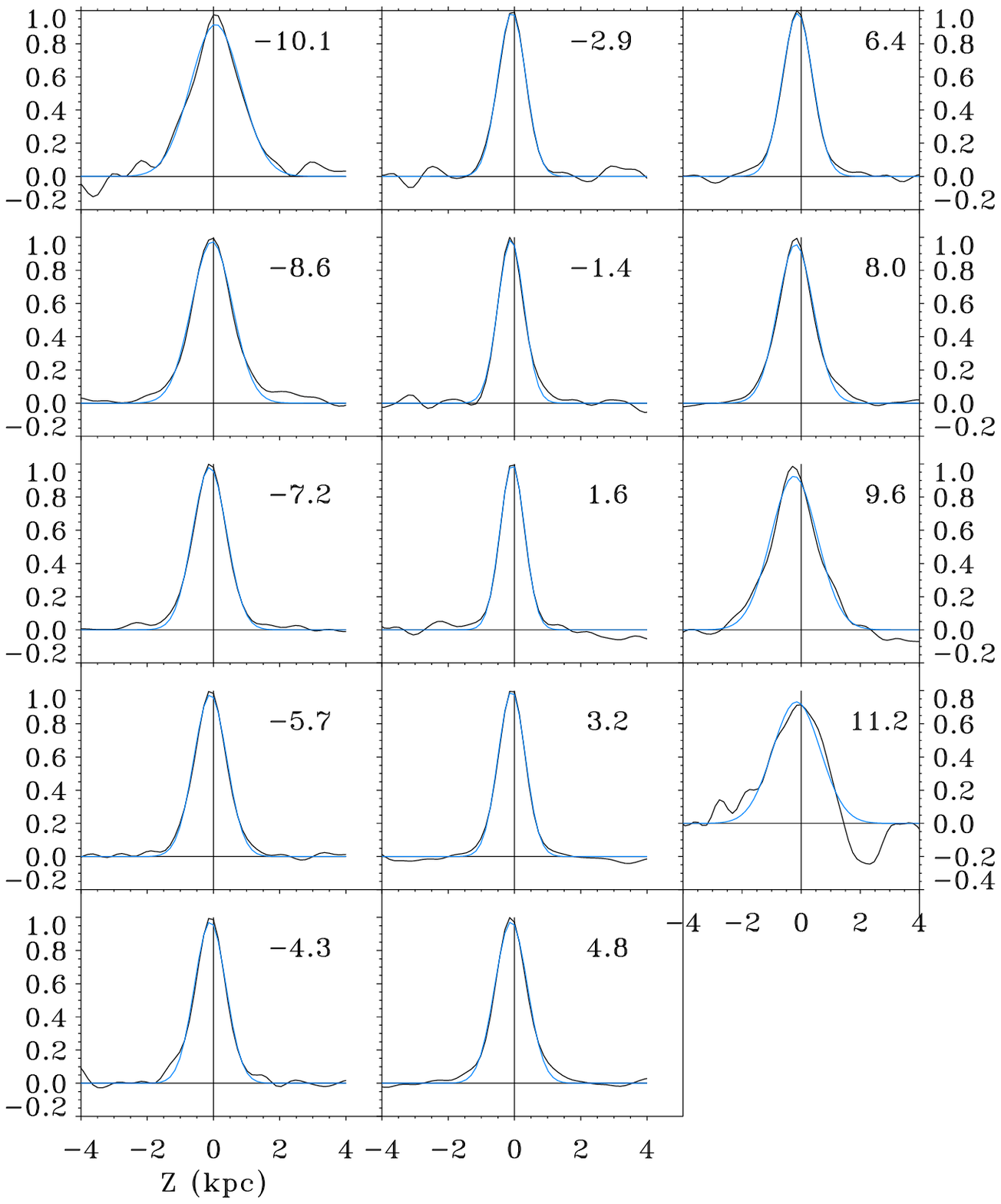}
  \caption[UGC7321: All vertical profiles of the \HI\ in each radial
  bin]{Left -- All the vertical profiles in a sample of galactocentric radial
    bins in UGC7321. The radius of the bin is given in the top right-hand
    corner of each plot. Each radial bin
    is $145.4$ pc wide -- the same size as the image pixels along the
    major and minor axes of UGC7321. The ordinate axis of each plot
    shows the \HI\ column density in units of \HI\ atoms cm$^{-2}$ in
    each vertical slice. Right --  The same as on the left-hand side, but now
we show average profiles, which were obtained after aligning the profile centres
    and vertically rescaling all profiles to a peak of $1$. Overlaid in
    blue is the Gaussian fit to each average profile. }
  \label{fig:ch6-u7321-all-profiles}
\end{figure*}

\subsection{ESO074-G015 (IC5052)}

The FWHM$_z$ vertical thickness of ESO074-G015 expanded from FWHM$_z
=0.4$ kpc at a galactocentric radius of $1.5$ kpc to FWHM$_z = 2.5$
kpc at $\approx 7.0$ kpc, with an additional thicknesses of $3.3$
and $4.0$ kpc measured at the edge of the \HI\ disk at radii 8.0-8.1
kpc (Fig.~\ref{fig:ch6-e074-g015-flaring}). These outermost two
measurements are consistent with the flaring shape of the disk, but 
the measurement has a large uncertainty of 0.5-0.6 kpc and were only measured 
in one channel map of the \HI\ data cube.  The flaring at $8$ kpc comprises the
warped eastward extension at a radius of $4$\arcmin\ in the \HI\
column density map (see Fig.~1b in paper I), and is detected as very
diffuse edge emission in the outer channel maps
 on the receding side. The channel maps show the emission at 2 $\times$ rms noise of
$1.83\times 10^{19}$ \atomscmsq.

The integrated flux of ESO074-G015 was $30\%$ less than that measured
from the \HI\ Parkes All Sky Survey (HIPASS) line spectrum obtained
from the online HIPASS Public Data 
Release\footnote{http://www.atnf.csiro.au/research/multibeam/release}. 
This suggested that there is additional extended \HI\ emission that was not
observed with the long ATCA array configurations. Indeed, a low resolution \HI\ column density
map available in the online Local Volume \HI\ Survey
(LVHIS\footnote{http://www.atnf.csiro.au/research/LVHIS/ic5052/ATCAresults}),
showed large scale emission around the warped disk extending to $\sim$12
kpc.  The smallest array configuration observations included in our
analysis were low sensitivity snapshot observations with the ATCA $375$
array configuration. The LVHIS image was
obtained with a $12$ hour integration using the ATCA EW367 array
configuration. Additional low resolution data of ESO074-G015 may detect \HI\ flaring 
to thicknesses greater than $4.0$ kpc.

Despite the gross asymmetry of the ESO074-G015 rotation curve, the flaring is
remarkably symmetric (see Fig.~\ref{fig:ch6-e074-g015-flaring}).
This \HI\ disk does thicken by $\sim$200 pc at radii inside $4$ kpc
particularly on the east side. This appears to be due to the
increased \HI\ velocity dispersion from $\sim$6 to 8 \kms at the same radii, 
and is more pronounced on the east side.
The inner west side of the disk from 2-4 kpc shows intense star
formation and radio continuum emission, but the east side is
optically more quiescent. The flaring profile is clearly
well-resolved by the synthesised \HI\ beam FWHM of $292$ pc. However,
the low peak signal-to-noise of $14$ in the \HI\ data cube restricts measurement of 
flaring in the faint outer \HI\ disk.
Despite being well-resolved spatially, the apparent mass asymmetry
indicated by the rotation curve precludes this galaxy from dark halo
flattening modelling.

\subsection{ESO109-G021 (IC5249)}

Like UGC7321, ESO109-G021 is renowned for being a ``superthin''
galaxy, exhibiting a scalelength-to-scaleheight ratio $h_R/h_z$ of $11$.
However, although ESO109-G021 has a similar maximum rotation speed to
UGC7321, its radial scale is approximately $2.5$ times larger, with a
long scalelength of $7$ kpc and a sharp stellar truncation at $17$
kpc. We find that the \HI\ disk flares from a FWHM$_z$ of $800$ pc at
$2$ kpc radius out to $2.8$ kpc at $20$ kpc (see
Fig.~\ref{fig:ch6-e109-g021-flaring}). The thickness of the gas disk
appears to be roughly constant from $2$ to $5$ kpc and begins to rise
exponentially from a radius of 6-7 kpc.  The steep flaring of the
gas disk at 17-18 kpc is consistent with the edge of the stellar
disk.

 The gas layer thickness was also been measured by
\citet{vdkjvkf2001} from the total \HI\ intensity map, as the signal-to-noise
of the channel maps was too low as they had only the original $12$ hour ATCA
6.0C observation by Carignan \& Beaulieu. They found the
FWHM$_z$ thickness rose from $1.3$ kpc near
the centre to $4.9$ kpc at $15$ kpc. Our flaring measurements were obtained
using an additional $30$ hrs of ATCA time split over EW352, 750D and
6.0A array configurations, yielding a peak channel map signal-to-noise
of $10.1$

The \HI\ disk was possibly unresolved at radii less than $2$ kpc,
as the large synthesis beam of $1.18$ kpc made it difficult to
resolve spatial features smaller than $\sim$800 pc. An unresolved
central gas thickness would explain the relatively thick central gas
layer, comparable in thickness to the stellar disk.

\subsection{ESO115-G021}

ESO115-G021 shows strong flaring (see Fig.~\ref{fig:ch6-e115-g021-flaring})
from a central FWHM$_z$ thickness of
$0.8$ kpc to over $2$ kpc at the outermost measured point, $6.1$ kpc
from the galactic centre. The \HI\ emission of the galaxy extends
to a galactocentric radius of $8.2$ kpc on the eastward side;
however, this emission is very diffuse and too faint for \HI\ thickness measurement.

The peak signal-to-noise of the channel maps used to measure the
flaring was only $18.6$ and the FWHM beam size was $194$ pc. Each
radial fit was the average of all vertical thickness fits at major
axis positions of a channel map that deprojected to this radius. Some
thickness measurements were averaged from measurements to 5-7 channel maps; 
however, most were obtained from two to three channel maps.

Despite the low average \HI\ velocity dispersion of $\langle \sigma_v
\rangle = 5.5 \pm 0.8$ \kms, the \HI\ disk flares to a vertical
HWFM$_z$ above the plane of $> 2$ kpc near the \HI\ edge.

\subsection{ESO138-G014}
\label{subsection:ESO138-G014}

ESO138-G014 was the only galaxy in our sample in which it was not 
possible to satisfactorily measure the \HI\ gas layer thickness. The measured
flaring thickness are shown in Fig.~\ref{fig:ch6-e138-g14-flaring}; 
however, the derived FWHM$_z$ are ballpark estimates only due to
missing large scale structure that was not measured with the large ATCA array
configurations used. ESO138-G014 and ESO274-G001
were the only two galaxies in our sample with extended \HI\ disks
($\sim$10\arcmin\ diameter) for which we were unable to obtain compact
array \HI\ observations. Although they both had similar peak
channel map signal-to-noise of $\sim$15, the low spatial resolution of
ESO138-G014 prevented correction of the fitted flaring
by iterative modelling. The irregular measured flaring profile, which displays
a decreasing disk thickness at radii outside of $15$ kpc, is probably due to 
the missing large scale structure.

Under a constant hydrostatic pressure, a fall in gas thickness of
$\sim$0.5-1.0 kpc could be explained by a decrease in \HI\ velocity
dispersion of 15-30\%\ or 1.5-3.0 \kms. Although a $3$ \kms\
decrease in velocity dispersion is observed between $7$ and $10$ kpc
radius, this decrease is significantly inside the onset radius of the
decline in gas layer thickness. Due to the irregular gas flaring
profile we decided not to model the dark halo of ESO138-G014.

\subsection{ESO146-G014}

The \HI\ disk of ESO146-G014 flares dramatically from a FWHM$_z$
thickness of $0.4$ kpc at radii of $3$ kpc, to FWHM$_z=1.6$ kpc at
$9.0$ kpc from the galactic centre (see
Fig.~\ref{fig:ch6-e146-g014-flaring}). The flaring is highly
asymmetric around the galactic centre, probably due to the low peak
\HI\ cube signal-to-noise of $11.3$ which
limited the number of fits at each radius.  Despite the noisy
data, the measured flaring profile still showed a steep exponential
rise from $4$ kpc radius to the last measured point at $8$ kpc radius.
This steep inner flaring, occuring at $1.5$ to $2.5$ stellar
scalelengths, is consistent with a low stellar surface density as
inferred from the B-band face-on surface brightness of $23.8$
\magarcsecsq\ and the low rotation speed. The low stellar surface
brightness and large flaring make this galaxy an excellent candidate
for measuring the shape of the dark halo.

\subsection{ESO274-G001}

As mentioned above, the \HI\ imaging of ESO274-G001 is also missing
some extended large scale structure due to a lack of \HI\ observations
using array configurations shorter than $1.5$ km. This missing
structure was clearly apparent in the flaring measurement process as the 
measurements of the \HI\ cube modeled with the initial measured flaring
were much lower than the flaring measured from the observed \HI\ channel maps.
The best fit model displayed in
Fig.~\ref{fig:ch6-e274-g001-flaring} shows that the flaring rises
roughly linearly from a FWHM$_z$ thickness of $200$ pc at less than
$1$ kpc radius to $1.9$ kpc at $5.5$ kpc radius.

\subsection{ESO435-G025 (IC2531)}

IC2531 displays a very large outer and remarkably symmetric flare (see
Fig.~\ref{fig:ch6-e435-g025-flaring}), rising from a gas thickness
of $800$ pc at $4.5$ kpc radius to 3-4 kpc at $30$ kpc. The central
gas thickness was initially partially unresolved given the
FWHM$_{\theta}$ beamwidth of $1.3$, and was decreased to the final
thickness by iterating. The large gas thickness uncertainty of between
$100$ and $500$ pc was due to the low peak channel map signal-to-noise
of $10.9$. 

\begin{figure}[t]
  \centering
\includegraphics[width=9cm]{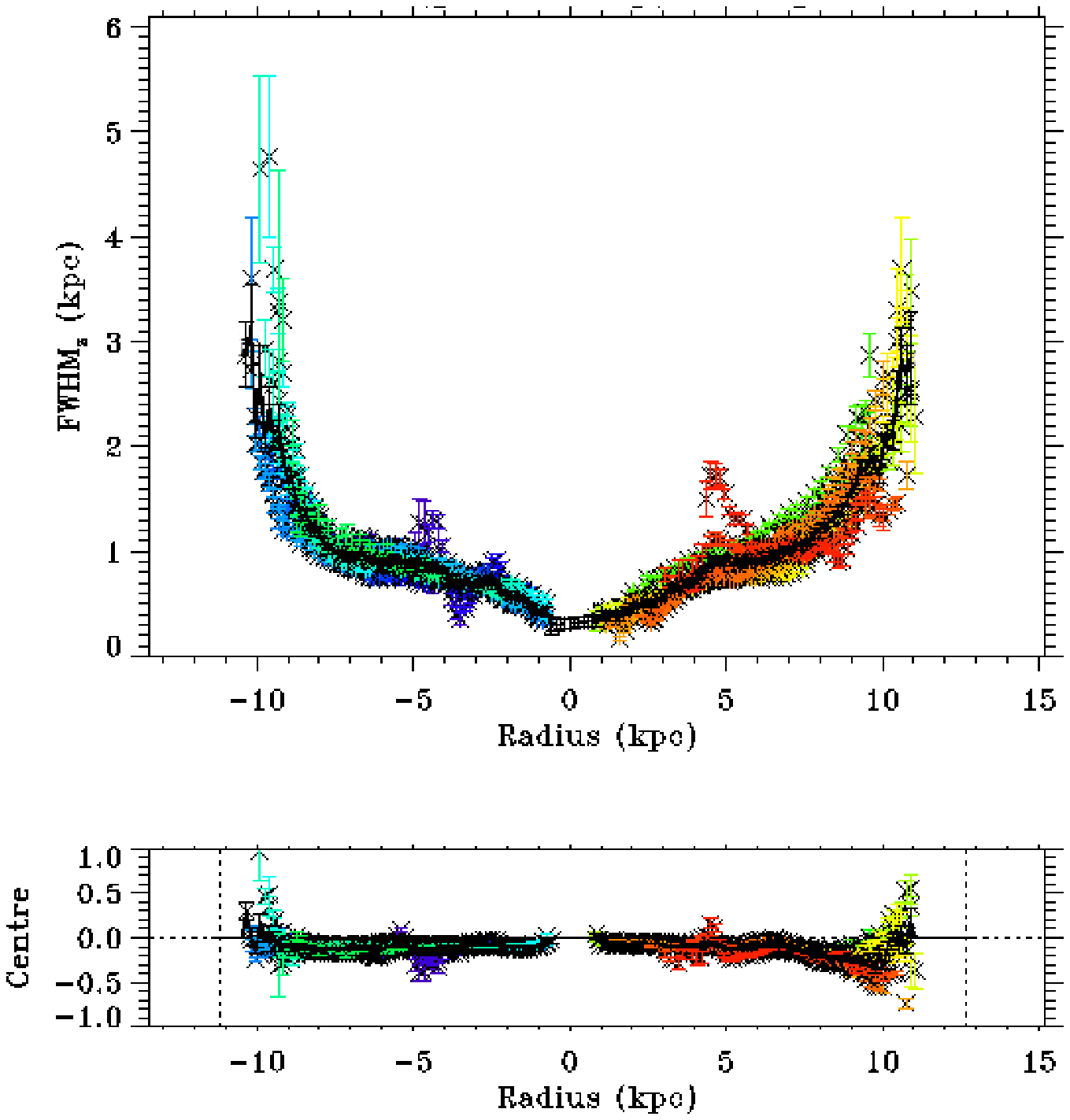}
  \caption[UGC7321: Measured \HI\ FWHM$_z$ thickness and \HI\ midplane
  location as a function of radius, where the measurements are
  colour coded by the velocity of each channel map]{Measured \HI\
    FWHM$_z$ thickness (upper panel) and \HI\ midplane location 
(lower panel) of
    UGC7321 as a function of radius, where the measurements are colour
    coded by the line-of-sight velocity of each channel map.
The most bluewards points were measured from the extreme approaching
velocities, and those most redwards were measured from the extreme
receding velocities. }
  \label{fig:ch6-u7321-deviation}
\end{figure}

\begin{figure}[ht]
  \centering
\includegraphics[width=8cm]{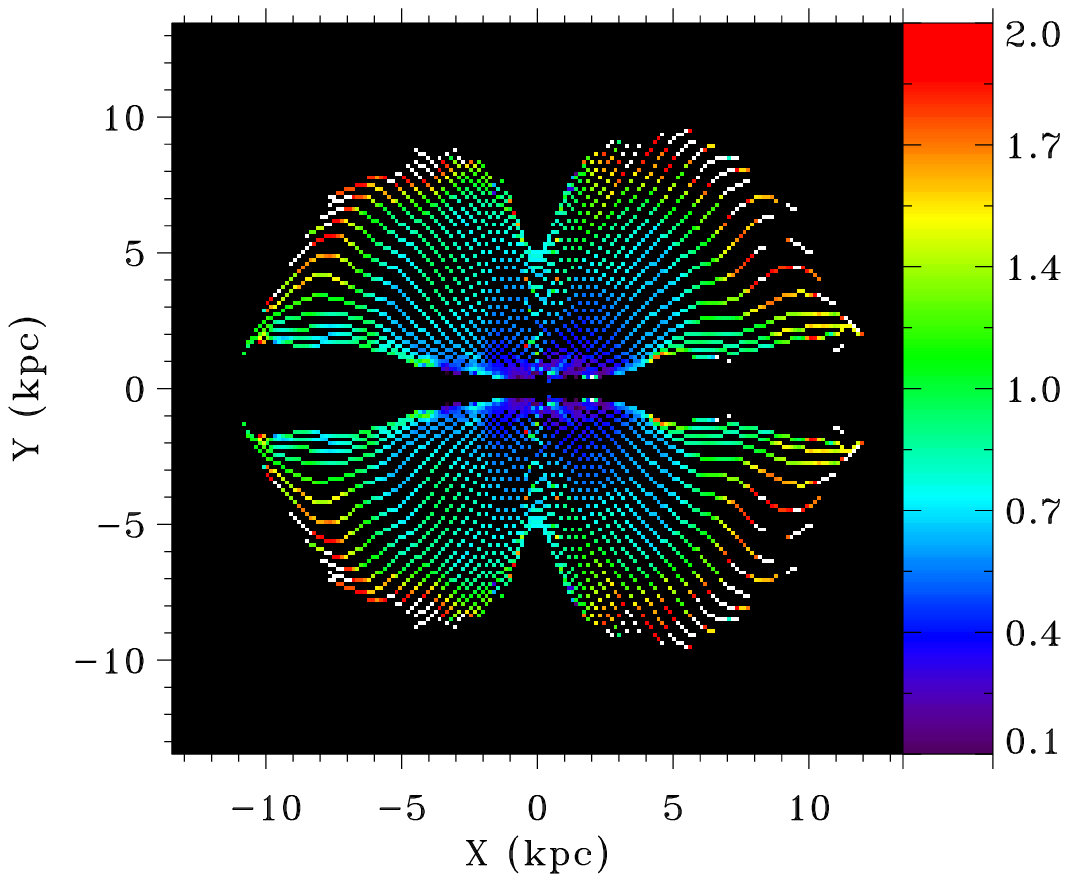}
\includegraphics[width=8cm]{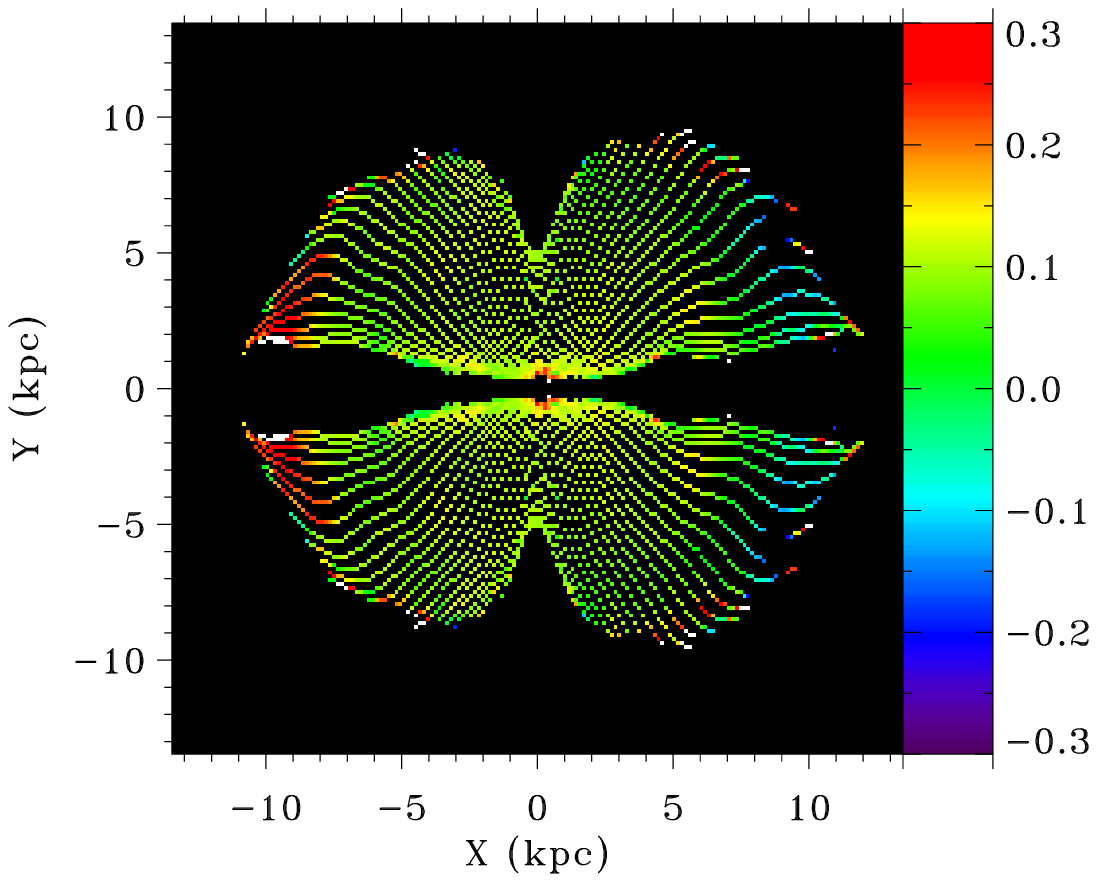}
  \caption[UGC7321: Measured \HI\ FWHM$_z$ thickness and \HI\ midplane
  location deprojected onto the X-Y plane]{Measured FWHM$_z$
    thickness (upper panel) and midplane displacement (lower panel) 
of the \HI\ disk
    of UGC7321 deprojected onto the X-Y plane, where the color scale
    on each plot shows the respective FWHM$_z$ or midplane
    displacement measurements.  Measurements in white are off the top
    end of the color scale.  Each line of measurements forming a
    coloured track shows the measurements from a single \HI\ channel
    map. The measurements are the same on each side of the $Y=0$ line,
    as due to the edge-on orientation of the gas disk, we can't
    determine if \HI\ emission comes from the near or far side of the
    disk.}
  \label{fig:ch6-u7321-azimuth-structure}
\end{figure}

\subsection{UGC7321}
\label{sec:ch6-u7321}

The measured flaring of UGC7321 is our most accurate determination of
the flaring in our sample, rising from a FWHM$_z$ thickness of $350$
pc at a radius of $0.5$ kpc to roughly $4$ kpc at $11$ kpc from the galactic 
centre. The $16$
hours of VLA time obtained by Matthews enabled the \HI\ channel
maps to be imaged with much better sensitivity than that acquired for the other galaxies
observed with the ATCA. As a result the error bars on the UGC7321 FWHM$_z$
thickness measurements are only 15-50 pc, except in the outermost
2-3 kpc. The relatively flat rotation curve also allowed the flaring
at most radii to be independently measured in over 10-20 different
channel maps.

The flaring profile of UGC7321 exhibits the high degree of symmetry 
of the gas thickness about the galactic center as expected for an 
undisturbed, isolated galaxy. At all radii the flaring profiles are symmetric
about the galactic centre to within the bounds of the errors bars 
(1$\sigma$ errors are $\Delta
{\rm FWHM}_z < 50$ pc within a radius of $9$ kpc). As the \HI\ velocity
dispersion as a function of radius is also very symmetric, this shows 
that the total vertical force must also be symmetric. 
Given a symmetric luminous density distribution,
this implies that the vertical force and density of the halo are
symmetric about the galaxy centre.

The most striking feature of the \HI\ flaring profile of UGC7321 is the change in the
shape of the flaring profile near the edge of the stellar disk. In the inner disk the 
\HI\ thickness appears to increase linearly with radius, 
while outside the stellar disk it
displays an exponential profile. In UGC7321, this ``break'' occurs at $7.8$ kpc, 
which is between the stellar truncation radius of $6.7$ kpc \citep{pbld2003} and the 
maximum stellar disk radius of $8.1$ kpc \citep{mw2003}. 

The vertical thickness of the \HI\ disk of UGC7321 has also been
studied by \citet{mw2003} who, as in the study of NGC2403 by \citet{fvmso2002},
tried to discern the \HI\ flaring from the \HI\ total intensity
map. \citet{mw2003} found that the best fit to the observed \HI\ total
intensity map was achieved using a thin \HI\ disk that flares from a
FWHM$_z$ of $260$ pc at the center to $1.64$ kpc at $11.8$ kpc, and 
an additional low density halo 
\HI\ component of constant FWHM$_z$ of $3.3$ kpc that lagged 
behind the disk gas rotation.  \citet{mw2003} parameterised the thin \HI\ disk 
using $h_f = h_1 + h_{f0} r^{5/4}$. This produced a smooth flaring profile, but 
lacked the steep exponential increase we measured in gas thickness at radii outside 
of the stellar disk of UGC7321 and most of the galaxies in our sample (see
Sect.~\ref{sec:ch6-disc}).The addition by \citet{mw2003} of a low density second gas
component with a large constant FWHM$_z$ seems to simulate the thick
gas layer at the outer radii. The filamentary nature of
this structure suggests that it is \HI\ ejecta, originating in the inner
disk, rather than a pervasive diffuse thick \HI\ layer. 
As shown below, our direct flaring measurements of UGC7321
   show that the gas layer can be well represented by a single gaussian
   gas component with a FWHM that varies smoothly with radius: modelling
   the gas layer thickness by two gas components appears to be
   unnecessary.

The high signal-to-noise of these observations allowed
inspection of the intrinsic shape of the vertical gas thickness. The left-hand
panel of Fig.~\ref{fig:ch6-u7321-all-profiles} displays the observed vertical
profiles in a sample of galactocentric radial bins on both sides of
the galactic centre. The profiles show the number of independent
measurements in each bin and the flaring of the gas thickness on both
sides of the galaxy. In order to inspect the vertical shape more
closely, we averaged the profiles in each bin after aligning the
profile centres and vertically rescaling all profiles to a peak of $1$
(see Fig.~\ref{fig:ch6-u7321-all-profiles}, right-hand panel). 
This shows that the profiles are indeed Gaussian in nearly all radii bins, with 
small high-$z$ wings seen only in annuli at radii of $\sim$4 and $\sim$8 kpc.

\begin{figure*}[t]
  \centering
\includegraphics[width=16cm]{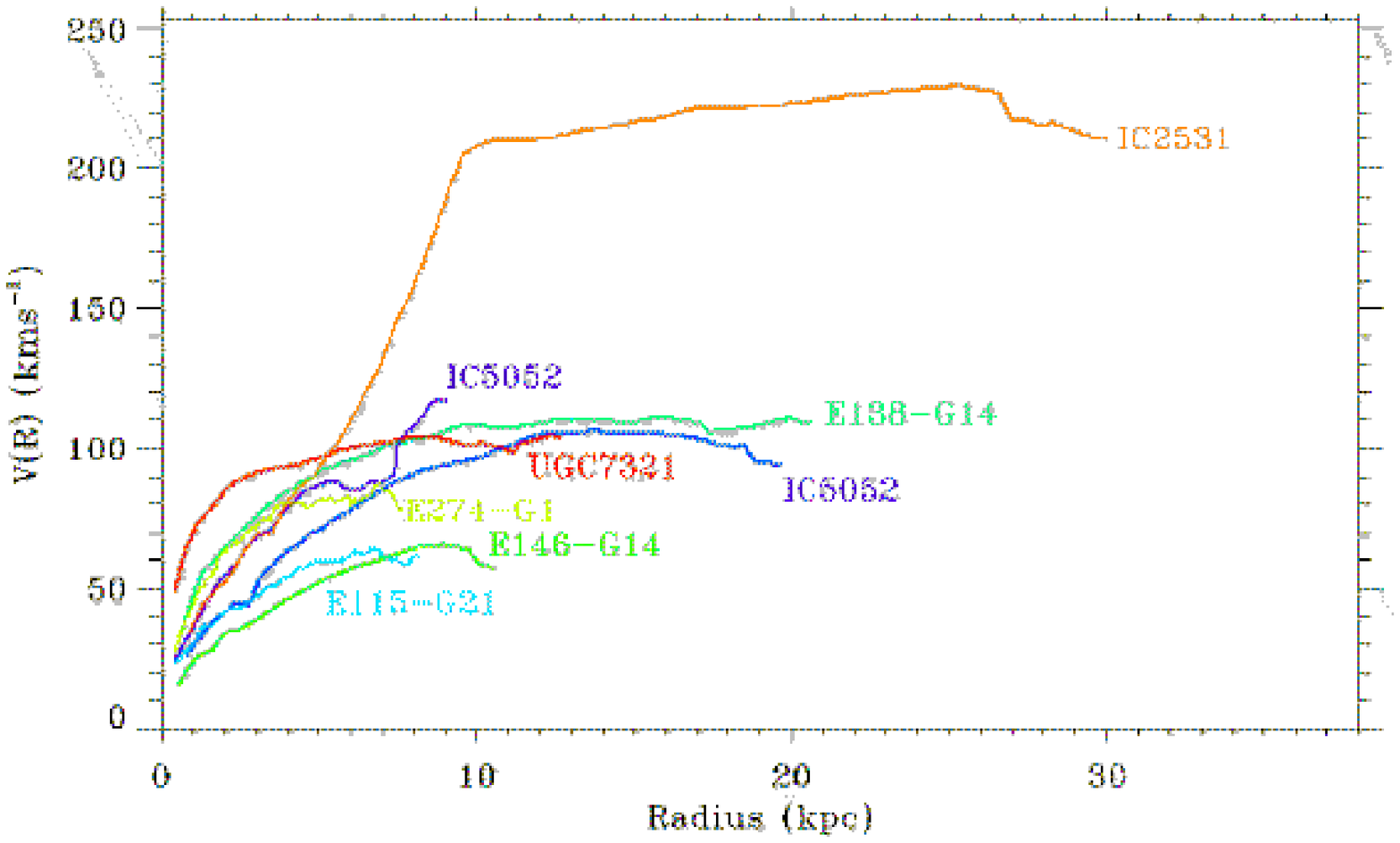}
  \caption[The rotation curves of all galaxies overlaid
  on each other. Likewise for the \HI\ velocity dispersion as a
  function of radius, and face-on \HI\ surface density for all galaxies
  in our sample]{The
\HI\ rotation curves 
for all galaxies in our
    sample overlaid on each other for comparison.}
  \label{fig:ch5-sample-all_vabs}
\end{figure*}

In Fig.~\ref{fig:ch6-u7321-deviation} we show the direct FWHM$_z$
(upper plot) and midplane position (lower plot) fits of the \HI\ disk.
The FWHM$_z$ shows a clear increase in 
thickness at radii between $3.5$ and $5.5$ kpc on both sides
occuring on both sides of the galactic centre. The colors show that
this occurs only in the extreme velocity channels, corresponding to
\HI\ in the line of nodes. The lower plot shows that the \HI\ midplane
also deviates at this radius and azimuth in the X-Y plane, dropping
below $\sim$200 pc below the average midplane position on the left-hand
side, and rising approximately $150$ pc above the galactic plane on
the right-hand side.

In Fig.~\ref{fig:ch6-u7321-azimuth-structure} we show the FWHM$_z$
thickness (upper panel) and midplane position (lower panel) measurements
deprojected onto the X-Y plane, where the color scale on each plot
shows the respective FWHM$_z$ or midplane displacement measurements.
Each
line of measurements forming a coloured ``track'' shows the measurements
from a single \HI\ channel map. The measurements are the same on each
side of the Y=0 line as, due to the edge-on orientation of the gas
disk, we can't determine if \HI\ emission comes from the near or far
side of the disk. Below we briefly describe the \HI\ warp of the
midplane before discussing the unusual azimuthal variation of both
the thickness and midplane displacement mentioned
above.

The midplane displacement plot (lower panel) shows that disk plane warps by
300-500 pc in the plane of the sky on the left-hand side centred
around Y=0, while on the right-hand side the warp amplitude is similar,
peaking at a small angle of $\sim$30$^{\circ}$ to the line-of-nodes. This
implies that the measured thickness is uneffected by the warp on the
left-hand side of the centre, while on the right-hand side the measured
thickness of the direct loop 1 fit may have been slightly
overestimated due to the projection of the warp. These effects of
warping are corrected by iterating the measurement as explained in
Sect.~\ref{sec:ch6-method}.

Also visible is a pronounced deviation of the midplane and
FWHM$_z$ thickness of gas in the line of nodes at radii between
2.5-5.0 kpc on the left-hand side of the galaxy and at radii from 4-5.5
kpc on the right-hand side. This midplane displacement is seen most clearly
in the lower plot of Fig.~\ref{fig:ch6-u7321-deviation}; while this
deviation of the gas thickness from the mean flaring can be seen in
both the upper plot of Fig.~\ref{fig:ch6-u7321-deviation} and the
upper panel of Fig.~\ref{fig:ch6-u7321-azimuth-structure}. One
explanation for the increased broadening and tilting of the disk
midplane along the Y=0 axis at this radius, is that the gas is
responding to the local perturbation of the total potential by the
suspected bar. Peanut-shaped deviations from elliptical isophotes in
the projected R-band image \citep{pbld2003} favour a bar oriented in the
plane of the sky (ie. along the Y=0 axis), rather closer to end-on
which would align the bar axis with the line-of-sight.  The measured 
enhanced local gas thickening along the Y=0 axis 
probably is the best observational evidence of
a bar in an edge-on galaxy.\footnote{We thank Lia Athanassoula for
pointing this out.} Other explanations for localised regions
of increased thickness are infalling material from a companion or
disruption by a companion, however UGC7321 is an isolated galaxy.  The
high symmetry of the flaring on both sides of the centre, and the high
stellar disk major-to-minor axis ratio ($a/b=10.3$ at $\mu_R=25.0$
\magarcsecsq, Matthews, Gallagher \&\ van Driel 1999) 
render disruption by an undetected neighbouring galaxy rather implausible.

Another noticable feature of
Fig.~\ref{fig:ch6-u7321-azimuth-structure} is that the tracks
denoting the location of the measurements in the X-Y plane display
wiggles which systematically vary over a range of azimuthal angles.
As our deprojection method is
based on the velocity field, these wiggles could be due to deviations
from circular rotation in the velocity field causing small errors in
the deprojection\footnote{Private communication with Agris Kalnajs.}. 
Kinks are typically seen in the velocity fields
of less inclined galaxies with strong spiral arms \citep[e.g.
M81,][]{rots1974} suggesting that, if seen face-on, UGC7321 may
exhibit weak spiral arms in the gas distribution.

\section{Discussion}
\label{sec:ch5-disc}

\subsection{\HI\ rotation curves}

In Fig.~\ref{fig:ch5-sample-all_vabs} the rotation curves
of our galaxies are overlaid. 
IC 2531 stands out as a massive, large galaxy compared to the rest
of the sample. We have two very slow rotators with maximum rotation
speeds of $\sim $60 km/s in our sample, while the others form a group of 
galaxies rotating at $\sim $100 km/s.
The shape of the rotation curves are in good agreement with the
rotation curves of 60 less inclined late-type dwarf galaxies measured
by \citet{swaters1999}. Although low
mass LSB galaxies exhibit a shallower inner rotation gradient, our
high resolution observations clearly resolve differential rotation, in
agreement with other observations by \citet{swaters1999} and \citet{dbb2002}.

\begin{figure*}[ht]
  \centering
\includegraphics[width=16cm]{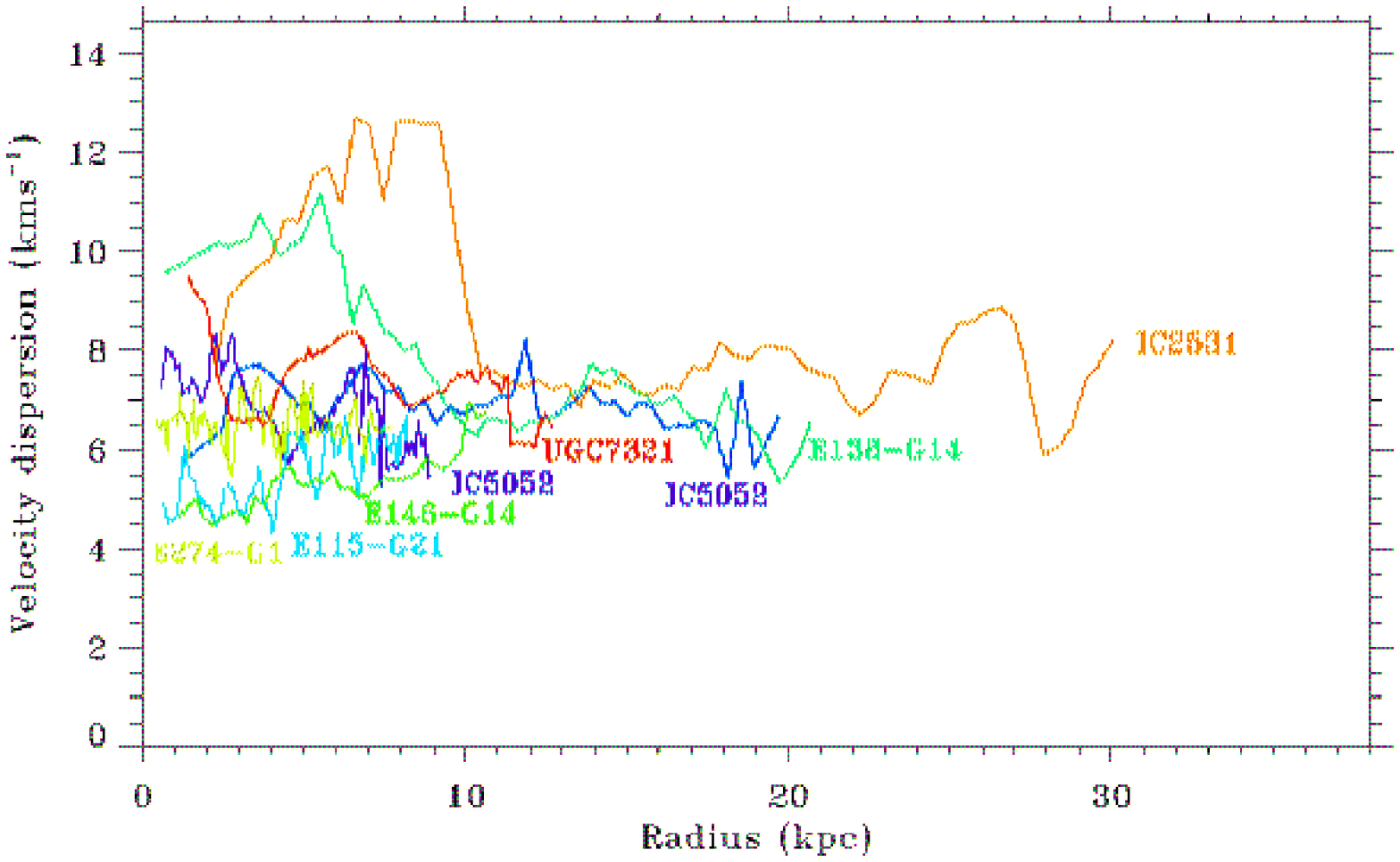}
  \caption[The rotation curves of all galaxies overlaid
  on each other. Likewise for the \HI\ velocity dispersion as a
  function of radius, and face-on \HI\ surface density for all galaxies
  in our sample]{The
\HI\ velocity dispersion profiles
for all galaxies in our
    sample overlaid on each other for comparison. }
  \label{fig:ch5-sample-all_disp}
\end{figure*}

\subsection{\HI\ velocity dispersion}

The \HI\ velocity dispersion measurements presented in this paper
comprise the highest spatial resolution measurements of velocity dispersion 
to date and the first measurements using data that span the
radial extent of each galaxy by more than $10$ \HI\ synthesised
beams. Fig.~\ref{fig:ch5-sample-all_disp} shows that the
\HI\ velocity dispersion of our galaxies ranges from $4.5$ \kms\ to
$12.5$ \kms, with most galaxies displaying a \HI\ velocity dispersion
of $6.5$ to $7.5$ \kms. Of the eight galaxies in our sample, all display 
radial structure in \HI\ velocity dispersion, except
ESO274-G001 which has an approximately constant \HI\ velocity dispersion of
$6.5$ \kms.

\begin{figure}[t]
  \centering
\includegraphics[width=9cm]{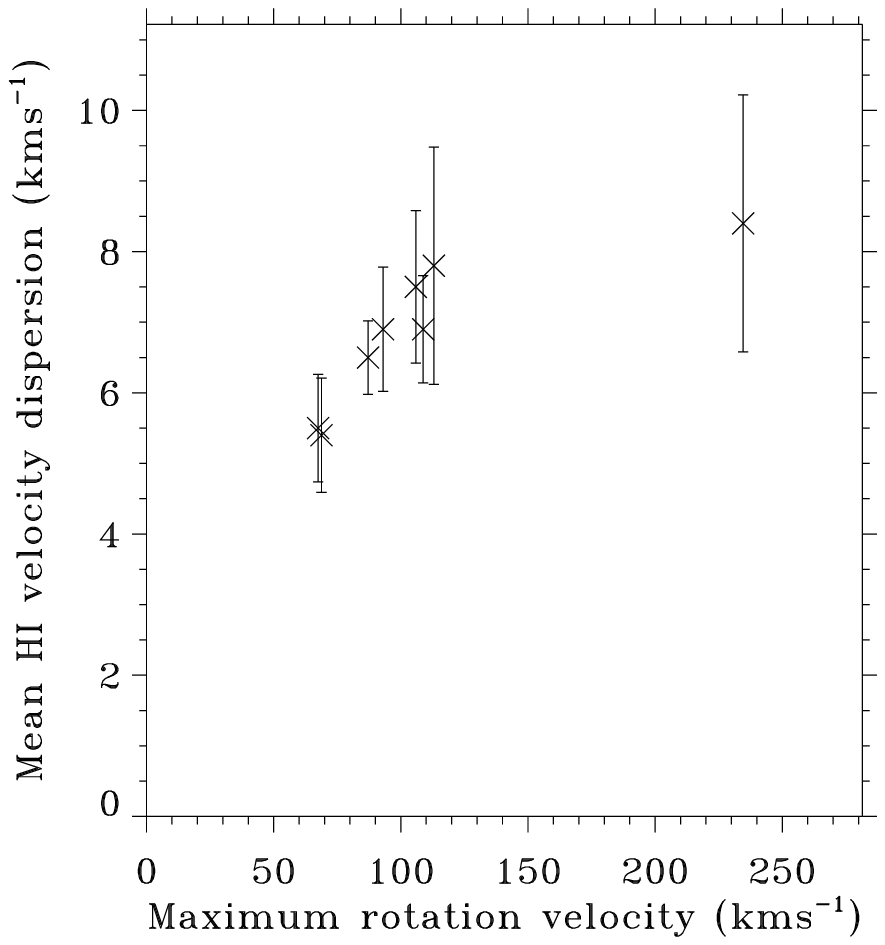}
  \caption[Mean \HI\ velocity dispersion vs. maximum rotation speed for
  each galaxy]{Mean \HI\ velocity dispersion $\langle \sigma_{v,\HI}(R)
    \rangle$ against the maximum rotation speed $V_{max}$ of each
    galaxy. The error bars display the standard deviation of the \HI\
    velocity dispersion for each galaxy, showing that more massive
    galaxies tend to show larger radial variation of the \HI\ velocity
    dispersion.}
  \label{fig:ch5-sample-mean_disp_vmax}
\end{figure}

Fig.~\ref{fig:ch5-sample-mean_disp_vmax} shows the mean \HI\
velocity dispersion as a function of the maximum
rotation velocity of the \HI\ disk. There is a definite increase in
the mean velocity dispersion with rotation speed from about $5$ to $8$
\kms\ for rotation velocities $70$ to $120$ \kms, but IC2531 has the same velocity
dispersion of approximately $8$ \kms\ although its maximum rotation speed is
$230$ \kms. \citet{dyb2004} found that edge-on galaxies with rotation velocity
greater than $120$ \kms\ have well-defined dustlanes, while those with lower
rotation do not. Dalcanton
et al. suggested that well-defined dust lanes are related to
lower gas turbulence and lower scale heights for the dust layer
in the more rapidly rotating galaxies. Do our data support this
suggestion?

From Eqn. (1), the scaleheight $h$ of a Gaussian dust layer is
related to the ISM velocity dispersion $\sigma$ by
\begin{equation}
\frac{\sigma^2}{h^2} = - \frac{\partial K_z}{\partial z}
= 4\pi G \rho_\circ,
\end{equation}
where $\rho_\circ$ is the mid-plane total density which is
dominated by baryons in normal spirals. The density $\rho_\circ =
\Sigma/2h_*$ where $\Sigma$ is the typical surface density of a
disk and $h_*$ is the baryon scale height.  It turns out that
$\Sigma$ and the stellar scaleheight are both approximately
proportional to the maximum circular velocity $V_c$ \citep{gfjp09,kvdkdg02}, 
so the typical value of $\rho_\circ$
is independent of $V_c$ and we expect the scale height of the dust
layer to be directly proportional to the ISM velocity dispersion
$\sigma$. In our sample of galaxies, the galaxy with the largest
$V_c$ (IC2531) shows a clean well-defined dustlane as expected
from the Dalcanton et al. sample. However Fig. 22 shows no evidence
for a decrease in $\sigma$ as $V_c$ increases, so our data do not
suppport the variable turbulence explanation for the change in dust
lane morphology with $V_c$.

Two of the larger galaxes, IC2531 and ESO138-G014, display \HI\
velocity dispersions 2-3 \kms\ larger at small radii coinciding with
the inner stellar disk. IC2531 is a large Sc galaxy with a strong bar,
while the \HI\ XV map of ESO138-G014 indicates a possible bar.  The
other barred galaxy in our sample, UGC7321, displays the next highest
central \HI\ velocity dispersion.
These three galaxies have a relatively large standard deviation in 
the \HI\ velocity dispersion as often observed in more face-on
strongly barred galaxies like NGC1365, presumably caused
by gas shocks and increased star formation. Unfortunately, 
ESO138-G014 was not in our optical sample due to its low galactic latitude, so we
were unable to confirm the existence of a bar by investigating the
stellar isophotes. 

The velocity dispersion profile of UGC7321 exhibits three broad
localised peaks with velocity dispersion 1-2.5 \kms\ higher than the
intervening medium. These occur at the galactic centre, and at radii
of $6$ kpc and $10$ kpc.  Stellar observations of UGC7321 by
\citet{pbld2003} show a central peanut-shaped stellar over-density with a
radius of $2.0$ kpc, which could be caused by
resonant off-plane thickening via bar buckling \citep[e.g.][]{cdfp1990}.
The rise of \HI\ velocity dispersion from $6.5$ \kms\ to
$9.5$ \kms\ in this region suggests that processes associated with bar
buckling could be providing the additional energy heating the gas. The
second \HI\ velocity dispersion peak, from $4$ kpc to $7$ kpc,
corresponds to the radius of the bar shown in the ``figure-8'' \HI\
signature and the stellar isophotes analysis by \citet{pbld2003}. The
third \HI\ velocity dispersion is of lower amplitude and occurs outside
the stellar disk. These results suggest that at least part of the
additional gas heating seen in the high inner \HI\ velocity dispersions
of barred galaxies in our sample could be due to star formation along
radial flows in the bar, and bar buckling. 

Observations of the face-on galaxies NGC3938 \citep{vdks1982c}, 
NGC628 \citep{svdk1984}, NGC1058 \citep{vdks1984,pr2006}, NGC6946
\citep{bv1992} and NGC5474 \citep{rdh1994} have shown that the \HI\ velocity
dispersion ranges from $4$ to $16$ \kms, and has generally been assumed to
be 6-8 \kms\ in the absence of significant star formation. In NGC 6946, 
detailed observations by \citet{ks1993} and \citet{bofvdhs2008}
indicate the presence of widespread high velocity gas with significant kinetic
energy, while the radial \HI\ velocity dispersion profile shows a peak of 
more than 10 \kms\ within $R_{25}$, followed by a gradual decline. Large-scale
turbulence probably plays a role in this. The cause of the \HI\ velocity
dispersions is complicated and not well-understood, but turbulence driven by 
star formation, supernovae, shearing motions and magnetic instabilities all
may play a role \citep{trlmlwkbdb2009,altmmr2009}.

The
typical velocity dispersion of $6.5$ to $7.5$ \kms\ of galaxies in our
sample is in agreement with these observations, which is to be
expected given the LSB nature of all the galaxies in our sample (except
IC2531 and IC5052). On small scales it is assumed that increased \HI\ velocity
dispersion is due to heating of the ISM by stellar winds from hot OB
stars, supernovae and expanding \HII\ regions 
\citep[e.g.][]{mcko1977,spitzer1978}. Star formation is known to increase
the turbulence of the ISM \citep{os1955}, as the newly formed stars
eject gas from the cloud within which they formed by radiation
pressure and stellar winds.  Although limited by a small sample size, 
our results suggest that increased gas velocity dispersion in disk galaxies is linked 
to the presence or prior formation of bars, which could cause the widespread
star formation required to form strong dust lanes. However, it could also 
be caused by star formation in spiral arms.

\begin{figure*}[t]
  \centering
\includegraphics[width=16cm]{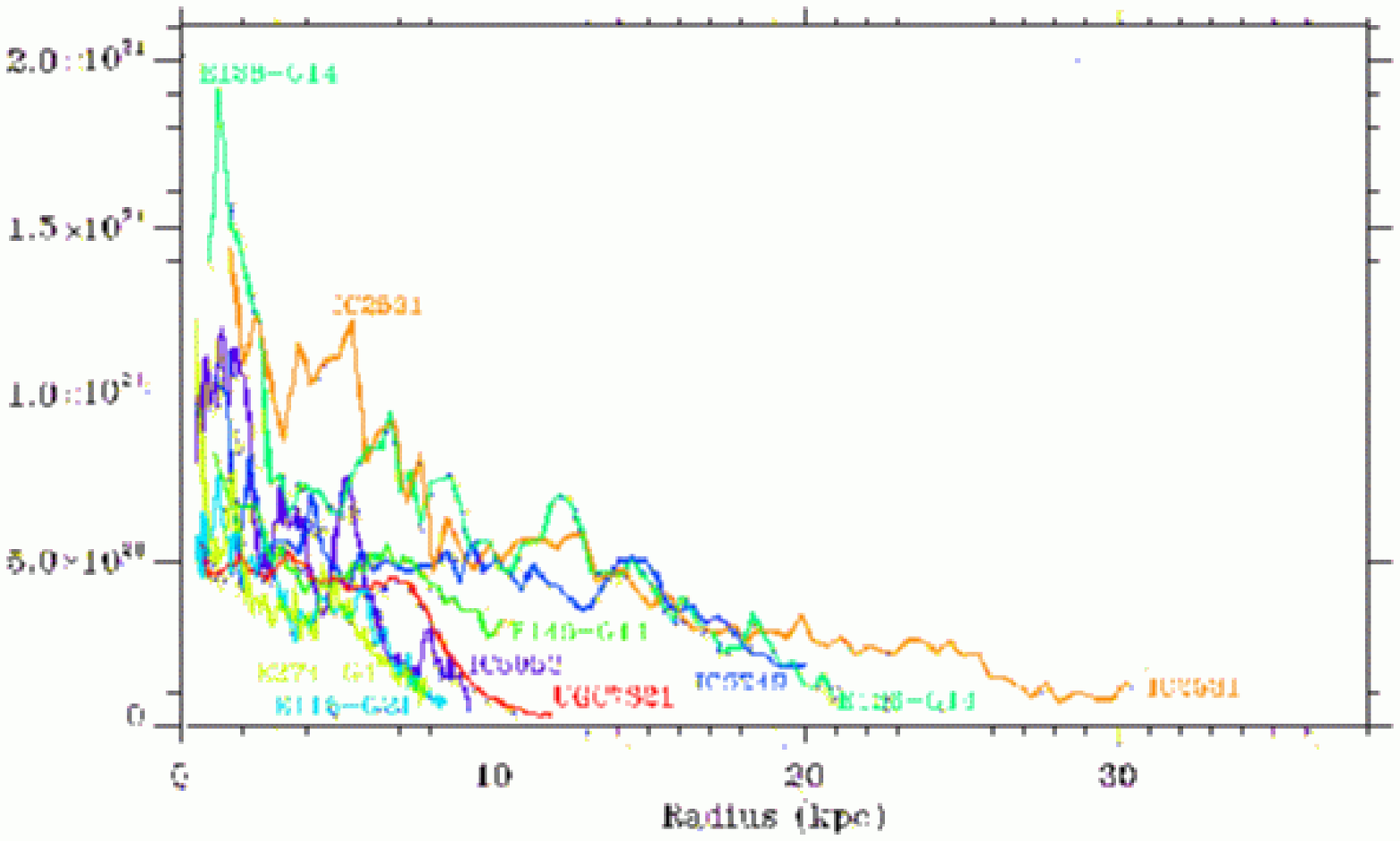}
  \caption[The rotation curves of all galaxies overlaid
  on each other. Likewise for the \HI\ velocity dispersion as a
  function of radius, and face-on \HI\ surface density for all galaxies
  in our sample]{The
face-on surface density profiles
for all galaxies in our
    sample overlaid on each other for comparison.}
  \label{fig:ch5-sample-all__surf}
\end{figure*}

The most recent observations of NGC1058 by \citet{pr2006} are the only
\HI\ velocity dispersion study of a face-on galaxy with high spectral
resolution ($2.58$ \kms). They found a general radial trend of a
declining \HI\ velocity dispersion
which was punctuated by many peaks and saddles where 
it ranged from $4$ to $14$ \kms. Two of
the three  regions with high \HI\ velocity dispersion did not coincide
with star formation tracers such as \Ha\ or radio continuum
emission. For the third one it did, but that region
was located at the galactic center. The high spatial
frequency of \HI\ velocity dispersion variation measured in our sample
also suggests that the small scale velocity structure of the neutral ISM is likely to be
quite complex.

If the canonical \HI\ velocity dispersion of $7$ \kms\ found in the
outer disks of galaxies were due solely to thermal motion, the gas
temperature would need to be $T_k \approx 6000$ K. In the outer Galaxy
where it is possible to measure the kinetic temperature and the
velocity dispersion, \citet{spitzer1978} found a thermal
contribution to the \HI\ velocity dispersion of only $\sim$1 \kms,
indicating that most of the broadening is non-thermal in origin. 

One possible explanation is that differential galactic rotation
transforms rotational energy into turbulent motions, perhaps via
shocks resulting from disk instabilities
\citep{quiroga1983,rk1987,sb1999}. Shocks in galactic bars
could also generate  gas heating. High resolution
hydrodynamical simulations, including gaseous self-gravity, realistic
radiative cooling, and rotation by \citet{wmn2002} show that large
scale rotation combined with gas self-gravity could be the source of
the additional non-thermal energy broadening the gas velocity
dispersion.  Their models of turbulent rotating disks are consistent
with the velocity dispersion of the extended \HI\ disk in the dwarf galaxy
NGC2915, which appears to have no prominent star formation. Their
numerical simulations also show that the stellar bar in NGC 2915
enhances the \HI\ velocity dispersion, which supports our finding of
higher velocity dispersion in the inner disks of IC2531, ESO138-G14
and UGC7321

\subsection{\HI\ radial surface density}

Fig.~\ref{fig:ch5-sample-all__surf} shows the measured
radial \HI\ surface density of each galaxy in our sample. UGC7321,
ESO274-G001 and ESO115-G021 all show a steep exponential decline at
large radii, while IC5052 and ESO138-G014 appear to exhibit a roughly
linear decline. With the exception of IC5052, most galaxies appear to
show roughly smooth \HI\ surface densities.

\subsection{Flaring of \HI\ layers}
\label{sec:ch6-disc}

\begin{figure*}[t]
 \centering
\includegraphics[width=16cm]{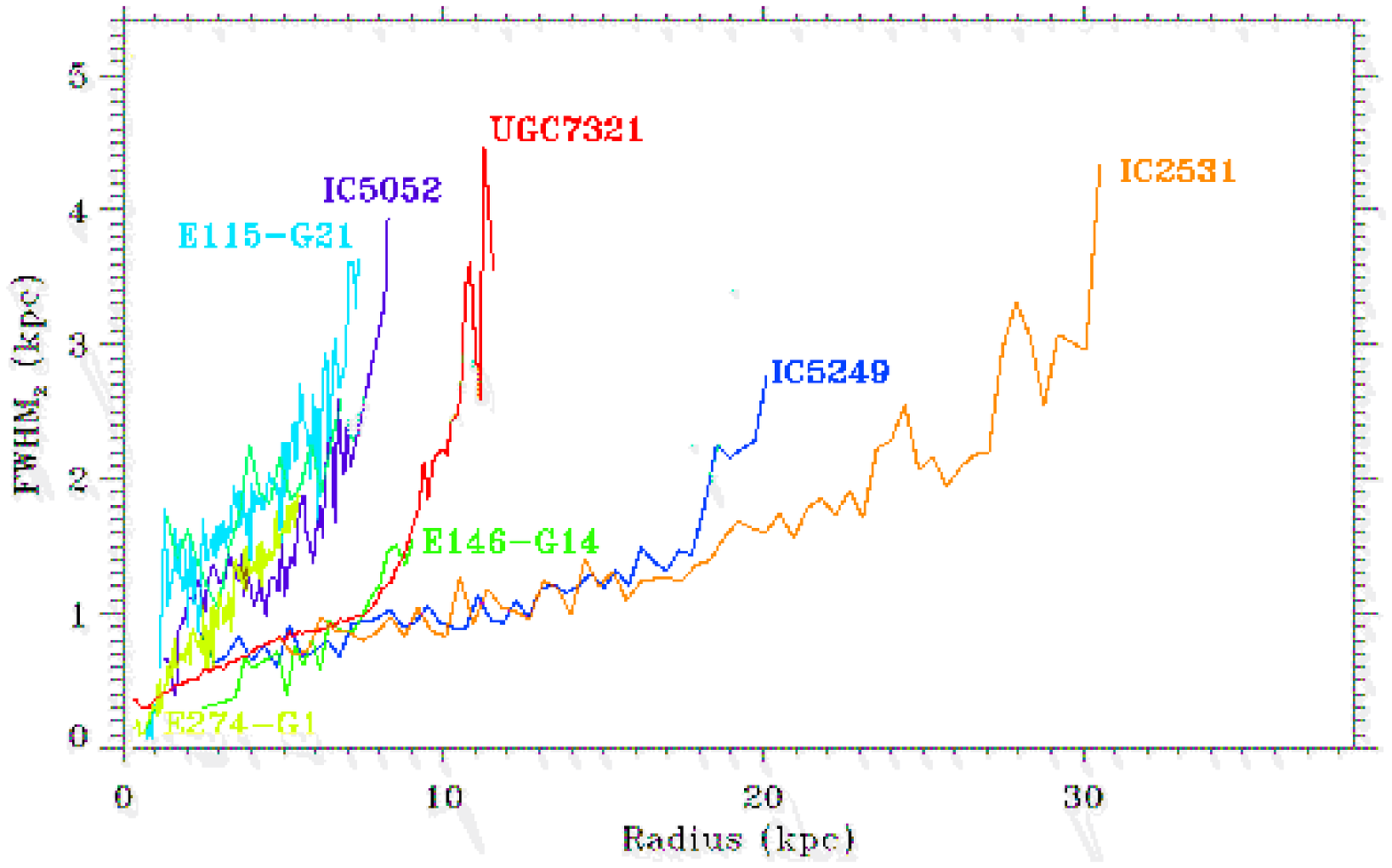}
  \caption[The FWHM thickness as a function of radius of all
  galaxies in the sample overlaid on each other]{Measured FWHM$_z$
    thicknesses of the \HI\ gas disks. ESO 138-G14 has been omitted from
this figure, as explained in Sect.~\ref{subsection:ESO138-G014}.
}
  \label{fig:ch6-sample-all_flares_vabs}
\end{figure*}

In Fig.~\ref{fig:ch6-sample-all_flares_vabs} we display a plot of
all the flaring profiles for each galaxy in our sample. The \HI\
flaring profiles appear to obey a typical shape, with the flaring
increasing linearly with radius where the stellar disk dominates the
local gravitational potential, and steepening to an exponential
profile in the outer disk where the potential is probably dominated by
the halo. This break in the change of the flaring profile is seen most clearly in UC7321. 
The onset of exponential curvature appears to occur just
outside the stellar truncation radius in the more massive galaxies with
maximum rotation speed $V_{max} > 100$ \kms. In smaller, lower surface 
brightness disk galaxies with $V_{max} < 100$ \kms, the break possibly occurs 
at smaller radii due to dark matter dominating the inner gravitational potential.

All of the galaxies in our sample display this flaring profile, except 
ESO138-G014 which is considered unreliable due to missing large-scale \HI\ structure 
(see Sect.~\ref{subsection:ESO138-G014}).
Unfortunately, we were unable to detect this radial change in flaring profile shape in 
ESO274-G001, the galaxy with the highest spatial resolution in our sample, as we 
were unable to measure the gas thickness at radii outside the stellar disk, due to a lack of 
short-spacing observations. Excluding ESO138-G014, all the more
massive galaxies ($V_{max} > 100$ \kms) display a shallower flaring
profile over the inner disk. With the exclusion of ESO146-G014, the
three remaining small disk galaxies ($V_{max} < 100$ \kms) all display
steep linear inner flaring profiles. 

The flaring profile of ESO146-G014 shows remarkable similarities to the 
flaring profile of UGC7321. Apart from a similar central face-on B-band stellar
surface brightness (respectively $23.4$ and $23.8$
\magarcsecsq), these two galaxies appear quite different --- ESO146-G014 has a much lower rotation speed and \HI\ velocity dispersion.

The three well-measured large
galaxies (IC2531, IC5249 and UGC7321) and the unusual ESO146-G014,
each display one common profile with a small FWHM$_z$. A
common flaring profile in the inner disk suggests a common ratio of
the \HI\ velocity dispersion to the stellar density, assuming the
stellar density dominates at low radii. The ratio of mean \HI\ velocity
dispersion to maximum rotation speed is very similar in IC5249,
UGC7321 and ESO146-G014, the value for IC2531 is roughly half that of
the others.

The radial gradient of the flaring is more varied in the smaller
galaxies. The coalignment of the inner flaring slopes of UGC7321,
IC5249 and IC2531 is probably a coincidence, as a comparison of
IC2531's flaring profile to that of the Galaxy \citep{om2000} shows
that the gas disk of IC2531 is approximately twice as thick as the
Galactic \HI\ layer.

\begin{figure*}[t]
 \centering
\includegraphics[width=16cm]{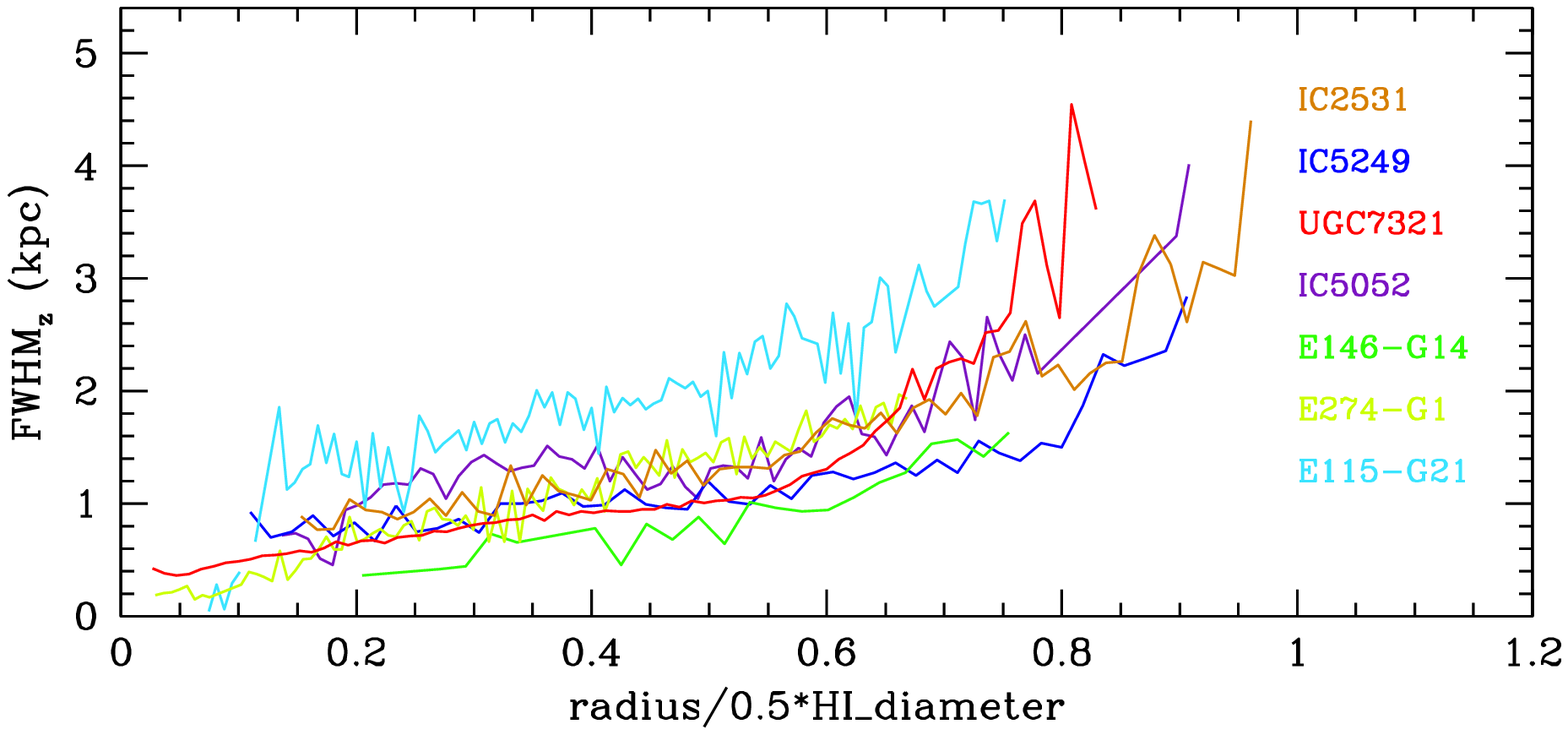}
  \caption[The FWHM thickness as a function of radius of all
  galaxies in the sample overlaid on each other]{Measured FWHM$_z$
    thicknesses of the \HI\ gas disks. The horizontal axis is radius scaled by the
half the \HI\ diameter from paper I.
}
  \label{fig:ch6-sample-all_flares_scaled}
\end{figure*}

In Fig.~\ref{fig:ch6-sample-all_flares_scaled} we show the flaring
profiles with the horizontal axis scaled with (half) the diameter of the \HI\
distribution, taken from paper I. This does make the profiles more comparable
in shape,  although still there is quite some variation of the thickness of the 
\HI\ in kpc from galaxy to galaxy.
We have done the same by scaling to (half) the optical radius, taken from
NED\footnote{This NASA/IPAC Extragalactic Database is operated by the Jet 
Propulsion Laboratory, California Institute of Technology, under contract with 
the National Aeronautics and Space Administration.}. Not surprisingly, the 
result is similar since the optical and \HI\ diameters do 
in general correlate among galaxies, 
although the relative positions of the lines with respect to each other is different.
For comparison, the flaring profile of NGC 891 \citep{vdkruit1981} would be on
the scale of Fig.~\ref{fig:ch6-sample-all_flares_scaled} run from about 0.3 kpc at 
0.2 scaled galactocentric radius to 1.9 kpc at 1.0. This is thinner
than in our systems, but that is not surprising since our sample was
selected to be made up of galaxies with low contributions of the disks to
the gravitational potential and the thickness of the \HI\ layer
is to a larger extent determined
by that of the dark matter halo, which would be less flattened than the disk.


In this work we have measured the vertical thickness of the \HI\ disk
by fitting a single Gaussian gas component. We have not attempted to
fit the thickness by a two Gaussian components as others have done,
because the single Gaussian model was a good fit. Earlier higher
resolution VLA observations of NGC891 yielded an exponentially flaring
thin \HI\ layer. More recent studies of its gas thickness, using
exceptionally sensitive lower resolution WSRT observations 
with $\approx350$ hrs integration time, recover
extended filamentary high latitude \HI\ with unusual velocity
structure comprising $20\%$ of the total \HI\ mass
\citep{foss2005,fobss2004}.
By applying this dual gas component modelling process to the initial
$144$ hrs of WSRT data on NGC891, \citet{ssvdh1997} found that
the traditional warped and flared gas layer did not simulate the high
latitude inner disk gas. Indeed the best models of NGC891 tested in
this fashion used two disks each of constant thickness, with
FWHM$_{z,thin}(R) \sim 0.5$ kpc and FWHM$_{z,thick}(R) \sim 6$ kpc,
for the thin and thick disk respectively, where the thick disk
rotation lags that of the planar gas, and also displays some radial
infall motion. 

Similar two-component \HI\ modelling has been undertaken on several
less-inclined large galaxies: NGC2403 \citep{fvmso2002}, NGC4559
\citep{bfobbs2005}, NGC253 \citep{bofvdhs2005} and NGC6946
\citep{fobss2004,bofvdhs2005}.  Each of these galaxies is actively
star-forming with significant fractions of their total \HI\ mass
(10-30\%) exhibiting anomalous velocities forbidden by circular
rotation. As discussed in Sect.~\ref{sec:ch6-galaxies},
\citet{mw2003} also applied this model to UGC7321, a low surface
brightness galaxy in our sample.  In each of these galaxies, modelling
has found a thick gas layer combined with a thin constant thickness or
flaring thin gas layer.  With the exception of UGC7321, the high
stellar density, star formation rate and stellar disk area of these
galaxies are all conducive to ejecting a greater \HI\ mass fraction to
high latitudes, whether by discrete star formation or large scale
galactic fountains.  Despite the massive amount of \HI\ in the halos of
these galaxies, this inner high latitude \HI\ is probably more akin to
Galactic high velocity clouds than a continuous broad \HI\ component.

As discussed in paper I, a small amount of
high latitude filamentary \HI\ is observed in the larger galaxies in
our sample, with $V_{max} > 100$ \kms, at projected radii corresponding
to the inner disk. However this gas makes up only a small fraction ($\ll
5\%$) of the total gas mass and may not be in hydrostatic
equilibrium. For this reason we have not attempted to measure its mass
distribution. 

\subsection{Open issues}

As mentioned above, the \HI\ velocity dispersion measured in
face-on galaxies were all undertaken with low
spatial resolution of $1$ kpc, much larger than the scale of most
processes in the ISM which are likely to cause local variation in the
\HI\ velocity dispersion. Deep high spatial and spectral \HI\ observations
of such systems are lacking.

In theory, given high spatial and spectral resolution, it is possible
to determine the intrinsic shape of the \HI\ line spectrum, and whether
the velocity structure is best represented by one or more Gaussians,
or a non-Gaussian function. 
\citet{pr2006} found that the \HI\ line profile was best fit by a
two-Gaussian shape where the ratio of the integrated flux of the thick
and thin Gaussians was constant at all radii with good
signal-to-noise. The only edge-on galaxy with bright \HI\ that is close
enough to image at the required resolution is the small Scd galaxy, NGC4244,
studied by \citet{olling1996a}.

It is also important to test the isothermality of the \HI\
distribution by fitting the \HI\ velocity dispersion to XV maps formed
from gas at different heights above the galactic plane. This is important;
without further information about the $z$-dependence of the velocity
dispersion, we would need to assume $z$-isothermality when using the
equations of hydrostatic
equilibrium to derive the vertical potential of the dark matter halo.
Although
the spatial resolution of ESO274-G001 is sufficient to independently
measure the gas at several heights above the plane, additional \HI\
observations in compact array configurations would be needed.

Our \HI\ observations in paper I show
significant optical depth (assuming \HI\ spin temperatures of 200-400
K) in isolated locations of several galaxies in our sample. 
It would
be interesting to generalise the XV modelling to also fit the $z$
direction, as this would allow the \HI\ opacity to be accurately measured
across the galaxy.  Such a generalised model would measure the gas
density $\rho(R,z)$. By integrating over the line of sight, the total
\HI\ column density and \HI\ opacity consistent with the kinematic model
could be calculated and
compared to the observed \HI\ column density to check the calculated
\HI\ opacity. This requires higher spatial resolution data than we have
available in this study.

\section{Summary and conclusions}
\label{sec:ch6-summ}

Using our new XV modelling method detailed in paper II
we have successfully fit the
rotation curves of all eight galaxies in our sample; and the \HI\
velocity dispersion and surface density in all the galaxies except for
the inner disk of IC2531. The rotation curve was the most reliably fit
achieving an accuracy of roughly 2-4 \kms. The surface density was
also very reliably measured. The \HI\ velocity dispersion was more
sensitive to local variations and noise in the XV map, however despite
these challenges the uncertainty in the derived \HI\ velocity
dispersion is only about 1--1.5 \kms, which is comparable to the accuracy
obtained from NGC4244 by \citet{olling1996a}. The prevalence of bars in the 
larger galaxies and the increase in the mean velocity dispersion with maximum 
rotation speed suggests that bar buckling may be a common cause of 
widespread gas heating in disk galaxies. 

The measured rotation curves show that all galaxies in our sample
display differential rotation, even the smallest galaxies IC5052 and
ESO115-G021 which have peak rotation speeds of only $60$ \kms. The
increased velocity dispersion by 1--2.5 \kms\ observed in galaxies
with bars suggests that shocks caused by disk instabilities in bars
can cause gas heating.

Iterative fitting of the \HI\ layer thickness enabled us to reliably
measure the \HI\ flaring in 7 of the 8 galaxies in our sample, despite
FWHM$_{\theta}$ beamwidths of up to $1.3$ kpc and a low peak channel
map signal-to-noise of around $13$ for most of the galaxies in our
sample. The only galaxy we were unable to reliably fit was ESO138G-014
which was missing large scale \HI\ structure due to an absence of short
baseline observations, and also suffered low peak signal-to-noise
($15$) and a large telescope beam ($0.965$ kpc). For galaxy
observations with only two of these three handicaps it was possible to
iteratively recover a flaring model that fit the observed \HI\ channel
map distribution. UGC7321 is the most accurately measured flaring
profile in our sample with errors of only 15-50 pc over most of the
radial extent. 

The flaring profiles show an impressive degree of the symmetry between 
flaring measurements on either side of the galaxy --- this is most clearly 
seen in UGC7321. Axial symmetry of the gas layer
thickness is expected as in the absence of local variations of the
stellar surface density or gas velocity dispersion, the vertical
flaring should reflect the smooth symmetric gravitational potential of
the dark matter halo. The exceptionally smooth flaring profile of
UGC7321 shows that, on the $0.7$ kpc scale of telescope beam, the
stellar surface density and dark matter halo gravitational potential
are also smooth.

The most remarkable feature of \HI\ flaring
profiles is a common shape displaying a linear
increase at low radii and steepening to an exponential increase in the
outer disk. The steepness of the inner flaring should be dominated by
the radial surface density, while at the stellar disk truncation
radius the flaring becomes exponential and is dominated by the dark
halo shape. This typical flaring shape is found in all of the galaxies
except ESO138-G014 which was incompletely imaged, and ESO274-G001
which appears to have a more linear flaring profile. With the
exception of ESO146-G014, all of the less massive galaxies ($V_{max} <
100$ \kms) display steep inner flaring profiles.  The four larger
galaxies ($V_{max} > 100$ \kms), excluding ESO138-G014, display much
shallower inner flaring profiles.

The high sensitivity of the UGC7321 observations allowed us to discern
variations from azimuthal symmetry of both the \HI\ midplane position
and the \HI\ thickness. We show that at radii from
approximately $2.5$ to $6$ kpc along the Y=0 axis, the FWHM$_z$ gas thickness
varies by up to $500$ pc from the azimuthally averaged flaring in
those radial annuli. At this same locations along the Y=0
axis the gas midplane tilts, dropping $\sim$200 pc below the average
disk midplane on the left-hand side, and rising $\sim$200 pc above the
midplane on the right-hand side. These local deviations are considered to
be best observational evidence for a bar in an edge-on disk galaxy.

The next paper in this series will be devoted to an analysis of the data
on UGC7321 as a first attempt at deriving constraints on the
flattening of a dark halo. 

\begin{acknowledgements}

We are very grateful to Albert Bosma who contributed greatly to
initiating this project. He pointed out that \HI\ flaring studies are
best done on edge-on galaxies with low maximum rotational velocities,
and we used an unpublished Parkes \HI\ survey of edge-on galaxies by
Bosma and KCF when selecting our galaxies.  JCO thanks E. Athanassoula,
M.  Bureau, R.  Olling, A. Petric and J. van Gorkom for helpful
discussions.  JCO is grateful to B. Koribalski, R. Sault, L.
Staveley-Smith and R.  Wark for help and advice with data reduction and
analysis.
We thank the referee, J.M. van der Hulst, for
his careful and  thorough reading of the manuscripts of this series of papers and
his helpful and constructive remarks and suggestions.

\end{acknowledgements}

\bibliographystyle{aa}

\end{document}